\documentclass[pdflatex,sn-mathphys-num]{sn-jnl}% Math and Physical Sciences Numbered Reference Style
\usepackage{graphicx}%
\usepackage{multirow}%
\usepackage{amsmath,amssymb,amsfonts}%
\usepackage{amsthm}%
\usepackage{mathrsfs}%
\usepackage[title]{appendix}%
\usepackage{xcolor}%
\usepackage{textcomp}%
\usepackage{manyfoot}%
\usepackage{booktabs}%
\usepackage{algorithm}%
\usepackage{algorithmicx}%
\usepackage{algpseudocode}%
\usepackage{listings}%

\usepackage[capitalize]{cleveref}
\usepackage{physics}
\usepackage{mathtools}
\usepackage{ulem}
\theoremstyle{thmstyleone}%
\newtheorem{theorem}{Theorem}%  meant for continuous numbers
\newtheorem{proposition}[theorem]{Proposition}% 
\newtheorem{lemma}{Lemma}

\newtheorem{conjecture}{Conjecture}

\newtheorem{corollary}{Corollary}

\theoremstyle{thmstyletwo}%
\newtheorem{remark}{Remark}%

\theoremstyle{thmstylethree}%
\newtheorem{definition}{Definition}%

\begin{document}

\title[Article Title]{Computational Certified Deletion Property of Magic Square Game and its Application to Classical Secure Key Leasing}


\author*[1]{\fnm{Duo} \sur{XU}}\email{xu.duo.x3@s.mail.nagoya-u.ac.jp}

\author[2]{\fnm{Yuki} \sur{Takeuchi}}\email{Takeuchi.Yuki@bk.MitsubishiElectric.co.jp}


\affil*[1]{\orgdiv{Graduate School of Informatics}, \orgname{Nagoya University}, \orgaddress{\street{Furo-cho}, \city{Nagoya-City}, \postcode{464-8601}, \state{Aichi-Prefecture}, \country{Japan}}}

\affil[2]{\orgdiv{Information Technology R\&D Center}, \orgname{Mitsubishi Electric Corporation}, \orgaddress{\street{5-1-1 Ofuna}, \city{Kamakura}, \postcode{247-8501}, \state{Kanagawa}, \country{Japan}}}

% \affil[3]{\orgdiv{Department}, \orgname{Organization}, \orgaddress{\street{Street}, \city{City}, \postcode{610101}, \state{State}, \country{Country}}}


\abstract{We present the first construction of a computational Certified Deletion Property (CDP) achievable with classical communication, derived from the compilation of the non-local Magic Square Game (MSG). We leverage the KLVY compiler to transform the non-local MSG into a 2-round interactive protocol, rigorously demonstrating that this compilation preserves the game-specific CDP. Previously, the quantum value and rigidity of the compiled game were investigated. We emphasize that we are the first to investigate CDP (local randomness in [Fu and Miller, Phys. Rev. A 97, 032324 (2018)]) for the compiled game. Then, we combine this CDP with the framework [Kitagawa, Morimae, and Yamakawa, Eurocrypt 2025] to construct Secure Key Leasing with classical Lessor (cSKL). SKL enables the Lessor to lease the secret key to the Lessee and verify that a quantum Lessee has indeed deleted the key. In this paper, we realize cSKL for PKE, PRF, and digital signature. Compared to prior works for cSKL, we realize cSKL for PRF and digital signature for the first time. In addition, we succeed in weakening the assumption needed to construct cSKL.}




\maketitle

\maketitle

\section{Introduction}
Non-local games have been a powerful tool to design cryptographic protocols such as device-independent quantum key distribution\cite{ABG07,VV14,ARV19,JMS20}, delegation of quantum computation\cite{RUV12,FV14,GKW15,McK16,NV16,FHM18}. We refer the curious readers to \cite{GKK19}. Non-local games are non-interactive cooperative games that comprise a Referee and two players, Alice and Bob. The Referee samples a pair of questions $q_A,q_B$ and sends $q_A$ to Alice, $q_B$ to Bob. Then, Alice and Bob produce the answers $a,b$, respectively. The Referee checks whether Alice and Bob win the game by checking a predicate $V(q_A,q_B,a,b)$. Alice and Bob are not allowed to communicate with each other during the game. Non-local games provide ways to self-testing the quantum device's statistical correlation among $q_A, q_B, a, b$. 
    \footnote{In this paper, we restrict the non-local games to $2$-player non-local games. General $k$-player non-local games can be defined similarly, but we do not use them in this work, and to avoid confusion, we omit them.}

However, non-local games require the two players to be spatially separated, which is hard to enforce in practice. Fortunately, there is a proposal to use cryptographic separation instead of spatial separation\cite{KLVY22}. Their compiler compiles a non-local game into a 2-round interactive protocol consisting of 2 parties, where the verifier is classical polynomial-time bounded and the prover is quantum polynomial-time bounded. Many works were done to prove that the prover can not make the verifier accept with significantly higher probability than the winning probability in the original non-local game \cite{NZ23,CMM+24,MPW24,CFNZ25,KPR＋25,KMP+25}. A line of work demonstrated that the compiled game is useful for building delegation in quantum computation, certified randomness, etc\cite{NZ23,CMM+24,MNZ24}. 

In this paper, we prove that the compilation preserves the so-called Certified Deletion Property for a specific non-local game, the Magic Square Game. Then, we show that the compiled game is capable of building secure key leasing with a classical lessor, which is new to the previous work.

Secure key leasing (SKL) is a quantum cryptographic primitive proposed recently\cite{KN22}. The primitive usually consists of a Lessor and a Lessee. The Lessor is the owner of the secret keys, who wants to lease the key to an untrusted Lessee. After the Lessee gets the secret key, it can use the key to perform tasks that are otherwise impossible without the secret key. For example, SKL for the public key encryption (PKE) enables the Lessee to decrypt the ciphertexts generated by the corresponding public key. At a later point in time, the Lessor can ask the Lessee to return the secret key or delete the secret key, where the retrieval or the deletion of the key can be verified by the Lessor. The Lessee lost the ability to use the secret key after it returned or deleted the secret key. 

SKL enables the Key Leasing without a key update. For example, Key Leasing for PKE can be achieved by the following method with key update in classical cryptography. Let $(pk,sk)$ be a pair of public key and secret key. The owner of the secret key $sk$ notifies every holder of $pk$ that $pk$ is obsolete and does not use it to encrypt any plaintext. This mechanism becomes inefficient when the public key is distributed to a large number of parties.

SKL has received much attention since its proposal\cite{KN22,APV23,AKN+23,AHH24,MPV23,CGJL23,PWY+25,KNP25}. In this section, we want to emphasize two essential aspects of SKL.

{\bf SKL with classical lessor} In \cite{CGJL23,PWY+25}, they showed that SKL can be implemented between a Lessor with only classical computers and a Lessee with quantum computers. The classical Lessor uses their cryptographic protocol to enforce the Lessee to prepare a quantum state without knowing any information about the secret key. The quantum state serves the role of the secret key. When the Lessor asks the Lessee to delete the key, the Lessee measures the quantum state and can no longer use the secret key. SKL with classical lessor is essential for practical use. Since the quantum computer and the quantum communication are expensive, it is always desirable for the Lessor to get access to SKL with only classical communication and classical computers.

{\bf Modular construction of SKL protocol} In \cite{KMY24}, they proposed a framework to build SKL protocols. They proposed a novel concept, the Certified Deletion Property, in the work. This change separates the protocol into the part to certify the deletion of the secret key and the part to act as a valid secret key. Due to the simplicity of their protocols, they realized SKL for public key encryption (PKE), pseudo-random functions (PRFs) with minimal assumptions, and SKL for digital signature (DS) assuming the hardness of the short interger solution (SIS) problem. 

\subsection{Our Results}
\begin{figure}[tbp]
    \centering
    \includegraphics[width=0.5\linewidth]{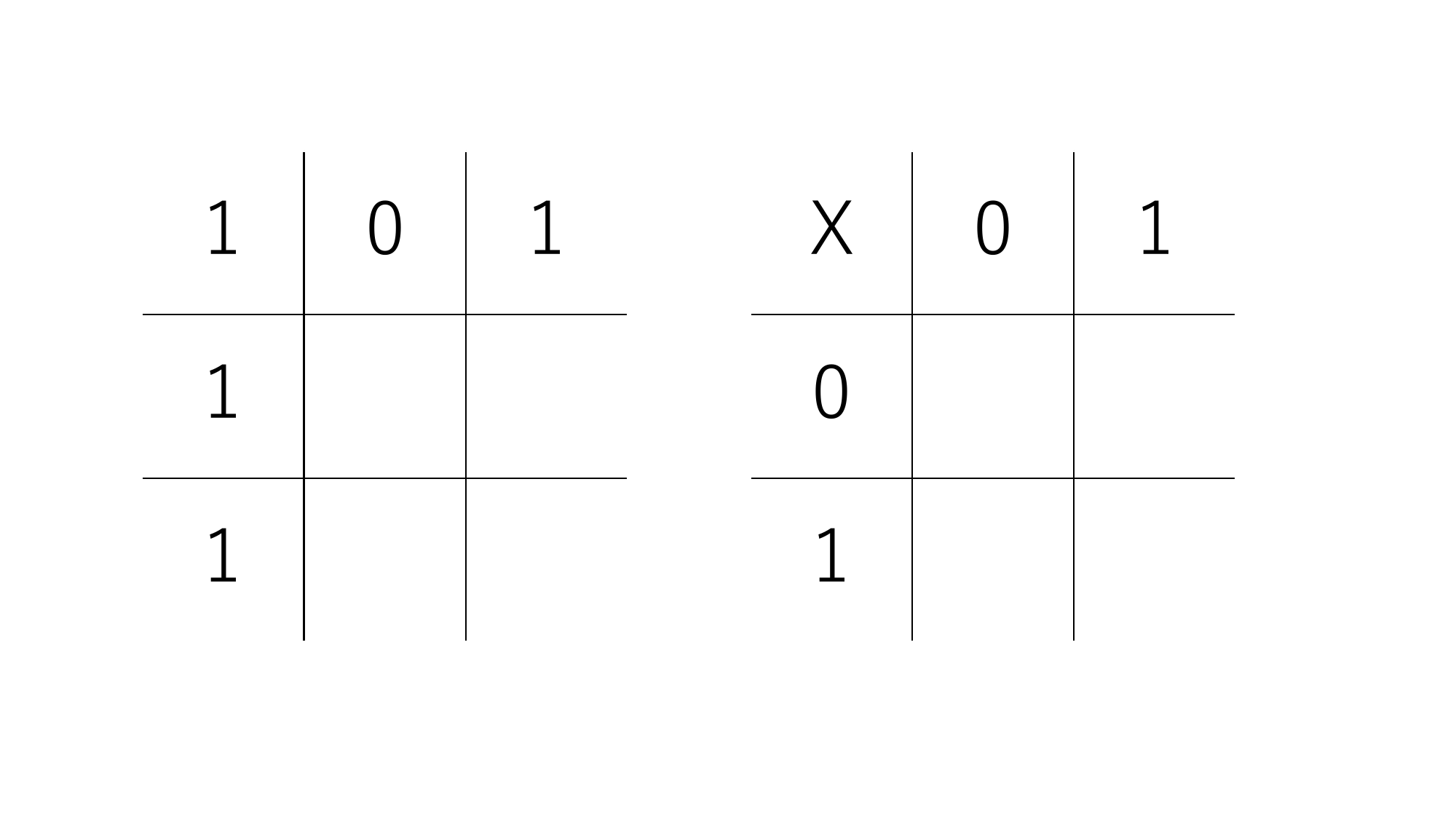}
    \caption{An example of the magic square game when $r=1$ and $c=1$. We show a valid answer to the left. To the right, we show an invalid answer. The reason the right is invalid is that Alice fills $X = 0$ and Bob fills $X=1$, where their answers do not match on the upper-left grid.}
    \label{fig:magic-square}
\end{figure}
{\bf The first computational certified deletion property:} We obtain the first computational certified deletion property with classical communication. We obtain the computational certified deletion property by compiling the magic square game (MSG) using the compiler from \cite{KLVY22}. MSG is a game as follows:
\begin{enumerate}
    \item The game utilizes a 3x3 magic square (see \cref{fig:magic-square}). The Referee samples $r,c \in \{1,2,3\}$ uniformly at random, where $r$ indicates a row in the magic square and $c$ indicates a column in the magic square.
    \item The Referee sends $r$ and $c$ to Bob and Alice, respectively. Bob and Alice are not allowed to communicate during the game.
    \item Bob sends its answer $b\in \{0,1\}^3$ to the Referee, which corresponds to the $3$ grids of the $r$-th row. Alice sends its answer $a\in \{0,1\}^3$ to the Referee, which corresponds to the $3$ grids of the $c$-th column.
\end{enumerate}
Alice and Bob win the game if the parity of $a$ is $1$ and the parity of $b$ is $0$, and $a[r] = b[c]$, where $a[r]$ is the $r$-th bit of $a$ and $b[c]$ is the $c$-th bit of $b$. Then, the following property, which can be viewed as a non-local certified deletion property, holds.
\footnote{The name ``non-local certified deletion property'' states that the certified deletion property appears in a non-local game. The original name ``local randomness'' from \cite{FM18} focuses on the fact that $b$ is visible to only Bob. Thus $b$ is local in the game.}

\begin{lemma}[Informal, local randomness\cite{FM18}]
    Given the information of $r$ after the game is over, Alice cannot guess $b$ correctly with probability $1$ conditioned on Alice and Bob winning the magic square game with probability $1$. 
\end{lemma}
The original MSG consists of two Provers, Alice and Bob. We use the KLVY compiler\cite{KLVY22} to obtain a $2$-round post-quantum argument with a single Prover. The compiled game is as follows:
\begin{itemize}
    \item The verifier encrypts $r$ with Quantum Fully Homomorphic Encryption (QFHE) and sends the ciphertext $ct$ to the prover. Then, the prover sends an encrypted answer $ct_b$ to the verifier. QFHE allows the prover to execute circuits and obtain an encrypted output, without knowing the underlying plaintext.
    \item The verifier then sends $c$ to the prover. The prover sends an answer $a$ in plain.
    \item The verifier decrypts $ct_b$ to obtain $b$. Then, the verifier decides whether to accept based on $r,c,b,a$ as in MSG.
\end{itemize}
We proved that the certified deletion property is preserved after compilation.
\begin{lemma}[Informal, one-shot computational certified deletion property]
    \label{lem:one-shot-comp-cdp}
    The verifier reveals $r$ to the prover after the compiled game; the prover cannot guess $b$ correctly with probability $1$. 
\end{lemma}
Finally, we obtain a computational certified deletion property with ${\rm negl}(\lambda)$ winning probability using the parallel repetition heuristic. We point out that the parallel repetition for post-quantum arguments is hard to prove, so we call this a ``heuristic''.
\begin{conjecture}[Informal version of Conjecture \ref{conj:formal-parallel-rep}]
    \label{conj:parallel-rep}
    If we repeat the compiled game in \cref{lem:one-shot-comp-cdp} for $k$ times in parallel, the winning probability of any QPT adversary is $\rm{negl}(k)$.
\end{conjecture}

Previously, many studies have been done to investigate the winning probability of the QPT adversary in the compiled game and the rigidity in the compiled game \cite{NZ23,CMM+24,BVB+24,MPW24,CFNZ25,KMP+25,KPR＋25}. But we are the first to investigate the local randomness/certified deletion property in the compiled game.

{\bf New Classical Secure Key Leasing Protocols:} We combine the computational certified deletion property with the framework from \cite{KMY24} to obtain PKE-cSKL, PRF-cSKL, DS-cSKL. {\bf We apply the compiled game technique from \cite{KLVY22} to obtain SKL protocols with classical lessor for the first time.} Previously, the compiled game technique has been applied only to obtain the delegation of quantum computation, proof of quantumness, certified randomness, etc.

\begin{lemma}[Informal, classical secure key leasing]
    Assuming the existence of claw-state generators (CSGs) and Conjecture \ref{conj:parallel-rep}, we have Secure Key Leasing for PKE, PRFs, and DS, with a classical lessor.
\end{lemma}

\begin{table}[!tbp]
    \centering
    \begin{tabular}{c|c|c|c|c|p{4.5cm}}
         Method & PKE-SKL & PRF-SKL & DS-SKL & Lessor & Cryptographic Assumptions \\
         \hline 
         Ours & Yes & Yes & Yes & Classical & Claw-State Generators + PKE(for PKE-SKL)/OWFs(for PRF-SKL)/SIS(for DS-SKL) \\
         \cite{KMY24} & Yes & Yes & Yes & Quantum & PKE(for PKE-SKL)/OWFs(for PRF-SKL)/SIS(for DS-SKL) \\ 
         \cite{CGJL23} & Yes & No & No & Classical & the hardness of LWE with exopential modulus \\
         \cite{PWY+25} & Yes & No & No & Classical & the hardness of LWE with polynomial modulus
    \end{tabular}
    \caption{Comparison between our SKL protocols with prior works \cite{KMY24,CGJL23,PWY+25}}
    \label{tab:results-skl}
\end{table}
Then, we compare our SKL protocols with those in the prior works in \cref{tab:results-skl}. We realized SKL for the same primitives as in \cite{KMY24}, but our protocols require only classical lessor. As for \cite{CGJL23,PWY+25}, we realized PRF-cSKL and DS-cSKL, which are not implemented in their works. However, we have to point out that their PKE-cSKL is actually an FHE-cSKL, which is not realized in \cite{KMY24} or this work. {\bf To conclude, our protocols combine the flexibility of the framework in \cite{KMY24} and the merit of having a classical lessor}. Then, we want to compare our assumptions with the prior works. We point out that Claw-State Generators (CSGs) can be constructed from a wide variety of assumptions, including LWE with both polynomial and exponential moduli\cite{BCM+18,PWY+25} and the hard problems on cryptographic group actions \cite{AMR22,GV24}. So, our protocol requires weaker assumptions than \cite{CGJL23,PWY+25} do. On the other hand, \cite{KMY24} implements PKE-cSKL and PRF-cSKL with the minimal assumptions. 
    \footnote{PRF-cSKL implies the plain PRFs without secure key leasing. It is well known that PRFs are equivalent to one-way functions.}
Our protocol uses stronger assumptions than \cite{KMY24} does. We argue that this strengthening of assumptions is pretty much unavoidable, as the classical Secure Key Leasing implies Proof of Quantumness. To construct a proof of quantumness using PKE or one-way functions remains an open problem. {\bf Thus, our protocols give an improvement over the cryptographic assumptions.}

\subsection{Technical Overview}
Our method can be divided into two parts. 

{\bf A computational certified deletion property:} We note that to build SKL, it is sufficient for the adversary to be unable to recover $b$ generated by computational basis measurement, which is produced when $r = 2$.
\begin{lemma}[Informal, one-shot computational certified deletion property]
    The verifier reveals $r$ to the prover after the compiled game; the prover cannot guess $b$ correctly with probability $1$ conditioned on $r=2$. 
\end{lemma}
Informally, the malicious prover can ignore $r\neq 2$. Thus, the security game of the computational certified deletion property can be compiled from the following non-local game $\sf CCD$:
\begin{enumerate}
    \item The Referee samples $r,c \in \{1,2,3\}$ uniformly at random where $r$ indicates a row in the magic square and $c$ indicates a column in the magic square.
    \item The Referee sends $r$ and $c$ to Bob and Alice, respectively. Bob and Alice are not allowed to communicate during the game.
    \item Bob sends its answer $b\in \{0,1\}^3$ to the Referee, which corresponds to the $3$ grids of the $r$-th row. Alice sends its answer $a\in \{0,1\}^3$ , which corresponds to the $3$ grids of the $c$-th column, and $b^\prime \in \{0,1\}^3$ to the Referee.
\end{enumerate}
Alice and Bob win the game iff $MSG(r,c,a,b) = \top$ and $b = b^\prime$ when $r=2$. $MSG(r,c,a,b)$ is the winning condition for the original MSG. We prove that the above game's winning probability is less than $1$, which we combine with the following lemma
\begin{lemma}[Theorem 6.1 from \cite{KMP+25}]
    Let $G$ be any two-player non-local game and let $S$ be any QPT strategy for the compiled game $G_{comp}$. We denote the quantum commuting value of $G$ as $w_{qc}(G)$, which can be considered as the maximum winning probability, informally. We denote the winning probability of $G_{comp}$ using quantum strategy $S$ as $w_{\lambda}(G_{comp},S)$, when the security parameter is $\lambda$. Then it holds that
    \begin{equation}
        \lim_{\lambda \rightarrow \infty} \sup w_{\lambda}(G_{comp}, S) \leq w_{qc}(G)
    \end{equation}
\end{lemma}
We can take a constant $\lambda_c$ as the security parameter for QFHE such that the compiled game's winning probability is less than a constant $w < 1$. 

The one-shot certified deletion property is not sufficient for cryptographic applications. So, we strengthen it using the parallel repetition heuristics. Unfortunately, we are unable to prove the parallel repetition formally. As a result, we assume Conjecture \ref{conj:parallel-rep} to obtain the soundness for the parallel repeated certified deletion property.

\begin{proposition}
    Assuming Conjecture \ref{conj:parallel-rep}, the parallel repeated certified deletion game of the Magic square game is sound. It is in the sense that for any QPT adversary, the probability that the adversary wins is negligible.
\end{proposition}

{\bf Plug the computational CDP into \cite{KMY24}'s framework:} To utilize the framework from Kitagawa et al. \cite{KMY24} as adapted in this paper, we leverage two main components: (1) The computational certified deletion property (CDP), which is detailed in earlier sections. (2) Key hiding for computational basis. We explain the second requirement using PKE-SKL from\cite{KMY24} as an example. They use key state $\ket{sk_1} \otimes \cdots \otimes \ket{sk_{n}}$ and a string $\theta \in \{0,1\}^n$ to indicate the basis for each $\ket{sk_i}$. $\ket{sk_i}$ is either $\ket{0 \sf PKE.sk_{i,0}}$ or $\ket{1 \sf PKE.sk_{i,1}}$ for $\theta[i]=0$, or $\frac{1}{\sqrt 2}(\ket{0 \sf PKE.sk_{i,0}} + \ket{1 \sf PKE.sk_{i,1}})$ or $\frac{1}{\sqrt 2}(\ket{0 \sf PKE.sk_{i,0}} - \ket{1 \sf PKE.sk_{i,1}})$ for $\theta[i]=1$. The Lessee cannot know both $\sf PKE.sk_{i,0}$ and $ \sf PKE.sk_{i,1} $ at the same time. Their proof relies on this fact to produce a special challenge without affecting the winning probability of their security game. However, our protocol uses classical communication and cannot control the form of the leased key strictly. Thus, we have developed an alternative way to make this proof work.

To realize the key hiding, we incorporate Claw-State Generators (CSGs). As defined in \cite{BK24}, CSGs enable a classical Lessor (verifier) to remotely prepare a quantum state for a quantum Lessee (prover), for instance, $\frac{1}{\sqrt 2}(\ket{0 x_0}+(-1)^z\ket{1 x_1})$, without explicitly revealing the classical strings $x_0, x_1$ to the Lessee. A key security property of CSGs, known as Search Security, guarantees that any malicious Lessee cannot simultaneously obtain both $x_0$ and $x_1$. This ensures that if the Lessee, through its quantum state, gains knowledge related to $x_0$, it remains ignorant of $x_1$, and vice-versa.
During our key generation protocol (PKE-cSKL.KG), the Lessor transfers the classical secret key components to the Lessee in a masked form. This is achieved by generating a pair of $x_{j,0}, x_{j,1}$ for each $j$ and sending values like ${\sf Ext}(x_{j,0}, r_{j}) \oplus {\sf PKE.sk}_{j,0}$ and ${\sf Ext}(x_{j,1}, r_{j}) \oplus {\sf PKE.sk}_{j,1}$
    \footnote{$r_{j}$ is a random seed.}
Here, ${\sf Ext}$ is a randomness extractor, defined as part of the CSG properties, used to transform the $x$ values into uniformly random strings, effectively masking the PKE.sk components.

With these two requirements effectively satisfied—by leveraging the compiled Magic Square Game for CDP and CSGs (along with classical blind quantum computing) for key hiding we can robustly adapt the security proofs from Ref. \cite{KMY24} to our PKE-cSKL (and other cSKL) protocols.

\begin{table*}[tbp]
    \centering
    \begin{tabular}{p{2cm}|p{4.5cm}|p{4.5cm}}
       Feature  &  This work & Kitagawa et al. \cite{KLYY25}\\
       \hline
       Computation-al Assumption  & Claw-State Generators (CSG) + PKE(for PKE-SKL)/SIS(for DS-SKL) & The hardness of LWE \\
       Unproven Conjecture & The soundness of parallel repetiton & None \\
       Round Complexity & Polynomial Rounds & 1 round for Setup followed by Non-interactive KeyGen (for PKE-cSKL)/1-round KeyGen (for PRF-cSKL and DS-cSKL)
    \end{tabular}
    \caption{The comparison between this work and \cite{KLYY25}}
    \label{tab:concurrent-work}
\end{table*}
\subsection{The Concurrent Work}
In Ref. \cite{KLYY25}, they propose PKE-cSKL, PRF-cSKL, and DS-cSKL as well. Both this work and their work utilize the framework by \cite{KMY24}, which makes the key generation protocol modular. We compare the difference in \cref{tab:concurrent-work}.

\subsection{The Organization of This Paper}
In \cref{sec:prem}, we introduce the notions and preliminaries that will be used in the paper. In \cref{sec:compiled-CD-MSG}, we propose a helpful tool, the compiled MSG (see \cref{dfn:compiled-msg}), and define the computational Certified Deletion Property (see \cref{dfn:comp-CDP}). In the same section, we give a proof of the computational Certified Deletion Property. In \cref{sec:dfn-cSKL}, we state the syntax of public-key encryption with classical Secure Key Leasing (PKE-cSKL) and the security definition. In the same section, we state our PKE-cSKL construction. In \cref{sec:pke-ow-vra-proof}, we prove the security of our PKE-cSKL. For pseudo-random functions with classical Secure Key Leasing (PRF-cSKL) and digital signature with classical Secure Key Leasing (DS-cSKL), we state the syntax, the security definition, and the proof in \cref{apdx:PRF-DS-cSKL}.

\section{Preliminary}
\label{sec:prem}

\subsection{Notions}
\begin{itemize}
\item[] {$\bf h(x)$} For any $x\in\{0,1\}^*$, $h(x)$ denotes the number of $1$s appeared in $x$.

\item[] {$\bf \mathrm{par}(x)$} For any $x\in\{0,1\}^*$, $par(x) = h(x)\mod 2$.

\item[] {$\bf [n]$} For any $n\in\mathbb N$, $[n] = \{i\in\mathbb N| i<n\}$. {\bf Note that our definition is different from the normal definition!} This makes our notion more neat.

\item[] {$\bf \approx_\alpha$} We use $a \approx_\alpha b$ as a shorthand for 
\begin{equation}
    |a - b| \leq \alpha
\end{equation}
\item[] {$s[i]$} Let $s$ be a bit string and $i \in \mathbb N$, $s[i]$ represents the $i$-th bit of $s.$
\item[] {$X(e_x)$ and $Z(e_z)$} We use $X(e_x)$ to denote the operator that applies Pauli $X$ to the $i$-th qubit if $e_x[i]=1$ and being trivial on the other qubits. $Z(e_z)$ is defined similar to $X(e_x)$, except for substituting $X$ with Pauli $Z$.
\end{itemize}

\subsection{Magic Square Game}
\label{sec:prem-msg}
The magic square game is a 2-player non-local game described as follows. 
\begin{definition}[Magic Square Game]
    \label{dfn:MSG}
    The game is played as follows:
    \begin{enumerate}
        \item The referee samples $x,y\in\{1,2,3\}$. Then, he sends $x, y$ to Alice and Bob, respectively.
        \item Alice sends $a\in\{0,1\}^3$ to the referee. Bob sends $b\in\{0,1\}^3$ to the referee.
    \end{enumerate}
    Alice and Bob win the game if and only if 
    \begin{equation}
        \mathrm{par}(a) = 0 \quad 
        \mathrm{par}(b) = 1\quad
        a[y] = b[x]
    \end{equation}
    {\bf We use $MSG(x,y,a,b)$ to denote the conditions above in the remaining part of this paper.}
\end{definition}

\begin{definition}[The maximum winning probability of non-local games]
    The quantum value $w_{q}$ is defined as the maximum winning probability in the following situation:
    \begin{itemize}
        \item Alice and Bob share a quantum state $\ket{\psi} \in {\cal H}_A \otimes {\cal H}_B$ in advance.
        \item Alice (resp. Bob) generates its answer $a$ (resp. $b$) by using measurements $\{{\cal M}_x\}_x$ (resp. $\{{\cal N}_y\}_y$). The measurements $\{{\cal M}_x\}_x$ and $\{{\cal N}_y\}_y$ act on ${\cal H}_A$ and ${\cal H}_B$, respectively.
    \end{itemize}
    The quantum commuting value $w_{qc}$ is defined as the maximum winning probability in the following situation:
    \begin{itemize}
        \item Alice and Bob share a quantum state $\ket{\psi} \in {\cal H}$ in advance.
        \item Alice (resp. Bob) generates its answer $a$ (resp. $b$) by using measurements $\{{\cal M}_x\}_x$ (resp. $\{{\cal N}_y\}_y$). The measurements $\{{\cal M}_x\}_x$ and $\{{\cal N}_y\}_y$ act on ${\cal H}$. \bf The measurements are commutative instead of tensored.
    \end{itemize}
\end{definition}

For any classical Alice and Bob, they cannot win the game with probability larger than $8/9$. For quantum Alice and Bob, they can win the game with probability $1$ using the strategy in Table \ref{tab:msg-strategy}. Though the strategy is not unique, we will use it for the sake of convenience.
\begin{table}[btp]
    \begin{tabular}{c|c|c}
        $XI$ & $IX$ & $XX$ \\
        \hline
        $IZ$ & $ZI$ & $ZZ$ \\
        \hline
        $-XZ$ & $-ZX$ & $YY$ 
    \end{tabular}
    \label{tab:msg-strategy}
    \caption{When Alice (resp. Bob) receives $x$ (resp. $y$) as their question, they measure the observables on $x$-th column (resp. $y$-th row) and outputs the outcomes as their answer.}
\end{table}

\subsection{Cryptographic Tools}
\label{sec:prem-crypto}

\begin{definition}[The definition of PKE with IND-CPA security\cite{IntroToModernCrypt}]
    \label{dfn:IND-CPA-PKE}
    Let $\lambda$ be the security parameter. A public-key encryption (PKE) scheme is a tuple of algorithms $({\sf PKE.KG}, {\sf PKE.Enc}, {\sf PKE.Dec})$ as follows:
    \begin{itemize}
        \item {${\sf PKE.KG}(1^\lambda) \rightarrow ({\sf pk}, {\sf sk})$} is a PPT algorithm which takes as input the security paramter. It outputs a pair of secret key {\sf sk} and public key {\sf pk}.
        \item {${\sf PKE.Enc}({\sf pk}, b)\rightarrow {\sf ct}_b$} is a PPT algorithm. The algorithm takes as input a public key ${\sf pk}$ and a single bit $b \in \{0,1\}$. It outputs the ciphertext ${\sf ct}_b$.
        \item {${\sf PKE.Dec}({\sf sk}, {\sf ct}_b) \rightarrow b^\prime$} is a PPT algorithm. The algorithm takes as input a secret key ${\sf sk}$ and a ciphertext ${\sf ct}_b$. It outputs a single bit $b^\prime$ as the decryption result.
    \end{itemize}

    The PKE scheme must satisfy the correctness as follows.
    
    {\bf correctness:} For $b \in \{0,1\}$, we have
    \begin{equation}
        \Pr[
        b \neq b^\prime:
        \begin{aligned}
            &({\sf pk}, {\sf sk}) \leftarrow {\sf PKE.KG}(1^\lambda) \\
            & {\sf ct}_b \leftarrow {\sf PKE.Enc}({\sf pk}, b)\\
            & b^\prime \leftarrow {\sf PKE.Dec}({\sf sk}, {\sf ct}_b)\\
        \end{aligned}
        ] = {\rm negl}(\lambda)
    \end{equation}
    Then, the PKE scheme must satisfy IND-CPA security as well.
    
    {\bf IND-CPA security:} For any QPT adversary $A^\lambda$
    \begin{equation}
        \begin{aligned}
            |
        &\Pr[A^\lambda({\sf pk}, {\sf ct}_0)=1:\begin{aligned}
            &({\sf pk}, {\sf sk}) \leftarrow {\sf PKE.KG}(1^\lambda) \\
            &ct_0 \leftarrow {\sf PKE.Enc}({\sf pk}, 0) \\
            &ct_1 \leftarrow {\sf PKE.Enc}({\sf pk}, 1) \\
        \end{aligned}] \\
        &- \Pr[A^\lambda({\sf pk}, {\sf ct}_1)=1:\begin{aligned}
            &({\sf pk}, {\sf sk}) \leftarrow {\sf PKE.KG}(1^\lambda) \\
            &ct_0 \leftarrow {\sf PKE.Enc}({\sf pk}, 0) \\
            &ct_1 \leftarrow {\sf PKE.Enc}({\sf pk}, 1) \\
        \end{aligned}]
        | = {\rm negl}(\lambda)
        \end{aligned}    
    \end{equation}
\end{definition}

We import the definition of claw-state generator from \cite{BK24}.
\begin{definition}[claw-state generator (CSG)]
    \label{dfn:CSG}
    A claw-state generator (CSG) is a protocol that takes place between a PPT sender $S(1^\lambda,n)$ and a QPT receiver $R(1^\lambda, n)$ as follows:
    \begin{equation}
        ((x_0,x_1,z), \ket{\psi}) \leftarrow \langle S(1^\lambda, n), R(1^\lambda, n) \rangle
    \end{equation}
    where $x_0,x_1 \in \{0,1\}^n$ and $z \in \{0,1\}$ are the sender's output, $\ket{\psi}$ is the receiver's output. CSG must satisfy the following properties:
    
    {\sf Correctness}: Let 
    \begin{equation}
        \begin{aligned}
            \Pi \coloneqq &\sum_{x_0 \neq x_1, z\in \{0,1\}} \ket{x_0,x_1,z}\bra{x_0,x_1,z} \\
            &\otimes \frac{1}{2}(\ket{0,x_0}+(-1)^z \ket{1, x_1}) (\bra{0,x_0}+(-1)^z \bra{1, x_1})
        \end{aligned}
    \end{equation}
    Then
    \begin{equation}
        E[\| \Pi \ket{x_0,x_1,z}\ket{\psi} \|: ((x_0, x_1, z), \ket{\psi}) \leftarrow \langle S(1^\lambda,n), R(1^\lambda, n) \rangle] = 1-{\rm negl}(\lambda)
    \end{equation}

    {\sf Search Security}: For any QPT adversary $\{{\sf Adv}_\lambda\}_{\lambda \in \mathbb N}$:
    \begin{equation}
            \Pr[ x_{\sf Adv} = (x_0, x_1): ((x_0, x_1, z), x_{\sf Adv}) \leftarrow \langle S(1^\lambda, n), {\sf Adv}_\lambda\rangle] = {\rm negl}(\lambda)
    \end{equation}
    
    {\sf Indistinguishability Security}: For any QPT adversary $\{{\sf Adv}_\lambda\}_{\lambda \in \mathbb N}$ and any $i \in [n]$:
    \begin{equation}
        |\Pr[b_{\sf Adv} = x_0[i] \oplus x_1[i]: ((x_0, x_1, z), b_{\sf Adv}) \leftarrow \langle S(1^\lambda, n), {\sf Adv}_\lambda\rangle] - 1/2 | = {\rm negl}(\lambda)
    \end{equation}

    {\sf Randomness Extraction}: There exists a deterministic classical algorithm ${\sf Ext}(x,r)$ where $x\in \{0,1\}^n, r\in \{0,1\}^{l(n)}$, for any QPT adversary  $\{{\sf Adv}_\lambda\}_{\lambda \in \mathbb N}$:
    \begin{equation}
        |\Pr[{\sf EXTRACT}(0)=1]-\Pr[{\sf EXTRACT}(1)=1]|={\rm negl}(\lambda)
    \end{equation}
    where ${\sf EXTRACT}(b)$ is as follows:
    \begin{enumerate}
        \item The challenger $C$ and the adversary ${\sf Adv}_\lambda$ runs $(x_0, x_1, z) \leftarrow \langle S(1^\lambda,n), R(1^\lambda,n) \rangle$. The challenger plays the role of the sender $S$. The adversary plays the role of the receiver $R$.
        \item The adversary sends $x^\prime \in \{0,1\}^n$ and $c \in \{0,1\}$ to the challenger.
        \item The challenger outputs $0$ as the output of the experiment and aborts, if and only if $x_c \neq x^\prime$.
        \item The challenger samples $r \in \{0,1\}^{l(n)}$ uniformly and computes ${\sf otp} = {\sf Ext}(x_{1-c}, r)$:
        \begin{enumerate}
            \item If $b = 0$, the challenger sends $(r,{\sf otp})$ to the adversary ${\sf Adv}_\lambda$.
            \item If $b = 1$, the challenger sends $(r,s)$ to the adversary ${\sf Adv}_\lambda$. $s$ is a binary string with the same length as ${\sf otp}$ chosen unifromly at random.
        \end{enumerate}
        \item The adversary outputs a single bit $b^\prime \in \{0,1\}$ as the output of the experiment.
    \end{enumerate}
\end{definition}\footnote{Our definition of CSG is different from the definition in \cite{BK24}. We add the additional requirement {\sf Randomness Extraction} to their definition. However, we observe that the additional requirement preserves the cryptographic assumptions needed. This is proved in \cref{appdx:CSG}.}

\begin{definition}[Classical Blind Quantum Computing from \cite{BK24}]
    \label{dfn:cbqc}
    Let $\lambda$ be the security parameter. Let $Q$ be a circuit which maps $\{0,1\}^* \times V$ to $W$. We want to point out that $V$ is the input quantum register and $W$ is the output quantum register of $Q$. An interactive protocol
    \begin{equation}
        (W, (e_x, e_z)) \leftarrow \langle S(1^\lambda, Q, V), C(1^\lambda, Q,s) \rangle 
    \end{equation}
    between 
    \begin{itemize}
        \item a QPT server with input the security parameter $1^\lambda$, the circuit $Q$, and a state on register $V$.
        \item a PPT client with input the security parameter $1^\lambda$, the circuit $Q$, and classical string $s$.
    \end{itemize}
    At the end of the protocol, the (honest) server outputs a state on register $W$ and the client outputs two classical strings $e_x, e_z$. The protocol must satisfy the two conditions:
    
    {\sf Correctness}: For any $Q$ and $s$, let $IDEAL[Q,s]$ be the map $V \rightarrow W$ defined by $Q(s, \cdot)$. Let $REAL[Q,s]$ be the map that runs
    \begin{equation}
        (W, (e_x, e_z)) \leftarrow \langle S(1^\lambda, Q, V), C(1^\lambda, Q,s) \rangle 
    \end{equation} first and applies $X(e_x)Z(e_z)$ to the register $W$. We have that for any $Q,s$,
    \begin{equation}
        \|IDEAL[Q,s] - REAL[Q,s]\|_{\diamond} = {\rm negl}(\lambda)
    \end{equation}
    
    {\sf Blindness}: For any circuit $Q$, two strings $s_0, s_1$, and QPT adversary $Adv_\lambda$:
    \begin{equation}
        \begin{aligned}
            | &\Pr[b= 1 : (b, (e_x, e_z)) \leftarrow \langle Adv_\lambda(1^\lambda,Q), C(1^\lambda, Q, s_0) \rangle] \\ &- \Pr[b= 1 : (b, (e_x, e_z)) \leftarrow \langle Adv_\lambda(1^\lambda,Q), C(1^\lambda, Q, s_1) \rangle] |={\rm negl}(\lambda)
        \end{aligned}
    \end{equation}
\end{definition}

\begin{theorem}[From \cite{BK24}]
    Assuming the existence of CSGs, there exists a Classical Blind Quantum Computing (CBQC) satisfying \cref{dfn:cbqc}.
\end{theorem}

We introduce the KLVY compiler as follows. Though the original compiler uses Quantum Fully Homomorphic Encryption (QFHE) to encrypt the first round, the central property being utilized is the Blindness. Thus we replace the QFHE with CBQC.
\begin{definition}[KLVY compiler\cite{KLVY22}]
    \label{dfn:KLVY-compiler}
    The compiled game is as follows:
    \begin{enumerate}
        \item The Verifier samples $q_A, q_B$ as the Referee does in the original non-local game. The Verifier (which is $C$) and the Prover (which is $S$) run $(W,(e_x,e_z))\leftarrow \Pi_{CBQC} = \langle S(1^\lambda), C(1^\lambda, M_{Bob}, q_B) \rangle $. $M_{Bob}(q_B)$ represents the circuit that Bob executes when given $q_B$ as the input.
        \item The Prover sends $ct_b$ to the Verifier. The Verifier computes $b = ct_b \oplus e_x$.
        \item The Verifier sends $q_A$ to the Prover in plain.
        \item The Prover sends $a$ to the Verifier in plain.
    \end{enumerate}
    The Verifier accepts if $V(q_A,q_B,a,b) = 1$ where $V$ is the predicate used to indicate whether Alice and Bob win the original game.
\end{definition}

\section{Compiled MSG and the Certified Deletion Property}
\label{sec:compiled-CD-MSG}
In this section, we define the compiled MSG and the computational certified deletion property. The computational certified deletion property is an analogue to the information-theoretic certified deletion property (\cite{KMY24}, Theorem 3.1).

\begin{definition}[Compiled MSG]
    \label{dfn:compiled-msg}
    The compiled MSG is an interactive protocol as follows:
    
    {{\sf SimBob}$\langle {\sf V}(1^\lambda, q_{B}), P(1^\lambda, \ket{\psi}) \rangle \rightarrow {{\sf b}, st_{q_B, {\sf b}}} $} is an interactive protocol between a QPT prover $P$ and a PPT verifier ${\sf V}$. The verifier ${\sf V}$ takes as input $q_B \in \{1,2,3\}$ and outputs ${\sf b} \in \{0,1\}^3$. The prover $P$ takes as input an arbitrary quantum state $\ket{\psi}$ and outputs $st_{q_B, {\sf b}}$ where $st_{q_B, {\sf b}}$ is the internal state conditioned on $q_B$ and ${\sf b}$.

    It must satisfy that
    
    {\sf Correctness}: Let $st_{{\sf real}, q_B, {\sf b}}$ be the prover's output state generated by $\langle {\sf V}(1^\lambda, q_{B}), P(1^\lambda, \ket{\psi}) \rangle \rightarrow {{\sf b}, st_{{\sf real}, q_B, {\sf b}}}$, conditioned on the input to ${\sf V}$ is $q_B$ and the output of ${\sf V}$ is $b$. Let $st_{{\sf ideal},q_B, {\sf b}}$ be the remaining state in Alice's register in the original MSG (see \Cref{dfn:MSG}), when the question to Bob is $q_B$ and the answer of Bob is ${\sf b}$. There exists an honest prover $P$ such that for any $q_B \in \{1,2,3\}, {\sf b}\in \{0,1\}^3$:
    \begin{equation}
        | st_{{\sf real}, q_B, \sf b} - st_{{\sf ideal},q_{B}, \sf b} | = {\rm negl}(\lambda)
    \end{equation}
\end{definition}

\begin{definition}[Computational Certified Deletion Property]
    \label{dfn:comp-CDP}
    Let $\tilde P$ be a QPT adversary. Let $n \in \mathbb N$ be a polynomial in $\lambda$. We define the following experiment ${\sf CCD}$:
    \begin{enumerate}
        \item {\sf V} samples $q_B^i \in \{1,2,3\}, q_A^i \in \{1,2,3\}$ for $i \in [n]$.
        \item For $i\in[n]$, {\sf V} and $\tilde P$ runs ${b_i, st} \leftarrow $ {\sf SimBob}$\langle$ ${\sf V}(1^\lambda, q_{B}^i), \tilde P(1^\lambda) \rangle$.
        \item {\sf V} sends $\{q_A^i \in \{1,2,3\} \}_{i \in [n]}$ to $\tilde P$. $\tilde P$ sends $\{a_i \in \{0,1\}^3 \}_{i \in [n]}$ to {\sf V}.
        \item The verifier {\sf V} outputs $0$ and aborts if for some $i\in[n]$
        \begin{equation}
            \label{eqn:CCP-first-round}
            MSG(q_A^i, q_B^i, a_i, b_i) = \bot
        \end{equation}
        \item The verifier sends $\{q_B^i \in \{1,2,3\}\}_{i \in [n]}$ to the prover $\tilde P$ in plain.
        \item The prover sends $\{b_i^\prime \in \{0,1\}^3\}_{i \in [n]}$ to the verifier. If for some $i\in[n]$ such that $q_B^i = 2$,
        \begin{equation}
            b_i^\prime \neq b_i
        \end{equation}
        the verifier outputs $0$. Otherwise, the verifier outputs $1$.
    \end{enumerate}
    A compiled MSG satisfies the computational certified deletion property if and only if
    \begin{equation}
        \Pr[{\sf CCD} = 1] = {\rm negl}(\lambda)
    \end{equation}
\end{definition}

In the remaining part of this section, we show the following theorem.
\begin{theorem}
    \label{thm:sec3-main}
    There exists a compiled MSG with the computational certified deletion property, assuming the existence of CSGs.
\end{theorem}

We give the construction of the compiled MSG as follows. The building blocks are
\begin{itemize}
    \item A secure CBQC protocol $\Pi_{CBQC}$ = $\langle S(1^\lambda,Q)$, $C(1^\lambda,Q,x) \rangle $.
    \item A sufficiently large constant $\lambda_c$.
\end{itemize}
The interactive protocol {\sf SimBob}$\langle {\sf V}(1^\lambda, q_{B}), P(1^\lambda) \rangle \rightarrow {b, st} $ is as follows:
\begin{enumerate}
    \item $V$ and $P$ engage in $(e_x,e_z) \leftarrow $ $\langle S(1^{\lambda_c},C_B)$, $C(1^{\lambda_c},C_{B},q_B) \rangle$, in which $V$ is the Client $C$ and $P$ is the Server $S$. $C_B$ is the circuit that $C_B(q_B, \cdot)$ applies the honest measurement by Bob in MSG, conditioned on receiving $q_B$ as the question. $V$ obtains $e_x \in \{0,1\}^3$ and $e_z \in \{0,1\}^3$.
    \item $P$ sends a string $r \in \{0,1\}^3$ to $V$. $V$ outputs $b = r \oplus e_x$ as the output of {\sf SimBob}$\langle {\sf V}(1^\lambda, q_{B}), P(1^\lambda)\rangle$. $P$ outputs its internal state as the output of {\sf SimBob}$\langle {\sf V}(1^\lambda, q_{B}), P(1^\lambda)\rangle$.
\end{enumerate}
We point out that the {\sf SimBob} above is exactly the FIRST round of a compiled Magic Square Game, compiled using the KLVY compiler (\cref{dfn:KLVY-compiler}). By showing the theorem below, we can prove \cref{thm:sec3-main}.

\begin{theorem}
    \label{thm:comp-CD-msg}
    The compiled MSG above satisfies the computational certified deletion property (see \cref{dfn:comp-CDP}).
\end{theorem}

Before we give the proof, we give an outline of the proof. First, we define a game ${\text{\sf comp-CD-MSG}}(\lambda)$ whose $n$-fold parallel repetition is the security game of the computational certified deletion property (see \cref{dfn:comp-CDP}). We show that for any QPT adversary $A^\lambda$, there exists a constant $c \in [0,1)$ such that the winning probability for $A^\lambda$ in ${\text{\sf comp-CD-MSG}}(\lambda)$ is less than or equal to $c$, for sufficiently large $\lambda$. Then, we use the parallel repetition heuristic to reduce the winning probability exponentially.

\begin{definition}[${\text{\sf comp-CD-MSG}}(\lambda)$]
    \label{dfn:one-shot-comp-CD-MSG}
    The game ${\text{\sf comp-CD-MSG}}(\lambda)$ is as follows:
    \begin{enumerate}
        \item {\sf V} samples $q_B \in \{1,2,3\}, q_A \in \{1,2,3\}$.
        \item $V$ and $P$ engage in $(e_x,e_z) \leftarrow \langle S(1^{\lambda_c},C_B), C(1^{\lambda_c},C_{B},q_B) \rangle$, in which $V$ is the Client $C$ and $P$ is the Server $S$. $C_B$ is the circuit that $C_B(q_B, \cdot)$ applies the honest measurement by Bob in MSG, conditioned on receiving $q_B$ as the question. $V$ obtains $e_x \in \{0,1\}^3$ and $e_z \in \{0,1\}^3$.
        \item $P$ sends a string $r \in \{0,1\}^3$ to $V$. $V$ outputs $b = r \oplus e_x$ as the output of {\sf SimBob}$\langle {\sf V}(1^\lambda, q_{B}), P(1^\lambda)\rangle$.
        \item {\sf V} sends $q_A$ to $P$. $P$ sends $a \in \{0,1\}^3$ to {\sf V}.
        \item The verifier {\sf V} outputs $0$ and aborts if
        \begin{equation}
            MSG(q_A, q_B, a, b) = \bot
        \end{equation}
        \item The verifier sends $q_B$ to the prover $P$ in plain.
        \item The prover sends $b^\prime \in \{0,1\}^3$ to the verifier. The verifier outputs $0$ if 
        \begin{equation}
            q_B = 2 \land b^\prime \neq b
        \end{equation}
         Otherwise, the verifier outputs $1$. 
    \end{enumerate}
\end{definition}
${\text{\sf comp-CD-MSG}}(\lambda)$ is the security game for one-shot computational Certified Deletion Property. We state the one-shot computational Certified Deletion Property below.

\begin{lemma}
    \label{lem:one-shot-comp-MSG-CD}
    There exists a constant $c < 1$, for any QPT adversary $P$ and any sufficiently large $\lambda \in \mathbb N$,  $\Pr[\text{\sf comp-CD-MSG}(\lambda) = 1] \leq c $.
\end{lemma}

\begin{proof}
    The outline of our proof is as follows. First, we define a security game ${\text{\sf comp-NI-CD-MSG}}(\lambda)$. Then, we show that any adversary $P$ for $\text{\sf comp-CD-MSG}(\lambda)$ can be converted to an adversary for ${\text{\sf comp-NI-CD-MSG}}(\lambda)$, where the winning probability remains unchanged before and after. Then, we will prove $\Pr[{\text{\sf comp-NI-CD-MSG}}(\lambda) = 1] < 1$ for all sufficiently large $\lambda \in \mathbb N$ (the threshold is a constant, though it may be huge, e.g., $10^{1000}$).

    We define ${\text{\sf comp-NI-CD-MSG}}(\lambda)$ below. We highlight the difference between ${\text{\sf comp-NI-CD-MSG}}(\lambda)$ and ${\text{\sf comp-CD-MSG}}(\lambda)$ is that the prover guesses the value of $b$ without receiving $q_B$.
    \begin{enumerate}
        \item {\sf V} samples $q_B \in \{1,2,3\}, q_A \in \{1,2,3\}$ .
        \item $V$ and $P$ engage in $(e_x,e_z) \leftarrow \langle S(1^{\lambda_c},C_B), C(1^{\lambda_c},C_{B},q_B) \rangle$, in which $V$ is the Client $C$ and $P$ is the Server $S$. $C_B$ is the circuit that $C_B(q_B, \cdot)$ applies the honest measurement by Bob in MSG, conditioned on receiving $q_B$ as the question. $V$ obtains $e_x \in \{0,1\}^3$ and $e_z \in \{0,1\}^3$.
        \item $P$ sends a string $r \in \{0,1\}^3$ to $V$. $V$ outputs $b = r \oplus e_x$ as the output of {\sf SimBob}$\langle {\sf V}(1^\lambda, q_{B}), P(1^\lambda)\rangle$.
        \item {\sf V} sends $q_A$ to $P$. $P$ sends $a \in \{0,1\}^3$ to {\sf V}.
        \item The verifier {\sf V} outputs $0$ and aborts if
        \begin{equation}
            MSG(q_A, q_B, a, b) = \bot
        \end{equation}
        \item The prover sends $b^\prime \in \{0,1\}^3$ to the verifier. The verifier outputs $0$ if 
        \begin{equation}
            q_B = 2 \land b^\prime \neq b
        \end{equation}
         Otherwise, the verifier outputs $1$. 
    \end{enumerate}
    
    \begin{proposition}
        For $\lambda $ $\in$ $ \mathbb N$, we have 
        \[
        \Pr[{\text{\sf comp-NI-CD-MSG}}(\lambda)=1] = \Pr[{\text{\sf comp-CD-MSG}}(\lambda)=1]
        \]
    \end{proposition}
    \begin{proof}
        $\Pr[{\text{\sf comp-NI-CD-MSG}}(\lambda)=1] \leq \Pr[{\text{\sf comp-CD-MSG}}(\lambda)=1]$ is trivial. Simulating the adversary in ${\text{\sf comp-NI-CD-MSG}}(\lambda)$ in round $1$ and round $2$ and using $b^\prime$ as the answer in the $3$rd round, we can win ${\text{\sf comp-CD-MSG}}(\lambda)$ with the same probability as in ${\text{\sf comp-NI-CD-MSG}}(\lambda)$.

        To prove the other side, we consider an adversary as follows. Let $A$ be the adversary in ${\text{\sf comp-CD-MSG}}(\lambda)$. The adversary for ${\text{\sf comp-NI-CD-MSG}}(\lambda)$ receives the first two messages from the verifier and uses the messages to simulate $A$. Then, the adversary simulates $A$ assuming $q_B = 2$ to obtain $b^\prime$ and uses it as the answer in the $3$rd round.
    
        When $q_B \in \{1, 3\}$, the $3$rd answer $b^\prime$ does not affect the winning probability of ${\text{\sf comp-CD-MSG}}(\lambda)$. Thus, $\Pr[{\text{\sf comp-NI-CD-MSG}}(\lambda)=1 | q_B\in \{1,3\}] = \Pr[{\text{\sf comp-CD-MSG}}(\lambda)=1 | q_B\in \{1,3\}]$. When $q_B = 2$, the $3$rd question to $P$ coincides with the underlying plaintext of the $1$st question. The adversary simulates $P$ perfectly in this case. Combining the two facts above, we have that $\Pr[{\text{\sf comp-NI-CD-MSG}}(\lambda)=1] \geq \Pr[{\text{\sf comp-CD-MSG}}(\lambda)=1]$. We complete the proof.
    \end{proof}

    Then, we propose the following non-local game, which compiles to ${\text{\sf comp-NI-CD-MSG}}(\lambda)$ using Ref. \cite{KLVY22}.
    \begin{definition}[${\text{\sf NI-CD-MSG}}$]
    \label{dfn:NI-CD-MSG}
    We define a non-local game
    \begin{enumerate}
        \item $R$ samples $q_B \in \{1,2,3\}, q_A \in \{1,2,3\}$ .
        \item $R$ sends $q_B$ to $B$ and $q_A$ to $A$, respectively.
        \item $A$ sends $a \in \{0,1\}^3$ to $R$ and $B$ sends $b \in \{0,1\}^3$ to $R$.
        \item $R$ outputs $0$ and aborts if
        \begin{equation}
            MSG(q_A, q_B, a, b) = \bot
        \end{equation}
        \item $A$ sends $b^\prime \in \{0,1\}^3$ to $R$. $R$ outputs $0$ if 
        \begin{equation}
            q_B = 2 \land b^\prime \neq b
        \end{equation}
         Otherwise, $R$ outputs $1$. 
    \end{enumerate}
    \end{definition}
    The game ${\text{\sf NI-CD-MSG}}$ above has quantum commuting value $w_{qc} < 1$. We can see that the proof in \cite{FM18} applies to prove $w_{qc} < 1$ for ${\text{\sf NI-CD-MSG}}$. We give another proof using the uncertainty relation in \cref{apdx:MSG}.
    \begin{lemma}
        \label{lem:NI-CD-MSG-qc}
        We have $w_{qc}<1$ for ${\text{\sf NI-CD-MSG}}$.
    \end{lemma}
    Then,  by the following lemma,  we can see that the winning probability converges to ${\text{\sf NI-CD-MSG}}$'s quantum commuting value. Thus there exists a $\lambda_c \in \mathbb N$ such that $\Pr[\text{\sf comp-CD-MSG}(\lambda) = 1] \leq c$ for each $\lambda > \lambda_c$. 
    \begin{lemma}[Theorem 6.1 from \cite{KMP+25}]
        \label{lem:compiled-value}
        Let $G$ be any two-player nonlocal game and let $S$ be any QPT strategy for the compiled game $G_{comp}$. Then it holds that
        \begin{equation}
            \lim_{\lambda \rightarrow \infty} \sup w_{\lambda}(G_{comp}, S) \leq w_{qc}(G)
        \end{equation}
    \end{lemma}
    We complete the proof of \cref{lem:one-shot-comp-MSG-CD}.  
\end{proof}

\begin{conjecture}[Parallel Repetition]
    \label{conj:formal-parallel-rep}
    Let $c$ be the maximum winning probability of ${\text{\sf comp-CD-MSG}}(\lambda)$ for any QPT adversary. The winning probability of its $k$-fold repetition, ${\text{\sf comp-CD-MSG}}(\lambda)^k$ is $O(c^{O(k)})$, which decreases exponentially as $k$ grows.
\end{conjecture}

By the conjecture above, we obtain \cref{thm:comp-CD-msg}. We emphasize that whether the parallel repetition reduces the winning probability of a post-quantum argument exponentially is an open problem. Thus, we rely on the conjecture to justify the parallel repetition.

\section{Public-key Encryption with Classical Secure Key Leasing}
\label{sec:dfn-cSKL}
In this section, we define the syntax for public key encryption with classical secure key leasing (PKE-cSKL). Our definition is analogous to the definition in \cite{KMY24}.

\begin{definition}[PKE-cSKL]
    \label{dfn:PKE-CSKL}
    Let ${\cal M}$ be the message space. A scheme of PKE-cSKL is a tuple of algorithms $({\sf Setup},{\sf KG}, {\sf Enc}, Dec, Del, {\sf DelVrfy})$:
    \begin{itemize}
        \item[] {${\sf Setup}{\langle Lessor(1^\lambda), Lessee(1^\lambda) \rangle} \rightarrow ({\sf msk}, st)/\bot$} is an interactive protocol between
            \begin{itemize}
                \item a QPT Lessee
                \item a PPT Lessor
            \end{itemize}
        When the protocol fails, the Lessor and the Lessee output $\bot$. When the protocol succeeds, the Lessor outputs a classical master secret key ${\sf msk}$. The (honest) Lessee outputs an internal state $st$.
        \item[] {${\sf KG}{\langle Lessor(1^\lambda, {\sf msk}), Lessee(1^\lambda, st) \rangle} \rightarrow (sk, {\sf pk}, {\sf vk})/\bot$} is an interactive protocol between
            \begin{itemize}
                \item a QPT Lessee
                \item a PPT Lessor
            \end{itemize}
        The Lessor takes as input the master secret key ${\sf msk}$, and the Lessee takes as input the internal state of the Lessee from {\sf PKE-cSKL.Setup}. The Lessor outputs a classical public key ${\sf pk}$ and a classical ${\sf vk}$. The (honest) Lessee outputs a quantum decryption key $sk$.
        \item[] {${\sf Enc}({\sf pk}, m) \rightarrow {\sf ct}_m$} is a PPT algorithm. ${\sf pk}$ is a classical public key and $m$ is a message from the message space. The algorithm outputs a classical ciphertext ${\sf ct}_m$.
        \item[] {$Dec(sk,{\sf ct}_m) \rightarrow m$} is a QPT algorithm. $sk$ is a quantum decryption key and ${\sf ct}_m$ is a classical ciphertext. The algorithm outputs a classical string representing the decryption result.
        \item[] {$Del(sk) \rightarrow {\sf cert}$} is a QPT algorithm. $sk$ is a quantum decryption key. The algorithm destroys the quantum key and outputs a valid classical certificate ${\sf cert}$.
        \item[] {${\sf DelVrfy}({\sf cert}, {\sf vk}) \rightarrow \top / \bot$} is a PPT algorithm. The algorithm takes as input a classical certificate ${\sf cert}$ and a classical verification key ${\sf vk}$. The algorithm outputs $\top$ if ${\sf cert}$ is a valid certificate of deletion. It outputs $\bot$ if ${\sf cert}$ is not a valid certificate.
    \end{itemize}
\end{definition}
{\bf Decryption correctness:} For every $m\in {\cal M}$, we have that
\begin{equation*}
    \label{eqn:PKE-cSKL-decryption-correctness}
    \Pr[m \neq m^\prime \lor res =\bot: 
    \begin{aligned}
        & res \leftarrow \text{\sf PKE-cSKL}.{\sf Setup}{\langle Lessor(1^\lambda), Lessee(1^\lambda) \rangle} \\
        & ({\sf msk}, {st}) \coloneqq res \\
        & ({sk}, {\sf pk}, {\sf vk}) \leftarrow  \text{\sf PKE-cSKL}.{\sf KG}{\langle Lessor(1^\lambda, {\sf msk}), Lessee(1^\lambda, st) \rangle}\\
        & {\sf ct}_m \leftarrow \text{\sf PKE-cSKL}.{\sf Enc}({\sf pk},m) \\
        & m^\prime \leftarrow \text{\sf PKE-cSKL}.Dec(sk,{\sf ct}_m)
    \end{aligned}
    ] = {\rm negl}(\lambda)
\end{equation*}

{\bf Deletion verification correctness:} We have
\begin{equation*}
    \Pr[b = \bot \lor res=\bot: 
    \begin{aligned}
        & res \leftarrow \text{\sf PKE-cSKL}.{\sf Setup}{\langle Lessor(1^\lambda), Lessee(1^\lambda) \rangle} \\
        & ({\sf msk}, {st}) \coloneqq res \\
        & ({sk}, {\sf pk}, {\sf vk}) \leftarrow  \text{\sf PKE-cSKL}.{\sf KG}{\langle Lessor(1^\lambda, {\sf msk}), Lessee(1^\lambda, st) \rangle}\\
        & {\sf cert} \leftarrow \text{\sf PKE-cSKL}.Del(sk) \\
        & {\sf b} \leftarrow \text{\sf PKE-cSKL}.{\sf DelVrfy}({\sf cert},{\sf vk})
    \end{aligned}
    ] = {\rm negl}(\lambda)
\end{equation*}

Then, we introduce the security definition.

\begin{definition}[IND-VRA security]
    \label{dfn:PKE-IND-VRA}
    The IND-VRA security for a PKE-cSKL scheme is formalized by the experiment ${\sf EXP}^\text{\sf ind-vra}_{\text{PKE-cSKL}, A}(1^\lambda,{\sf coin})$:
    \begin{enumerate}
        \item The challenger $C$ and the adversary $A$ runs $\text{\sf PKE-cSKL.Setup}\langle C(1^\lambda), A(1^\lambda) \rangle$. If the output is $\bot$ (${\sf Setup}$ aborted), the experiment ends and the output is $0$.
        \item The challenger $C$ and the adversary $A$ runs $\text{\sf PKE-cSKL.KG}\langle C(1^\lambda), A(1^\lambda) \rangle$. 
        \item The challenger sends {\sf pk} to the adversary.
        \item The adversary sends a classical string {\sf cert} and $(m_0,m_1) \in {\cal M}^2$ to the challenger. If $\text{\sf PKE-cSKL.DelVrfy}({\sf cert}, {\sf vk}) = \bot$, the challenger outputs $0$ and the experiment ends. Otherwise, the challenger computes ${\sf ct} \leftarrow \text{\sf PKE-cSKL}.{\sf Enc}({\sf pk},m_{\sf coin})$ and sends ${\sf ct}, vk,$ to the adversary.
        \item The adversary outputs a guess ${\sf coin}^\prime \in \{0,1\}$.
    \end{enumerate}
    {\sf PKE-cSKL} is IND-VRA secure if and only if for any QPT $A$
    \begin{equation}
        \begin{aligned}
            {\sf Adv}&^\text{\sf ind-vra}_{\text{PKE-cSKL},A}(\lambda) := |\Pr[{\sf EXP}^\text{\sf ind-vra}_{\text{PKE-cSKL}, A}(1^\lambda,{\sf 0})=1] \\
            &-\Pr[{\sf EXP}^\text{\sf ind-vra}_{\text{PKE-cSKL}, A}(1^\lambda,{\sf 1})=1]| = {\rm negl}(\lambda)
        \end{aligned}
    \end{equation}
\end{definition}

We can also define a one-way variant of the security above.

\begin{definition}[OW-VRA security]
    \label{dfn:PKE-OW-VRA}
    The OW-VRA security for a PKE-cSKL scheme is formalized by the experiment ${\sf EXP}^\text{\sf ow-vra}_{\text{PKE-cSKL}, A}(1^\lambda)$:
    \begin{enumerate}
        \item The challenger $C$ and the adversary $A$ run $\text{\sf PKE-cSKL.Setup}\langle C(1^\lambda), A(1^\lambda) \rangle$. If the output is $\bot$ (${\sf Setup}$ aborted), the experiment ends, and the output is $0$.
        \item The challenger $C$ and the adversary $A$ run $\text{\sf PKE-cSKL.KG}\langle C(1^\lambda), A(1^\lambda) \rangle$.
        \item The challenger sends {\sf pk} to the adversary.
        \item The adversary sends a classical string {\sf cert}. If $\text{\sf PKE-cSKL.DelVrfy}({\sf cert}, {\sf vk}) = \bot$, the challenger outputs $0$ and the experiment ends. 
        \item The challenger samples $m\in {\cal M}$ uniformly and computes ${\sf ct} \leftarrow \text{\sf PKE-cSKL}.{\sf Enc}({\sf pk},m)$ and sends ${\sf ct}$, $vk$ to the adversary.
        \item The adversary sends $m^\prime \in {\cal M}$ to the challenger. The challenger outputs $1$ if $m^\prime = m$. Otherwise, the challenger outputs $0$.
    \end{enumerate}
    {\sf PKE-cSKL} is OW-VRA secure if and only if for any QPT $A$
    \begin{equation}
        {\sf Adv}^\text{\sf ow-vra}_{\text{PKE-cSKL},A}(\lambda):= \Pr[{\sf EXP}^\text{\sf ind-vra}_{\text{PKE-cSKL}, A}(1^\lambda)=1]= {\rm negl}(\lambda)
    \end{equation}
\end{definition}

\begin{lemma}
    If there exists a PKE-cSKL with OW-VRA security (see \cref{dfn:PKE-OW-VRA}), then there exists a PKE-cSKL with IND-VRA security (see \cref{dfn:PKE-IND-VRA}).
\end{lemma}
We can prove the lemma above with almost the same proof as Lemma 3.12 from \cite{AKN+23},  Lemma 4.6 from \cite{KMY24}. So we omit the proof.

\begin{table}[tbp]
    \centering
    \begin{tabular}{c|c|c}
        $XI$ & $IX$ & $XX$ \\
        \hline
        $IZ$ & $ZI$ & $ZZ$ \\
        \hline
        $-XZ$ & $-ZX$ & $YY$ 
    \end{tabular}
    \label{tab:msg-strategy3}
    \caption{When Alice (resp. Bob) receives $x$ (resp. $y$) as their question, they measure the observables on $x$-th column (resp. $y$-th row) and output the outcomes as their answer.}
\end{table}

We will use the following function to cancel the $Z$ error introduced in the key generation process (we explain this in the Proof of Decryption correctness later in this section). Since the $Z$ error only affects Pauli $X$ observables, $postprocessing_A$ alters the input $a$ only for the bits generated by observables with at least one $X$, according to the optimal strategy of MSG as shown in \cref{tab:msg-strategy3}.
\begin{definition}[Software Error Correction for Z error]
    \label{dfn:postprocessing}
    We define the function $postprocessing_A(q_A,a, e_0 , e_1)$ such that
    \begin{equation}
         \left \{ 
        \begin{array}{ll}
           (a[0]\oplus e_0)(a[1]) (a[2]\oplus e_0 )  &  q_A = 0\\
           (a[0]\oplus e_1)a[1](a[2] \oplus e_1)  &  q_A = 1\\
           (a[0]\oplus e_0\oplus e_1)a[1](a[2]\oplus e_0 \oplus e_1)  &  q_A=2
        \end{array}
        \right .
    \end{equation}
\end{definition}
We note that the $postprocessing$ function preserves the parity of the input.

We use the following primitives as the building blocks
\begin{itemize}
    \item A IND-CPA secure PKE scheme {\sf PKE = (PKE.KG, PKE.Enc, PKE.Dec)} for one-bit message. Let $l_{pk}$ be the length of the public key, $l_{sk}$ be the length of the secret key.
    \item A secure classical blind quantum computing protocol $\Pi_{CBQC}$ (see \cref{dfn:cbqc}).
    \item A CSG $\Pi_{CSG}$ with Indistinguishability Security (see \cref{dfn:CSG}). Let ${\sf Ext}(x,r)$ be the extractor in Randomness Extraction.
    \item A compiled MSG ${\sf SimBob}$.
\end{itemize}

Let ${\cal M}$ be the message space for {\sf PKE-cSKL}. We define $J_B = \{j \in \mathbb N | \exists i\in\mathbb N, q_B^i=1 \land \lfloor j/2 \rfloor = i  \}$, which is important in our construction.
\begin{remark}
    $q_B, q_A \in \{1,2,3\}$ in the original MSG. However, to make sure the index, e.g. $b_i[q_A^i]$, starts with $b[0]$, we use $q_B^i,q_A^i \in \{0,1,2\}$ in our protocol construction and the proof.
\end{remark}

{{\sf PKE-cSKL}.{\cal Setup}$\langle$Lessor($1^\lambda$), Lessee($1^\lambda$)$\rangle$} We describe the interactive setup protocol between the Lessor and the honest Lessee below. Note that in the actual protocol, the malicious Lessee may not follow the protocol.
\begin{figure}[tbp]
    \centering
    \includegraphics[width=0.8\linewidth]{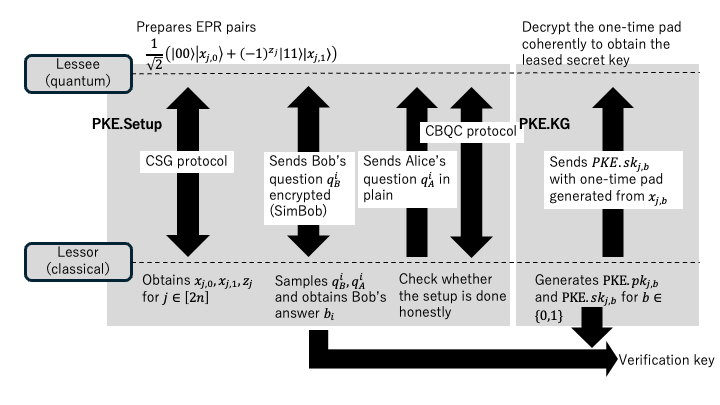}
    \caption{We summarize what the honest Lessee does during ${\sf PKE-cSKL}.Setup$ and ${\sf PKE-cSKL}.KG$ in the figure above. In ${\sf PKE-cSKL}.Setup$, the Lessor prepares a clawed EPR pair on the Lessee's quantum computer remotely. Then, they simulate Bob in the original MSG. The remaining state of the Lessee is passed down to ${\sf PKE-cSKL}.KG$ to generate the leased secret key. Note that the Lessor checks whether the Lessee cheats on $j \in J_B$ and aborts if the answer is yes. In ${\sf PKE-cSKL}.KG$, the Lessor samples secret keys and sends them encrypted to the Lessee to generate the final leased secret key. Besides the secret key, the Lessor generates the verification key with $q_B^i$, $q_A^i$, $b_i$, ${\sf PKE}.sk_{j,0}$, and ${\sf PKE}.sk_{j,1}$.} 
    \label{fig:CBQC-honest-lessee}
\end{figure}
\begin{enumerate}
    \item The Lessor repeats the following steps for $j \in [2n]$:
    \begin{enumerate}
        \item The Lessor and the Lessee engage in $\Pi_{CSG} = \langle S(1^\lambda,l_{sk}), R(1^\lambda, l_{sk}) \rangle$. The Lessor obtains $x_{j,0}, x_{j,1} \in \{0,1\}^{l_{sk}}$ and $z_j \in \{0,1\}$ for $j \in [2n]$. The honest Lessee prepares $\frac{1}{\sqrt 2}(\ket{0}_{{\sf B}_j}\ket{x_{j,0}}_{{\sf X}_j}+(-1)^{z_j}\ket{1}_{{\sf B}_j}\ket{x_{j,1}}_{{\sf X}_j})$ on {registers} ${\sf B}_j$ and ${\sf X}_j$. 
        \item The honest Lessee initializes ${\sf A}_j$ with $\ket{0}_{{\sf A}_j}$. The honest Lessee applies CNOT using ${\sf B}_j$ as the control qubit and registers ${\sf A}_j$ as the target qubit.
    \end{enumerate}
    \item The Lessor repeats the steps (a)-(c) for $i \in [n]$:
    \begin{enumerate}
        \item The Lessor samples $q_B^i,q_A^i \in \{0,1,2\}$ uniformly. 
        \item The Lessor and the Lessee engage in ${\sf SimBob}\langle {\sf V}(1^\lambda, q_B^i), P(1^\lambda) \rangle $, where the Lessor plays the role of ${\sf V}$ and the Lessee plays the role of $P$. The Lessee follows the honest prover in {\sf SimBob}. Let the output of the Lessor be $b_i$.
        \item The Lessor sends $q_A^i$ to the Lessee.
    \end{enumerate}
    \item The Lessor repeats the following steps for $j\in[2n]$:
    \begin{enumerate}
        \item The honest Lessee initializes the registers ${\sf R}_j$ and ${\sf R}_j^\prime$ to $\ket{0^{l_{claw}}}$.
        \item The honest Lessee applies CNOTs with each bit of ${\sf X}_j$ as the control qubit and each bit of ${\sf R}_j$ as the target qubit.
        \item The honest Lessee engages in CBQC with the Lessor and executes the circuit as follows:
        \begin{itemize}
            \item If $j \notin J_B$, the Lessee does nothing.
            \item If $j \in J_B$, the Lessee applies CNOT gates with each qubit of ${\sf R}$ as the control qubit and each qubit of ${\sf R}^\prime$ as the target qubit. The circuit ``copies'' ${\sf R}$ to ${\sf R}^\prime$.
        \end{itemize}
        Let the output of the Lessor be $(e_x^j,e_z^j)$. The Lessor parses $e_{x}^j = e_{x,0}^j||e_{x,1}^j$ and $e_{z}^j = e_{z,0}^j||e_{z,1}^j$. Note that CBQC hides the information of the circuit, thus the Lessee does not know whether $j \in J_B$.
        \item Let $i = \lfloor j/2 \rfloor $. For $j \in J_B$, the honest Lessee measures the register ${\sf R}_j^\prime$ to obtain outcome $\omega$, and the Lessor computes $x^\prime_j = \omega \oplus e_{x,1}^j$ and $c = b_i[j\mod 2]$. The Lessor checks whether $x^\prime_j = x_{j,c}$. It aborts the protocol and outputs $\bot$ if for some $j \in J_B$, the statement does not hold.
        \item Again, the honest Lessee applies CNOTs with each bit of ${\sf X}_j$ as the control qubit and each bit of ${\sf R}_j$ as the target qubit.
    \end{enumerate}
    \item The Lessor outputs a master key
    \begin{equation}
        {\sf{msk}} \coloneqq (\{ q_B^i, b_i^\prime, q_A^i \}_{i \in [n]}, \{ e_{z,1}^j, z_j, x_{j,0}, x_{j,1}\}_{j \in [2n]: j \notin J_B})
    \end{equation}
    where $b_i^\prime[q_A^i] = b_i[q_A^i]$ and the other bits are generated by xoring the same uniformly random bit to $b_i$'s corresponding bit. 
\end{enumerate}

{{\sf PKE-cSKL}.{\cal KG}$\langle$Lessor($1^\lambda$, \sf{msk}), Lessee($1^\lambda$, st)$\rangle$} We describe the interactive key generation protocol between the Lessor and the honest Lessee below. Note that in the actual protocol, the malicious Lessee may not follow the protocol.
\begin{enumerate}
    \item The Lessor generate $({\sf PKE.pk}_{j,0}$,${\sf PKE.sk}_{j,0})$$\leftarrow$ $\mathsf{PKE.KG}(1^\lambda)$ and $({\sf PKE.pk}_{j,1}$,${\sf PKE.sk}_{j,1})$ $\leftarrow$ $\mathsf{PKE.KG}(1^\lambda)$ for $j \in [2n]$.
    \item For $j \in [2n]$, the Lessor samples $r_{j,0}, r_{j,1}$ as the internal randomness of the classical algorithm ${\sf Ext}$ for CSG (see \cref{dfn:CSG}). Then, it computes $h_{j,0} = {\sf Ext}(x_{j,0},r_{j,0}) \oplus {\sf PKE.sk}_{j,0}$, $h_{j,1}={\sf Ext}(x_{j,1},r_{j,1}) \oplus {\sf PKE.sk}_{j,1}$.
    \item The Lessor sends {$h_{j,0}$, $h_{j,1}$, $r_{j,0}$, $r_{j,1}$} to the Lessee.
    \item The honest Lessee initializes the register ${\sf SK}_j$ to $\ket{0^{l_{sk}}}$. Then, the honest Lessee applies the unitary $U_{r_{j,0}, r_{j,1}}$ which satisfies
    \begin{equation}
        \begin{aligned}
            &\forall\ b\in\{0,1\}, y\in\{0,1\}^{l_{claw}}, s\in\{0,1\}^{l_{sk}}, \\
           & U_{r_{j,0}, r_{j,1}}\ket{b}_{{\sf A}_j}\ket{y}_{{\sf X}_j}\ket{s}_{{\sf SK}_j} = \ket{b}_{{\sf A}_j}\ket{y}_{{\sf X}_j}\ket{s\oplus {\sf Ext}(y, r_{j,b})}_{{\sf SK}_j}
        \end{aligned}
    \end{equation}
    Furthermore, the honest Lessee applies the unitary $U_{h_{j,0},h_{j,1}}$ which satisfies
    \begin{equation}
        \begin{aligned}
            &\forall\ b\in\{0,1\}, y\in\{0,1\}^{l_{claw}}, s\in\{0,1\}^{l_{sk}} \\
            &U_{h_{j,0}, h_{j,1}}\ket{b}_{{\sf A}_j}\ket{y}_{{\sf X}_j}\ket{s}_{{\sf SK}_j} = \ket{b}_{{\sf A}_j}\ket{y}_{{\sf X}_j}\ket{s\oplus h_{j,b}}_{{\sf SK}_j}
        \end{aligned}
    \end{equation}
    \item The Lessor outputs a public key
    \begin{equation}
        \mathsf{pk} \coloneqq ({\sf PKE.pk}_{j,0}, {\sf PKE.pk}_{j,1})_{j \in [2n]}
    \end{equation}
    and a verification key
    \begin{equation}
        \begin{aligned}
            &{\sf{dvk}} \coloneqq (\{ q_B^i, b_i^\prime, q_A^i \}_{i \in [n]}, 
            \\
            &\{ {\sf PKE.sk}_{j,0}, {\sf PKE.sk}_{j,1}, e_{z,1}^j, z_j, x_{j,0}, x_{j,1}\}_{j \in [2n]: j \notin J_B})
        \end{aligned}
    \end{equation}
    The Lessee outputs the result state on registers ${\sf A}_j{\sf X}_j{\sf SK}_j$ and $\{q_A^i\}_{i\in[n]}$ as the secret key.
\end{enumerate}
\begin{remark}
    For the sake of clarity, we present the ideal quantum secret key $\ket{sk}$ which consists of registers $({\sf A}_j, {\sf SK}_j, {\sf X}_j)$ as follows:
    \begin{equation}
        \label{eqn:ideal-key-state-pke}
        \begin{aligned}
            &\frac{1}{\sqrt 2}(\ket{0, {\sf PKE.sk}_{j,0}, x_{j,0}} \pm \ket{1, {\sf PKE.sk}_{j,1}, x_{j,1}})& q_B^i = 0 \\
            &\ket{0,{\sf PKE.sk}_{j,0}, x_{j,0}}\ {\rm or}\  \ket{1,{\sf PKE.sk}_{j,1} , x_{j,1}}& q_B^i = 1 \\
            &\frac{1}{2}(\ket{0, {\sf PKE.sk}_{j,0}, x_{j,0}}\bra{0, {\sf PKE.sk}_{j,0}, x_{j,0}}+  \\ &\ket{1, {\sf PKE.sk}_{j,1}, x_{j,1}}\bra{1, {\sf PKE.sk}_{j,1}, x_{j,1}})& q_B^i = 2 
        \end{aligned}
    \end{equation}
    We point out that the register $({\sf A}_{2i}, {\sf SK}_{2i}, {\sf X}_{2i})$
    and $({\sf A}_{2i+1}, {\sf SK}_{2i+1}, {\sf X}_{2i+1})$ are entangled for $q_B^i = 2$. The whole state is $ V_{2i}V_{2i+1} \ket{\Phi_{b[1]b[0]}}_{{\sf A}_{2i} {\sf A}_{2i+1}}$
    where $V_j$ is an isometry mapping from ${\cal H}_{{\sf A}_j}$ to ${\cal H}_{{\sf A}_j}\otimes {\cal H}_{{\sf SK}_j} \otimes {\cal H}_{{\sf R}_j}$ such that $V_j \ket{b}_{{\sf A}_j} = \ket{b, {\sf PKE.sk}_{j,b},x_{j,b}}_{{\sf A}_j {\sf SK}_j {\sf R}_j}$ and $\ket{\Phi_{ab}} \coloneqq \frac{1}{\sqrt 2}(\ket{0a}+(-1)^b\ket{1(1-a)})$ is one of the four Bell states. We present the honest Lessee who outputs a quantum state almost identical to the ideal state, except for ${\rm negl}(\lambda)$ trace distance, in the proof of Decryption correctness.
\end{remark}

{{\sf PKE-cSKL.Enc}$(\sf{pk},m)$}:
\begin{enumerate}
    \item Parse $\mathsf{pk} = ({\sf PKE.pk}_{j,0}, {\sf PKE.pk}_{j,1})_{j \in [2n]}$ and $m = m_0 || \dots || m_{2n-1}$. 
    \item Compute 
    \begin{equation}
        \begin{aligned}
            ct_{j,0} \leftarrow \text{\sf PKE.Eval}(\text{\sf PKE-cSKL.pk}_{j,0},m_j) \\ ct_{j,1} \leftarrow \text{\sf PKE.Eval}(\text{\sf PKE-cSKL.pk}_{j,1},m_j)
        \end{aligned}
    \end{equation}
     for $j \in [2n]$.
    \item Output ${\sf ct} = (\{ct_{j,0}, ct_{j,1}\})_{j \in [2n]}$ as the ciphertext for $m \in {\cal M}$.
\end{enumerate}

{{\sf PKE-cSKL}.Dec$(sk,{\sf ct})$}:
\begin{enumerate}
    \item Parse $sk \coloneqq (\{ q_A^i \}_{i \in [n]}, ({\sf A}_j, {\sf SK}_j, {\sf X}_j)_{j \in [2n]})$ and ${\sf ct} = (\{ct_{j,0}, ct_{j,1}\})_{j \in [2n]}$. The registers $({\sf A}_j, {\sf SK}_j, {\sf X}_j)_{j \in [2n]}$ are holding the key state.
    \item Let $U_{Dec, j}$ be a unitary on register $({\sf A_j}, {\sf SK}_j, {\sf OUT}_j)$ as follows:
    \begin{equation}
        \begin{aligned}
            &U_{Dec,j} \ket{b}_{{\sf A}_j}\ket{\sf PKE.sk_{j,b}}_{{\sf SK}_j}\ket{v}_{{\sf OUT}_j} 
            \\
            =& \ket{b}_{{\sf A}_j}\ket{\sf PKE.sk_{j,b}}_{{\sf SK}_j} \ket{v \oplus \mathsf{PKE.Dec(PKE.sk_{j,b}, ct_{j,b})}}_{{\sf OUT}_j}
        \end{aligned}
    \end{equation}
    The algorithm applies the unitary $U_{Dec,j}$ to register $({\sf A}_j, {\sf SK}_j, {\sf OUT}_j)$, where ${\sf OUT}_j$ is initialized to $\ket{0}$. Then, measure the register ${\sf OUT}_j$ in the computational basis and obtain the outcome $t_j^\prime$.
    \item Output $t_0^\prime || \dots || t_{2n-1}^\prime$.
\end{enumerate}

{{\sf PKE-cSKL}.Del$(sk)$} The algorithm is as follows.
\begin{enumerate}
    \item Parse $sk \coloneqq (\{ q_A^i \}_{i \in [n]}, ({\sf A}_j, {\sf SK}_j, {\sf X}_j)_{j \in [2n]})$. The registers $({\sf A}_j, {\sf SK}_j, {\sf X}_j)_{j \in [2n]}$ are holding the key state. 
    \item For $i\in[n]$, measure the register $(A_{2i}, A_{2i+1})$ with $\{\mathcal A_{q_A^i}^a\}$, where $\{\mathcal A_{q_A^i}^a\}$ is Alice's measurement in \cref{tab:msg-strategy} when the question to it is $q_A^i$. Set $a_i$ to be the outcome.
    \item For $j\in[2n]$, measure every qubit of registers ${\sf SK}_j$, ${\sf X}_j$ in Hadamard basis. Let the measurement outcome be $d_j, d_j^\prime$, respectively.
    \item Output ${\sf cert}= (\{ a_i \}_{i\in [n]}, \{d_j, d_j^\prime \}_{j \in [2n]})$.
\end{enumerate}

{{\sf PKE-cSKL}.{\sf DelVrfy}$({\sf cert}, {\sf dvk})$} The algorithm is as follows.
\begin{enumerate}
    \item Parse 
    \begin{equation}
        {\sf cert} = (\{ a_i \}_{i\in [n]}, \{d_j, d_j^\prime\}_{j \in [2n]})
    \end{equation}
    and 
    \begin{equation}
        \begin{aligned}
            {\sf{dvk}} \coloneqq (\{ q_B^i, b_i^\prime, q_A^i \}_{i \in [n]}, \{ {\sf PKE.sk}_{j,0}, {\sf PKE.sk}_{j,1}, \\
            e_{z,1}^j, z_j, x_{j,0}, x_{j,1}\}_{j \in [2n]: j \notin J_B})
        \end{aligned}
    \end{equation}
    \item Computes 
    \begin{equation}
        \label{eqn:z-error}
        \begin{aligned}
            e_{i,0} & = d_{2i} \cdot ({\sf PKE.sk}_{2i,0} \oplus {\sf PKE.sk}_{2i,1}) \\
            &\oplus (d_{2i}^\prime \oplus e_{z,1}^{2i}) \cdot (x_{2i,0} \oplus x_{2i,1}) \oplus z_{2i} \\
            e_{i,1} & = d_{2i+1} \cdot ({\sf PKE.sk}_{2i+1,0} \oplus {\sf PKE.sk}_{2i+1,1}) \\
            &\oplus (d_{2i+1}^\prime \oplus e_{z,1}^{2i+1}) \cdot (x_{2i+1,0} \oplus x_{2i+1,1}) \oplus z_{2i+1}
        \end{aligned}
    \end{equation}
    and $a_i^\prime = postprocessing(q_A^i, a_i, e_{i,0}, e_{i,1})$ for $i \in [n]$ such that $q_B^i \neq 1$. We remind the readers that $postprocessing$ is defined in \cref{dfn:postprocessing}.
    \item If $MSG(q_A^i,q_B^i,a_i^\prime, b_i^\prime) = 0$ for some $i\in[n]$, output $\bot$. Otherwise, output $\top$.

    \begin{lemma}
        The protocol described above satisfies the Decryption correctness and the Deletion verification correctness.
    \end{lemma}
    \begin{proof}
        The honest Lessee in {\sf PKE-cSKL}.$Setup$ and {\sf PKE-cSKL}.$KG$ outputs a key state $\ket{sk_1} \otimes \ket{sk_2} \dots \ket{sk_n}$. The process to generate the key state can be viewed as the parallel repetition of generating $\ket{sk_i}$ for each $i\in [n]$. We prove the correctness for the case $q_B^i = 1$ and $q_B^i \in \{0,2\}$, repsectively.  The readers can refer to \cref{fig:CBQC-honest-lessee} for a summarized description of how the key generation process works.
        \paragraph*{\bf The case for $q_B^i = 1$} After the Step $1$ of $\text{\sf PKE-cSKL}.Setup$, the honest Lessee obtains $\frac{1}{\sqrt 2}(\ket{00}_{{\sf A}_j {\sf B}_j}\ket{x_{j,0}}_{{\sf X}_j} + (-1)^{z_j}\ket{11}_{{\sf A}_j {\sf B}_j}\ket{x_{j,1}}_{{\sf X}_j})$ on register ${\sf A}_j {\sf B}_j {\sf X}_j$.

        In Step $2$ of {\sf PKE-cSKL}.$Setup$, the honest Lessee measures the register ${\sf B}_{2i}$ and ${\sf B}_{2i+1}$ in computational basis. The state collapses to
        \begin{equation}
            \ket{b_i[0], x_{2i, b_i[0]}}_{{\sf A}_{2i} {\sf X}_{2i}}\ket{b_i[1], x_{2i+1, b_i[1]}}_{{\sf A}_{2i+1} {\sf X}_{2i+1}}
        \end{equation}

        In Steps $3(c)$ and $3(d)$ of {\sf PKE-cSKL}.$Setup$, the honest Lessee actually measures the register ${\sf X}_{2i}{\sf X}_{2i+1}$ in the computational basis. Since the states on register ${\sf A}_{2i}{\sf X}_{2i}$ and ${\sf A}_{2i+1}{\sf X}_{2i+1}$ are already in the computational basis, the measurement will not disturb the key state.

        In {\sf PKE-cSKL}.$KG$, the honest Lessee applies the unitary $U_{r_{j,0}, r_{j,1}}$ to obtain 
        \[
        \ket{b_i[0], x_{2i, b_i[0]}, {\sf Ext}(x_{2i, b_i[0]}, r_{2i,b_i[0]})}_{{\sf A}_{2i} {\sf X}_{2i} {\sf SK}_{2i}}
        \]
        on register ${\sf A}_{2i} {\sf X}_{2i} {\sf SK}_{2i}$ and 
        \[
        \ket{b_i[1], x_{2i+1, b_i[1]},{\sf Ext}(x_{2i+1, b_i[1]}, r_{2i+1,b_i[1]})}_{{\sf A}_{2i+1} {\sf X}_{2i+1} {\sf SK}_{2i+1}}
        \]
        on register ${\sf A}_{2i+1} {\sf X}_{2i+1} {\sf SK}_{2i+1}$. Then, the honest Lessee applies $U_{h_{j,0}, h_{j,1}}$ for $j \in \{2i, 2i+1\}$. Since $h_{j,0} = {\sf Ext}(x_{j,0},r_{j,0}) \oplus {\sf PKE.sk}_{j,0}$ and $h_{j,1} = {\sf Ext}(x_{j,1},r_{j,1}) \oplus {\sf PKE.sk}_{j,1}$, the extracted randomness cancels on register ${\sf SK}_j$. The honest Lessee obtains
        \[
            \ket{b_i[0], x_{2i, b_i[0]}, {\sf PKE.sk}_{2i,b_i[0]}}_{{\sf A}_{2i} {\sf X}_{2i} {\sf SK}_{2i}}
        \]
        on register ${\sf A}_{2i} {\sf X}_{2i} {\sf SK}_{2i}$ and
        \[
            \ket{b_i[1], x_{2i+1, b_i[1]},{\sf PKE.sk}_{2i+1,b_i[1]})}_{{\sf A}_{2i+1} {\sf X}_{2i+1} {\sf SK}_{2i+1}}
        \]
        on register ${\sf A}_{2i+1} {\sf X}_{2i+1} {\sf SK}_{2i+1}$.

        \paragraph*{The verification correctness} In this paragraph, we show the deletion verification correctness when $q_B^i = 1$. The key state with the registers ${\sf A}_{2i}$ and ${\sf A}_{2i+1}$ is as follows
        \begin{equation}
            \ket{b_i[0]}_{{\sf A}_{2i}}\ket{b_i[1]}_{{\sf A}_{2i+1}}
        \end{equation}
        The state is the remaining state of Alice, when Bob has already measured its registers on the preshared EPR pairs. In {\sf PKE-cSKL}.$Del$, the honest Lessee measures the register ${\sf SK}_j {\sf X}_j$ in Hadamard basis, which has no effect on the register ${\sf A}_j$. In {\sf PKE-cSKL}.${\sf DelVrfy}$, the Lessor produces $a_i^\prime$, such that $a_i^\prime[1] = a_i[1]$. Then, we note that $q_A^i, q_B^i, a_i^\prime, b_i^\prime$ should be a valid answer of the magic square, which passes the verification for deletion.

        \paragraph*{The decryption correctness} In this paragraph, we show the decryption correctness. The key state is in the support of $\ket{0, {\sf PKE.sk}_{j,0}}\bra{0, {\sf PKE.sk}_{j,0}} + \ket{1, {\sf PKE.sk}_{j,1}}\bra{1, {\sf PKE.sk}_{j,1}}$. By the same calculation as in \cite{KMY24}, the reader should be able to verify the correctness.

        \begin{figure*}[tbp]
            \centering
            \includegraphics[width=0.7\linewidth]{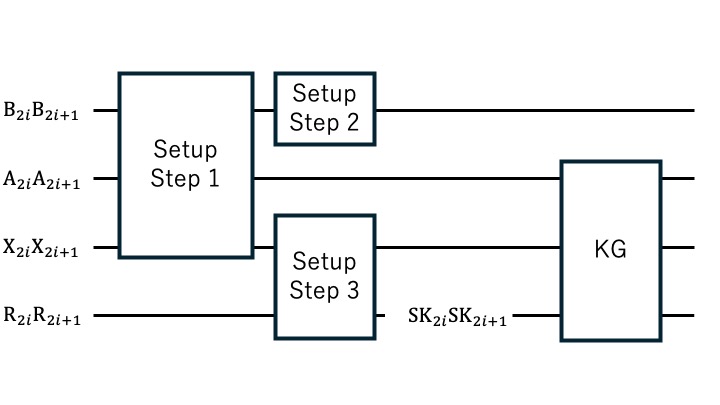}
            \caption{The circuit corresponding to {\sf PKE-cSKL}.KG and each step of {\sf PKE-cSKL}.Setup are illustrated as boxes. The figure shows the registers affected by each unitary.}
            \label{fig:circuit}
        \end{figure*}
        \paragraph*{\bf The case for $q_B^i \neq 1$} After the step $1$ of $\text{\sf PKE-cSKL}.Setup$, the honest Lessee obtains $\frac{1}{\sqrt 2}(\ket{00}_{{\sf A}_j {\sf B}_j}\ket{x_{j,0}}_{{\sf X}_j} + (-1)^{z_j}\ket{11}_{{\sf A}_j {\sf B}_j}\ket{x_{j,1}}_{{\sf X}_j})$ on register ${\sf A}_j {\sf B}_j {\sf X}_j$ for each $j \in [2n]$. 

        As shown in \cref{fig:circuit}, Step $2$ in {\sf PKE-cSKL}.$Setup$ commutes with the other operations. To make the proof brief, we move Step $2$ in {\sf PKE-cSKL}.$Setup$ to the last of the key generation process.

        In the step $3$ of $\text{\sf PKE-cSKL}.Setup$, we note the circuit executed by the Lessor during CBQC does not disturb the quantum state on register ${\sf A}_{2i}{\sf X}_{2i}{\sf A}_{2i+1}{\sf X}_{2i+1}$. However, the CBQC will introduce Pauli $X$ error and $Z$ error to the underlying qubits. We show that the $X$ error does not affect the register ${\sf A}_{2i}{\sf X}_{2i}{\sf A}_{2i+1}{\sf X}_{2i+1}$ and we can trace the $Z$ error as follows.

        \begin{proposition}
            \label{prop:err-analysis}
            For $i \in [n]$ such that $q_B^i = 0$ or $q_B^i = 2$, Step 3 in {\sf PKE-cSKL}.Setup applies a Pauli error $Z^{(x_{j,0} \oplus x_{j,1}) \cdot e_{z,0}^j}$ to the state on register ${\sf A}_{j}{\sf X}_j$, where $j \in \{2i, 2i+1\}$.
        \end{proposition}
        \begin{proof}[Proof of \cref{prop:err-analysis}]
            We show how the quantum state on register ${\sf A}_{2i}{\sf X}_{2i}$ evolves below. The quantum state on register ${\sf A}_{2i+1}{\sf X}_{2i+1}$ evolves similarly.
            \begin{itemize}
                 \item After the Steps 3(a) and 3(b) in {\sf PKE-cSKL}.Setup, the overall state becomes
                 \begin{equation}
                     \frac{1}{\sqrt 2}(\ket{00, x_{2i,0}, x_{2i,0}} + (-1)^{z_j} \ket{11, x_{2i,1}, x_{2i,1}})
                 \end{equation}
                 \item After the Steps 3(c) and 3(d) in {\sf PKE-cSKL}.Setup, the overall state becomes
                 \begin{equation}
                    \begin{aligned}
                        &\frac{1}{\sqrt 2}(\ket{00, x_{2i,0}}X(e_{x,0}^{2i})Z(e_{z,0}^{2i})\ket{x_{2i,0}} + (-1)^{z_j} \ket{11, x_{2i,1}}X(e_{x,0}^{2i})Z(e_{z,0}^{2i})\ket{x_{2i,1}}) \\
                        =& \frac{1}{\sqrt 2}((-1)^{x_{2i,0}\cdot e_{z,0}^{2i}}\ket{00, x_{2i,0}}\ket{x_{2i,0} \oplus e_{x,0}^{2i}} + (-1)^{z_j \oplus (x_{2i,1} \cdot e_{z,0}^{2i})} \ket{11, x_{2i,1}}\ket{x_{2i,1}\oplus e_{x,0}^{2i}}) \\
                        =& \frac{1}{\sqrt 2}(\ket{00, x_{2i,0}}\ket{x_{2i,0} \oplus e_{x,0}^{2i}} + (-1)^{z_j \oplus (x_{2i,0} \oplus x_{2i,1}) \cdot e_{z,0}^{2i}} \ket{11, x_{2i,1}}\ket{x_{2i,1}\oplus e_{x,0}^{2i}})
                    \end{aligned}
                 \end{equation}
                 We can see that a Pauli error $\sigma_z^{(x_{2i,0} \oplus x_{2i,1}) \cdot e_{z,0}^{2i}}$ is introduced to the register ${\sf B}_{2i}$.
                 \item After the Step 3(e) in {\sf PKE-cSKL}.Setup, the Lessee applies CNOT to disentangle the register ${\sf A}_{2i}{\sf X}_{2i}$ and the register ${\sf R}$. The state is 
                 \begin{equation}
                     \frac{1}{\sqrt 2}(\ket{00, x_{2i,0}} + (-1)^{z_j \oplus (x_{2i,0} \oplus x_{2i,1}) \cdot e_{z,0}^{2i}} \ket{11, x_{2i,1}})
                 \end{equation}
                 where a Pauli error $\sigma_z^{(x_{2i,0} \oplus x_{2i,1}) \cdot e_{z,0}^{2i}}$ is applied to the register ${\sf A}_{2i}$.
            \end{itemize}
        \end{proof}

        Then, in the {\sf PKE-cSKL}.$KG$, the honest Lessee applies the unitaries to develop the clawed EPR pair into
        \begin{equation}
             \frac{1}{\sqrt 2}(\ket{00, x_{2i,0}, \text{\sf PKE.sk}_{2i,0}} + (-1)^{z_j \oplus (x_{2i,0} \oplus x_{2i,1}) \cdot e_{z,0}^{2i}} \ket{11, x_{2i,1}, \text{\sf PKE.sk}_{2i,1}})
        \end{equation}

        \paragraph*{The decryption correctness} In this paragraph, we show the decryption correctness. After the honest Lessee runs {\sf SimBob} on registers ${\sf B}_{2i}{\sf B}_{2i+1}$, the state on register ${\sf A}_{2i} {\sf SK}_{2i}$ is still in the subspace spanned by $\ket{0, {\sf PKE.sk}_{2i,0}}$ and $\ket{1, {\sf PKE.sk}_{2i,1}}$. The same can be said about the state on register ${\sf A}_{2i+1} {\sf SK}_{2i+1}$. By simple calculation, the key state is able to decrypt the ciphertext.

        \paragraph*{The verification correctness} In this paragraph, we show the deletion verification correctness. In {\sf PKE-cSKL}.$Del$, the honest Lessee measures the register ${\sf X}_{2i}$ and ${\sf SK}_{2i}$ in Hadamard basis. Let the measurement outcomes be $d_j^\prime$ and $d_j^{\prime\prime}$, respectively. This introduce a Pauli error $\sigma_z^{d^\prime\cdot (x_{2i,0} \oplus x_{2i,1}) \oplus d'' \cdot ({\sf PKE.sk}_{2i,0} \oplus {\sf PKE.sk}_{2i,1})}$ to the register ${\sf A}_{2i}$. The resulting state on register ${\sf A}_{2i} {\sf B}_{2i}$ is an EPR pair with Pauli $Z$ error on ${\sf B}_{2i}$. The same can be said for the register ${\sf A}_{2i+1} {\sf B}_{2i+1}$.
        
        Then, the honest Lessee and the Lessor run {\sf SimBob}. When the Lessor asks the Lessee to delete its keys, the honest Lessee measures the register ${\sf A}_{2i}{\sf A}_{2i+1}$ like Alice does in the magic square game. Combining the Pauli error introduced in {\sf PKE-cSKL}.$Setup$ and the Pauli error introduced in {\sf PKE-cSKL}.$Del$, we can see that the Lessor correctly computes the Pauli error and corrects it in {\sf PKE-cSKL.DelVrfy}. The honest Lessee and the Lessor simulate the honest Alice and Bob in the magic square game. Thus, the honest Lessee should pass the verification correctly.
    \end{proof}
\end{enumerate}

\section{Security proof of 2-party classical {\sf PKE-cSKL}}
\label{sec:pke-ow-vra-proof}
\begin{theorem}
    \label{thm:PKE-OW-VRA}
    The 2-party PKE-cSKL in the previous section satisfies \cref{dfn:PKE-OW-VRA}.
\end{theorem}

Let ${\cal A}$ be an adversary in ${\sf EXP}^\text{\sf ow-vra}_{\text{PKE-cSKL}, A}(1^\lambda)$. We can prove the theorem using the following sequence of Hybrids.

{${\sf Hyb}_0:$} \begin{enumerate}
    \item For each $j \in [2n]$, the challenger and the adversary repeat the following steps: 
    \begin{enumerate}
        \item The challenger samples $({\sf PKE.pk}_{j,0}, {\sf PKE.sk}_{j,0}) $ $\leftarrow$ $ {\sf PKE.KG}(1^\lambda)$.
        \item Then, the challenger and adversary participate in the $\Pi_{CSG}$ protocol. The challenger obtains $x_{j,0}, x_{j,1} \in \{0, 1\}^{l_{sk}}$ and $z_j \in \{0, 1\}$ after the execution.
    \end{enumerate}
     
    \item For each $i \in [n]$, the challenger and the adversary repeats the following steps:
    \begin{enumerate}
        \item The challenger samples $q_B^i, q_A^i \in \{0, 1, 2\}$ uniformly.
        \item  The challenger (as Verifier) and the adversary (as Prover) then engage in the ${\sf SimBob}\langle {\sf V}(1^\lambda, q_B^i), P(1^\lambda) \rangle$ protocol, from which the challenger obtains $b_i$. The adversary computes $b_i^\prime$ such that $b_i^\prime[q_A^i] = a_i[q_B^i]$ and the other bits are generated uniformly at random, where ${\rm par}(b_i^\prime) = 1$.
    \end{enumerate}
     
    \item For each $j \in [2n]$, the challenger and the adversary repeat the following steps:
    \begin{enumerate}
        \item The challenger and the adversary engage in $\Pi_{CBQC}$, where the challenger acts as the Client (C) and the adversary acts as the Sender (S). If $j \in J_B$, the challenger and the adversary run the protocol
        \begin{equation}
            \Pi_{CBQC} = \langle S(1^\lambda, Q, V), C(1^\lambda, Q, 1)  \rangle 
        \end{equation}
        Otherwise, the challenger and the adversary run the protocol
        \begin{equation}
            \Pi_{CBQC} = \langle S(1^\lambda, Q, V), C(1^\lambda, Q, 0)  \rangle 
        \end{equation} After the execution, the challenger obtains the classical string $(e_x^j, e_z^j)$. The challenger parses $e_x^j = e^{j,0}_x || e^{j,1}_x$ and $e_z^j = e^{j,0}_z || e^{j,1}_z$.
        \item The adversary sends a classical string $\omega$ to the challenger.
        \item If $j \notin {J_B}$, the challenger checks if $\omega \oplus e^{j,1}_x = x_{j,c}$, where $c = b_{\lfloor j/2 \rfloor}[j \bmod 2]$. If this condition is not met, the experiment ends and outputs $0$.
    \end{enumerate}
    \item For each $j \in [2n]$, the challenger samples $r_{j,0}, r_{j,1}$. The challenger computes $h_{j,0} = {\sf Ext}(x_{j,0}, r_{j,0}) \oplus {\sf PKE.sk}_{j,0}$ and $h_{j,1} = {\sf Ext}(x_{j,1}, r_{j,1}) \oplus {\sf PKE.sk}_{j,1}$ (${\sf Ext}$ is the randomness extractor defined in \cref{dfn:CSG}). These values $(h_{j,0}, h_{j,1}, r_{j,0}, r_{j,1})$ are then sent by the challenger to the adversary.
    \item The challenger sends the classical public key \textbf{pk := $({\sf PKE.pk}_{j,0}, {\sf PKE.pk}_{j,1})_{j \in [2n]}$} to the adversary.
    \item The adversary sends a classical string ${\sf cert} = (\{a_i\}_{i \in [n]}, \{d_j, d'_j\}_{j \in [2n]})$ to the challenger.
    \item For each $i \in [n]$, the challenger computes $e_{i,0}$ and $e_{i,1}$ using the formulas:
    \begin{equation}
    \label{eqn:z-error-hyb0}
        \begin{aligned}
            e_{i,0} =& d_{2i} \cdot ({\sf PKE.sk}_{2i,0} \oplus {\sf PKE.sk}_{2i,1})  \\
                &\oplus (d'_{2i} \oplus e_{2i,1}^z) \cdot (x_{2i,0} \oplus x_{2i,1}) \oplus z_{2i} \\
            e_{i,1} =& d_{2i+1} \cdot ({\sf PKE.sk}_{2i+1,0} \oplus {\sf PKE.sk}_{2i+1,1}) \\
            &\oplus (d'_{2i+1} \oplus e_{2i+1,1}^z) \cdot (x_{2i+1,0} \oplus x_{2i+1,1}) \oplus z_{2i+1}
        \end{aligned}
    \end{equation}
    Then, the challenger computes $a'_i = {postprocessingA}(q_A^i, a_i, e_{i,0}, e_{i,1})$. If ${MSG}(q_A^i, q_B^i, a'_i, b_i^\prime) = 0$ for any $i \in [n]$, the experiment ends and the output is $0$.
    \item The challenger samples a message $m \in \mathcal{M}$ uniformly. For each $j \in [2n]$, the challenger computes $ct_{j,0} \leftarrow {\sf PKE.Enc}({\sf PKE.pk}_{j,0}, m[j])$ and $ct_{j,1} \leftarrow {\sf PKE.Enc}({\sf PKE.pk}_{j,1}, m[j])$.
    \item Let 
    \begin{equation}
        {{\sf ct}} = (\{ct_{j,0}, ct_{j,1}\})_{j \in [2n]}
    \end{equation}
    and 
    \begin{equation}
        \begin{aligned}
            dvk := (\{ q_B^i, b_i^\prime, q_A^i \}_{i \in [n]}, \{ {\sf PKE.sk}_{j,0}, {\sf PKE.sk}_{j,1}, \\
            e_{z,1}^j, z_j, x_{j,0}, x_{j,1}\}_{j \in [2n]: j \notin J_B})
        \end{aligned}
    \end{equation}
    The challenger sends the ciphertext ${{\sf ct}}$ and the verification key ${dvk}$ to the adversary.
    \item The adversary sends a message $m' \in \mathcal{M}$ to the challenger.
    \item The challenger outputs $1$ if $m' = m$. Otherwise, the challenger outputs $0$.
\end{enumerate}

{${\sf Hyb}_1:$} We define ${\sf Hyb}_1$ almost the same as ${\sf Hyb}_0$, except for the following differences. ${\sf Hyb}_1$ introduces a modification to the plaintext $m$ prior to encryption. Specifically, in Step $8$, for indices $j \in J_B$ , $m[j]$ is XORed with $b_i[j \mod 2]$ to form $\bar m[j]$. For $j \notin J_B$, $\bar m[j]$ = $m[j]$. This $\bar m$ is then encrypted using ${\sf PKE.Enc}({\sf PKE.pk}_{j,0}, \bar m[j])$ and ${\sf PKE.Enc}({\sf PKE.pk}_{j,1}, \bar m[j])$. The adversary's success condition is correspondingly updated to correctly recover $\bar m$ instead of $m$ in Step $11$.

\begin{lemma}
    \label{lem:pke-hyb0-hyb1-diff}
    We have that $|\Pr[{\sf Hyb}_0 = 1] - \Pr[{\sf Hyb}_1 = 1]|=0$.
\end{lemma}
\begin{proof}[\cref{lem:pke-hyb0-hyb1-diff}]
    The challenge $m$ in ${\sf Hyb}_0$ is sampled uniformly at random. The challenge $\bar m$ in ${\sf Hyb}_1$ is obtained by padding $b_i[j \mod 2]$ to $m[j]$ for each $j \in J_B$. Thus, $\bar m$ is also chosen uniformly at random. The input to the adversary is the same in ${\sf Hyb}_0$ and ${\sf Hyb}_1$, which completes the proof.
\end{proof}

{${\sf Hyb}_2:$} We define ${\sf Hyb}_2$ almost the same as ${\sf Hyb}_1$, except for the following differences. ${\sf Hyb}_2$ modifies the ciphertext generation in Step $8$ to leverage the IND-CPA security of the underlying PKE scheme. While ${\sf Hyb}_1$ computes $ct_{j,0} \leftarrow {\sf PKE.Enc}({\sf PKE.pk}_{j,0}, \bar m[j])$ and $ct_{j,1} \leftarrow {\sf PKE.Enc}({\sf PKE.pk}_{j,1}, \bar m[j])$, ${\sf Hyb}_2$ computes $ct_{j,b_i[j \mod 2]} \leftarrow  {\sf PKE.Enc}({\sf PKE.pk}_{j,b_i[j \mod 2]}, m[j] \oplus b_i[j \mod 2])$ and $ct_{j,1-b_i[j \mod 2]}$ $\leftarrow$\ ${\sf PKE.Enc}({\sf PKE.pk}_{j,1-b_i[j \mod 2]}, m[j] \oplus (1- b_i[j \mod 2]))$ for $j \in J_B$. 

\begin{lemma}
    \label{lem:pke-hyb1-hyb2-diff}
    We have that $|\Pr[{\sf Hyb}_1 = 1] - \Pr[{\sf Hyb}_2 = 1]|={\rm negl}(\lambda)$.
\end{lemma}
\begin{proof}[\cref{lem:pke-hyb1-hyb2-diff}]
    We will show the Lemma by arguing that the adversary has no information about ${\sf PKE.sk}_{j, (1- b_i[j \mod 2])}$. This is concluded in
    
    {${\sf Hyb}_1^\prime$:} This is basically the same Hybrid as ${\sf Hyb}_1$ except for:
    \begin{itemize}
        \item Let $i = \lfloor j/2 \rfloor$. In Step $4$, the challenger samples ${\sf rand}_j \in \{0,1\}^{l_{sk}}$ uniformly at random and computes $h_{j, (1-b_i[j \mod 2])} = {\sf rand}_j \oplus {\sf PKE.sk}_{j,(1-b_i[j \mod 2])}$, for each $j \in J_B$.
    \end{itemize}
    By the randomness extraction property of CSG (see \cref{dfn:CSG}), we obtain the following equation.
    \begin{equation}
        \label{eqn:alter-hj}
        |\Pr[{\sf Hyb}_1 = 1] - \Pr[{\sf Hyb}_1^\prime=1]| = {\rm negl}(\lambda)
    \end{equation}
    
    {${\sf Hyb}_1^{\prime\prime}$:} This is basically the same Hybrid as ${\sf Hyb}_1^\prime$ except for:
    \begin{itemize}
        \item Let $i = \lfloor j/2 \rfloor$. In Step $8$, the challenger generates $ct_{j,(1-b_i[j \mod 2])} \leftarrow {\sf PKE.Enc}({\sf PKE.pk}_{j, 1-b_i[j \mod 2]}, m[j] \oplus (1-b_i[j \mod 2]))$ instead of $ct_{j,(1-b_i[j \mod 2])} \leftarrow {\sf PKE.Enc}({\sf PKE.pk}_{j, 1-b_i[j \mod 2]}, \bar m[j])$, for each $j \in J_B$.
    \end{itemize}
    Note that the distribution of $h_{j,(1-b_i[j \mod 2])}$ is the same as the uniform distribution. The adversary in ${\sf Hyb}_1^\prime$ and ${\sf Hyb}_1^{\prime\prime}$ has no information about ${\sf PKE.sk}_{j, (1 - b_i[j \mod 2])}$. By the IND-CPA of PKE, we obtain the equation as follows:
    \begin{equation}
        \label{eqn:alter-the-challenge}
        |\Pr[{\sf Hyb}_1^\prime = 1] - \Pr[{\sf Hyb}_1^{\prime\prime}=1]| = {\rm negl}(\lambda)
    \end{equation}
    By the same argument as for \cref{eqn:alter-hj}, we obtain the following equation:
    \begin{equation}
        \label{eqn:alter-hj-back}
        |\Pr[{\sf Hyb}_1^{\prime\prime} = 1] - \Pr[{\sf Hyb}_2=1]| = {\rm negl}(\lambda)
    \end{equation}
    Combining \cref{eqn:alter-hj}, \cref{eqn:alter-the-challenge} and \cref{eqn:alter-hj-back}, we complete the proof of \cref{lem:pke-hyb1-hyb2-diff}.
\end{proof}

{${\sf Hyb}_3:$} We define ${\sf Hyb}_3$ almost the same as ${\sf Hyb}_2$, except for the following differences. ${\sf Hyb}_3$ removes the specific abort condition found in Step $3(c)$ of ${\sf Hyb}_2$. In ${\sf Hyb}_2$, if $j \notin J_B$, the challenger checks $\omega \oplus e_{x,1}^j \neq x_{j,c}$ where $c = b_{\lfloor j/2 \rfloor}[j \mod 2]$. If the condition were not met, the experiment would abort and output $0$. This verification check is entirely omitted in ${\sf Hyb}_3$, ensuring the experiment proceeds regardless of this outcome.

Since ${\sf Hyb}_3$ removes the abort condition in Step $3(c)$, the adversary can always win with higher probability using the same strategy as in ${\sf Hyb}_2$. We conclude this fact in the lemma below.
\begin{lemma}
    \label{lem:pke-hyb2-hyb3-leq}
    We have $\Pr[{\sf Hyb}_2 = 1] \leq \Pr[{\sf Hyb}_3 = 1]$.
\end{lemma}

{${\sf Hyb}_4:$} We define ${\sf Hyb}_4$ almost the same as ${\sf Hyb}_3$, except for the following differences. ${\sf Hyb}_4$ alters the input to the $\Pi_{CBQC}$ protocol in Step $3(a)$. In ${\sf Hyb}_3$, $\Pi_{CBQC}$ is run with input $1$ if $j \in J_B$ and $0$ otherwise. In ${\sf Hyb}_4$, the $\Pi_{CBQC}$ protocol is uniformly run with input $0$ for all $j \in [2n]$.

\begin{lemma}
    \label{lem:pke-hyb3-hyb4-diff}
    We have that $|\Pr[{\sf Hyb}_3 = 1] - \Pr[{\sf Hyb}_4 = 1]| = {\rm negl}(\lambda)$.
\end{lemma}
\begin{proof}[\cref{lem:pke-hyb3-hyb4-diff}]
    By the blindness of $\Pi_{CBQC}$ (\cref{dfn:cbqc}), we can see that changing the classical input from $1$ to $0$ does not affect the output distribution. This proves \cref{lem:pke-hyb3-hyb4-diff}.
\end{proof}

\begin{lemma}
    \label{lem:pke-hyb4-negl}
    We have that $\Pr[{\sf Hyb}_4 = 1] = {\rm negl}(\lambda)$.
\end{lemma}

\begin{proof}[\cref{lem:pke-hyb4-negl}]
    We prove \cref{lem:pke-hyb4-negl}. We bound $\Pr[{\sf Hyb}_4 = 1]$ using the Certified Deletion Property of the Magic Square Game (\cref{dfn:comp-CDP}). We can transform any Alice and Bob (the adversary for ${\sf Hyb}_4$) into an adversary $\tilde P =(A_0, A_1)$ against the Certified Deletion Property of the Magic Square Game \cref{dfn:comp-CDP}.
    
    {$A_0$}: The adversary $A_0$ (acting as Prover $\tilde{P}$ for the CCD game and simulating the challenger for an internal ${\sf Hyb}_4$ adversary ${\cal A}^{{\sf Hyb}_4}$) performs the following steps:
    \begin{enumerate}
    \item Initialize the ${\sf Hyb}_4$ adversary ${\cal A}^{{\sf Hyb}_4}$. 
    \item For each $j \in [2n]$: 
        \begin{enumerate} 
            \item Sample $({\sf PKE.pk}_{j,0}, {\sf PKE.sk}_{j,0}) \leftarrow {\sf PKE.KG}(1^\lambda)$. 
            \item Participate in the $\Pi_{CSG}$ protocol with ${\cal A}^{{\sf Hyb}_4}$ to obtain $x_{j,0}, x_{j,1} \in \{0, 1\}^{l_{sk}}$ and $z_j \in \{0, 1\}$. 
        \end{enumerate} 
        \item For each $i \in [n]$: 
        \begin{enumerate} 
            \item Engage in the ${\sf SimBob}\langle {\sf V}(1^\lambda, q_B^i), P(1^\lambda) \rangle$ protocol as the Prover (P). Forward the messages from the Verifier to ${\cal A}^{{\sf Hyb}_4}$, the answers from ${\cal A}^{{\sf Hyb}_4}$ to the Verfier.
            \item Receive $q_A^i \in \{0, 1, 2\}$ from $V$.
            \item Send $q_A^i$ to ${\cal A}^{{\sf Hyb}_4}$. 
        \end{enumerate} 
        \item For each $j \in [2n]$: 
        \begin{enumerate} 
            \item Engage in the $\Pi_{CBQC}$ protocol as Client (C) with ${\cal A}^{{\sf Hyb}_4}$ (Sender), using input $0$ for all $j \in [2n]$: $\Pi_{CBQC} = \langle S(1^\lambda, Q, V), C(1^\lambda, Q, 0) \rangle$. Obtain classical string $(e_x^j, e_z^j)$, parsing $e_x^j = e^{j,1}_z || e^{j,1}_x$ and $e_z^j = e^{j,0}_z || e^{j,1}_z$. 
            \item Receive a classical string $\omega$ from ${\cal A}^{\sf Hyb_4}$.
        \end{enumerate} 
        \item For each $j \in [2n]$, sample $r_{j,0}, r_{j,1}$. Compute $h_{j,0} = {\sf Ext}(x_{j,0}, r_{j,0}) \oplus {\sf PKE.sk}_{j,0}$ and $h_{j,1} = {\sf Ext}(x_{j,1}, r_{j,1}) \oplus {\sf PKE.sk}_{j,1}$. Send $(h_{j,0}, h_{j,1}, r_{j,0}, r_{j,1})$ to ${\cal A}^{\sf Hyb_4}$.
        \item Send the classical public key ${\sf pk := ({\sf PKE.pk}_{j,0}, {\sf PKE.pk}_{j,1})}_{j \in [2n]}$ to ${\cal A}^{\sf Hyb_4}$.
        \item Receive a classical string ${\sf cert} = (\{a_i\}_{i \in [n]}, \{d_j, d'_j\}_{j \in [2n]})$ from ${\cal A}^{\sf Hyb_4}$.
        \item For each $i \in [n]$:
            \begin{enumerate}
            \item Compute $e_{i,0}$ and $e_{i,1}$ according to \cref{eqn:z-error-hyb0}.
            \item Compute $a'_i$ = ${postprocessing_A}$($q_A^i$, $a_i$, $e_{i,0}$, $e_{i,1}$).
            \end{enumerate} 
        \item Output the internal state $st$ and $\{a^\prime_i\}_{i \in [n]}$ as the output. 
    \end{enumerate}
    
    {$A_1$}: The adversary $A_1$ (acting as Prover $\tilde{P}$ for the CCD game and continuing to simulate the challenger for ${\cal A}^{{\sf Hyb}_4}$) performs the following steps:
    \begin{enumerate}
        \item Receive the internal state $st$ from $A_0$ and the list $\{q_B^i\}_{i \in [n]}$ from the CCD Verifier $V$. 
        \item The adversary computes $b_i^\prime$ such that $b_i^\prime[q_A^i] = a_i[q_B^i]$ and the other bits are generated uniformly at random, where ${\rm par}(b_i^\prime) = 1$.
        \item Sample a message $m \in \mathcal{M}$ uniformly.
        \item For each $j \in J_B$, compute $ct_{j,0} \leftarrow {\sf PKE.Enc}({\sf PKE.pk}_{j,0}, m[j] )$ and $ct_{j, 1} \leftarrow {\sf PKE.Enc}({\sf PKE.pk}_{j,1}, m[j] \oplus 1)$. For $j \notin J_B$, compute $ct_{j,0} \leftarrow {\sf PKE.Enc}({\sf PKE.pk}_{j,0}, m[j])$ and $ct_{j,1} \leftarrow {\sf PKE.Enc}({\sf PKE.pk}_{j,1}, m[j])$.
        \item Let
        \begin{equation}
            {\sf ct} = ({ct_{j,0}, ct_{j,1}})_{j \in [2n]}
        \end{equation} 
        and 
        \begin{equation}
            \begin{aligned}
                {\sf dvk} := (\{q_B^i, b_i, q_A^i\}_{i \in [n]}, \{{\sf PKE.sk}_{j,0}, {\sf PKE.sk}_{j,1}, \\
                e^{j,1}_z, z_j, x_{j,0}, x_{j,1}\}_{j \in [2n]: j \notin {J_B}})
            \end{aligned}
        \end{equation}
         Send ${\sf ct}$ and ${\sf dvk}$ to ${\cal A}^{{\sf Hyb}_4}$. 
        \item Receive a message $m' \in \mathcal{M}$ from ${\cal A}^{{\sf Hyb}_4}$. 
        \item Output $m \oplus m'$. 
    \end{enumerate}

    First, we show that whenever the adversary ${\cal A}^{\sf Hyb_4}$ produces a valid ${\sf cert}$, $\{a^\prime_i\}_{i\in[n]}$ is a valid answer for the Magic Square Game. By definition of  {\sf PKE-cSKL.DelVrfy}, $MSG(q_A^i, q_B^i, a^\prime_i, b_i)=1$ for each $i\in[n]$. Thus, $A_0$ produces a valid answer for the Magic Square Game with the same probability as ${\cal A}^{\sf Hyb_4}$ produces a valid certificate of deletion.
    \begin{remark}
        The reader may notice that while $A_0$ outputs the postprocessed $a_i^\prime$ for every $i \in [n]$, the Lessor only postprocesses $a_i$ for $i \in [n]$ such that $q_B^i \neq 1$. For $i \in [n]$ such that $q_B^i = 1$, the challenger checks whether $a_i^\prime[1] = b_i[q_A^i]$ where $a_i[1]$ is untouched in the postprocessing. For other $i$, $A_0$ does exactly the same as the lessor does in {\sf PKE-cSKL.DelVrfy}. Thus, we can see that postprocessing every $a_i$ and output $a_i^\prime$ as the answer does not compromise the validity of the certificate.
    \end{remark}

    Then, we can see that whenever the adversary ${\cal A}^{\sf Hyb_4}$ produces the correct $m^\prime$, we have $m^\prime[j] \oplus m[j] = b_i[j \mod 2]$ for $j \in J_B$. For $i$ such that $q_i^B = 1$, the adversary obtains $b_i[0]$ and $b_i[1]$. With the information, the adversary can recover $b_i$ for $i$ such that $q_i^B = 1$. 
    \begin{remark}
        Since ${\rm par}(b)=1$ for the valid answer of MSG, it suffices to recover $b$ using only a single bit other than $b[q_A^i]$. We can see that at least one of $b_i[0]$ and $b_i[1]$ differs from $b[q_A^i]$.
    \end{remark}
    
    Combining the two facts above, we have
    \begin{equation}
        \Pr[{\sf Hyb}_4 = 1] \leq \Pr[{\cal A}^{\sf Hyb_4}\ wins\ {\sf CCD}]
    \end{equation}
\end{proof}
 
Now, we prove \cref{thm:PKE-OW-VRA}. 
     
\begin{proof}[\cref{thm:PKE-OW-VRA}]     
     Combining \cref{lem:pke-hyb0-hyb1-diff}, \cref{lem:pke-hyb1-hyb2-diff}, \cref{lem:pke-hyb2-hyb3-leq}, \cref{lem:pke-hyb3-hyb4-diff}, \cref{lem:pke-hyb4-negl}, we have that
    \begin{align}
        \Pr[{\sf Hyb}_0 = 1] = \Pr[{\sf Hyb}_1 = 1] \approx_{{\rm negl}(\lambda)} \Pr[{\sf Hyb}_2 = 1] \\
        \leq \Pr[{\sf Hyb}_3 = 1] \approx_{{\rm negl}(\lambda)} \Pr[{\sf Hyb}_4 = 1] = {\rm negl}(\lambda)
    \end{align}
\end{proof}

\begin{theorem}
    Assuming the existence of CSGs (see \cref{dfn:CSG}) and PKE and the soundness of parallel repetition, there exists a PKE-cSKL satisfying \cref{dfn:PKE-IND-VRA}.
\end{theorem}

\section{Conclusion}
In this paper, we revisited the KLVY compiler \cite{KLVY22}. We proved that the certified deletion property is preserved after the compilation. This is the first time that someone has investigated the property of the compiled game besides the quantum value and the rigidity. Then, we combine the certified deletion property with the framework from \cite{KMY24} to obtain classical SKL for PKE, PRF, and DS. Our second contribution is that we obtain the first SKL for PRF and DS with a classical lessor. Our protocol requires only the existence of CSGs, which is a primitive constructed from the hardness of LWE, the hardness of cryptographic group actions, etc.
\subsection{Open Problems}
In this section, we conclude the open problems:
\paragraph*{\bf Quantitative Bounds of the compiled certified deletion property} In this work, we have shown that for security parameter $\lambda$ larger than a constant $\lambda_c$, the winning probability is smaller than $w<1$. This suffices for our purpose. But a question is how fast the winning probability converges. Prior works have proposed methods to give the quantitative bounds of the compiled game's winning probability using semi-definite programming\cite{CFNZ25,KPR＋25}. However, the game $\text{\sf comp-CD-MSG}(\lambda)$ consists of too many inputs and outputs and thus the SDP program's size is so large that it becomes infeasible to solve.
\paragraph*{\bf Parallel repetition of the compiled certified deletion property} In this work, we conjectured that parallel repetition reduces the winning probability exponentially. We note that \cite{HK25,BQSY24} have shown the Parallel Repetition Theorem for (private-coin) {\bf three-message} arguments. However, the security game for the Certified Deletion Property consists of six messages. It suggests that we need to search for another way to show the parallel repetition.

\backmatter

\bmhead{Supplementary information}

If your article has accompanying supplementary file/s please state so here. 

Authors reporting data from electrophoretic gels and blots should supply the full unprocessed scans for key as part of their Supplementary information. This may be requested by the editorial team/s if it is missing.

Please refer to Journal-level guidance for any specific requirements.

\section*{Declarations}
% \subsection{Funding}
% DX is financially supported by JST SPRING, Grant Number 8 JPMJSP2125. YT is partially supported by the MEXT Quantum Leap Flagship Program (MEXT Q-LEAP) Grant Number JPMXS0120319794.
% \subsection{Competing Interests}
% The authors have no relevant financial or non-financial interests to disclose.
% \subsection{Ethics approval and consent to participate}
% Not applicable.
% \subsection{Data, Material and Code availability}
% No data, material, or code is used or generated during the research.
% \subsection{Consent for publication}
% All authors agree to publish.
% \subsection{Author{s'} contribution statements}
% The research was mainly conducted by D.X. {Y.T.} took part in the discussion and helped D.X{.} improve{d} the presentation of this work.

\bmhead{Acknowledgements}
This work was financially supported by JST SPRING, Grant Number 8 JPMJSP2125. The author (XU) would like to take this opportunity to thank the ``THERS Make New Standards Program for the Next Generation Researchers.'' YT is partially supported by the MEXT Quantum Leap Flagship Program (MEXT Q-LEAP) Grant Number JPMXS0120319794. Also, we thank Prof. Harumichi Nishimura, who is the supervisor of XU, for checking the writing of this paper.


\begin{appendices}

\section{Proof of \cref{lem:NI-CD-MSG-qc}}
\label{apdx:MSG}
First, we will state some useful lemmas.
\begin{lemma}[Maccone-Pati's Uncertainty Relation (\cite{MP14,Maz17})]
    \label{lem:UR}
    Let $\ket{\psi}$ be a quantum state and $\ket{\psi_\perp}$ be any normalized quantum state which is orthogonal to $\ket{\psi}$. For any two observables $A$ and $B$, the following inequality holds:
    \begin{equation}
        \label{eqn:UR}
        \Delta(A)+\Delta(B) \geq \max(L_1, L_2)
    \end{equation}
    where $\Delta(\cdot)$ is the variance and
    \begin{equation}
        \begin{aligned}
            L_1 =& 1/2 |\bra{\psi}(A \pm B) \ket{\psi_\perp}|^2 \\
            L_2 =& \pm i \bra{\psi}[A,B]\ket{\psi}+|\bra{\psi}(A \pm iB)\ket{\psi_\perp}|^2
        \end{aligned}
    \end{equation}
\end{lemma}

In \cite{MP14}, they have shown the following fact:
\begin{corollary}
    \label{cor:anticommute-UR}
    Let $\ket{\psi}$ be a quantum state. Let $A$ and $B$ be two observables such that $\bra{\psi}\{A,B\}\ket{\psi}=0$ and $AB\ket{\psi} \neq 0$. We have
    \begin{equation}
        \Delta(A)+\Delta(B) > 0
    \end{equation}
\end{corollary}

We prove MSG's certified deletion property for one-shot MSG.
\begin{lemma}[The certified deletion property of Magic Square Game]
    \label{lem:cd-msg}
    Let us consider the security game $\text{\sf CD-MSG}$ for the certified deletion property:
    \begin{enumerate}
        \item $R$ samples $q_A, q_B \in \{1,2,3\}$ uniformly. Then, it sends $q_A$ and $q_B$ to Alice and Bob, respectively.
        \item Alice sends $a \in \{0,1\}^3$ and Bob sends $b \in \{0,1\}^3$ to the $R$.
        \item $R$ outputs $0$ and aborts if
        \begin{equation}
            MSG(q_A, q_B, a, b) = \bot
        \end{equation}
        \item $R$ sends $q_B^\prime = q_B$ to Alice.
        \item Alice sends $b^\prime \in \{0,1\}^3$ to $R$.
        \item $R$ outputs $0$ if $q_B^\prime = 2 \land  b\neq b^\prime$. Otherwise, $R$ outputs $1$.
    \end{enumerate}
    Alice and Bob win the game if $R$ outputs $1$. Let Alice and Bob's strategy be $({\cal H}, \ket{\psi}\in{\cal H}, \{A_{q_A}\}_{q_A \in \{1,2,3\} }, \{B_{q_B}\}_{q_B \in \{1,2,3\}})$, where ${\cal H}$ is an arbitray Hilbert space, $\ket{\psi}$ is a quantum state on ${\cal H}$, $A_{q_A}$ and $B_{q_B}$ are commutable measurements on ${\cal H}$ for any $q_A, q_B \in \{1,2,3\}$. We define the quantum commuting value $w_{qc}$ as the maximum winning probability of Alice and Bob.

    We have 
    \begin{equation}
        w_{qc} < 1
    \end{equation}
\end{lemma}

Note that our lemma differs from that in \cite{KT20}. In the second round, we only require Alice to answer correctly when Bob measures its part in the computational basis. We prove \cref{lem:cd-msg} as follows.
\begin{proof}
First, we assume that $w_{qc} = 1$. Then, we show a contradiction and complete our proof.

Let $A_{q_A} = \{A_{q_A}^{a,b^\prime}\}_{a,b^\prime}$ and $B_{q_B} = \{ B_{q_B}^{b} \}_{b}$ be projective measurements. We define observables $F_{r,c}$ and $G_{r,c}$ as follows:
\begin{equation}
    F_{r,c} = \sum_{a,b^\prime} (-1)^{a[c]} A_{r}^{a,b^\prime} ,\quad 
    G_{r,c} = \sum_{b} (-1)^{b[r]} B_{c}^{b}
\end{equation}
When $w_{qc} = 1$, Alice and Bob must produce answers that pass Step $3$ with probability $1$. In other words, $\{F_{r,c}\}$ and $\{G_{r,c}\}$ is an optimal strategy for the Magic Square Game. By the rigidity of the Magic Square Game\cite{FM18,WBMS16}, we have 
\begin{equation}
    \bra{\psi}\{ G_{r,c},
    G_{r^\prime,c^\prime} \} \ket{\psi} = 0
\end{equation}

Let $\ket{\psi_{a,b^\prime,q_A}}$ be the post-measurement state. When we fix $q_A=1, q_B = 2$, the following equation holds:
\begin{equation}
    b[2] = b^\prime[2] 
\end{equation}
Combining the equation above with $w_{qc}=1$, we have $\Delta(G_{2,2}) \coloneqq \bra{\psi_{a,b^\prime,1}}G_{2,2}^2\ket{\psi_{a,b^\prime,1}}-(\bra{\psi_{a,b^\prime,1}}G_{2,2}\ket{\psi_{a,b^\prime,1}})^2 = 0$ where $\Delta(G_{2,2})$ is the variance. This states the fact that $b[2]$ is uniquely determined when $a,b^\prime,q_A,q_B$ are fixed.

By the winning condition of MSG, when we fix $q_A=1, q_B = 1$, we have $\Delta(G_{1,1})=0$. We have $\Delta(G_{1,1}) + \Delta(G_{2,2}) = 0$, which contradicts with \cref{cor:anticommute-UR}. When $G_{1,1}$ and $G_{2,2}$ are anti-commutable, $\ket{\psi_{a,b^\prime,1}}$ is not the common eigenstate. By construction, $G_{1,1}$ and $G_{2,2}$ have only eigenvalues $\pm 1$. If $\ket{\psi_{a,b^\prime,1}}$ is the common eigenstate, $\bra{\psi}\{ G_{1,1}, G_{2, 2} \} \ket{\psi} = \pm 2$. The fact that $\ket{\psi_{a,b^\prime,1}}$ is not the common eigenstate implies $\Delta(G_{1,1}) + \Delta(G_{2,2}) > 0$. We complete the proof.
\end{proof}

\section{PRF-cSKL and DS-cSKL}
\label{apdx:PRF-DS-cSKL}

\subsection{Pseudo-random Functions with Classical Secure Key Leasing}
In this subsection, we define the syntax for pseudo-random functions with classical secure key leasing (PRF-cSKL). Our definition is analogous to the definition in \cite{KMY24}.

\begin{definition}[PRF-cSKL]
    \label{dfn:PRF-CSKL}
    Let $D_{\sf prf}$ be the domain and $R_{\sf prf}$ be the range of the PRF. A scheme of PRF-cSKL is a tuple of algorithms $({\sf Setup}, {\sf KG}, {\sf Eval}, LEval, Del, {\sf DelVrfy})$:
    \begin{itemize}
        \item[] {${\sf Setup}{\langle Lessor(1^\lambda), Lessee(1^\lambda) \rangle} \rightarrow ({\sf mk}, st)/\bot$} is an interactive protocol between
            \begin{itemize}
                \item a QPT Lessee
                \item a PPT Lessor
            \end{itemize}
        When the protocol fails, the Lessor and the Lessee output $\bot$. When the protocol succeeds, the Lessor outputs a classical master key ${\sf mk}$. The (honest) Lessee outputs an internal state $st$.
        \item[] {${\sf KG}{\langle Lessor(1^\lambda, {\sf mk}), Lessee(1^\lambda, st) \rangle} \rightarrow (sk, {\sf msk}, {\sf dvk})$} is an interactive protocol between
            \begin{itemize}
                \item a QPT Lessee
                \item a PPT Lessor
            \end{itemize}
        The Lessor outputs a classical master secret key ${\sf msk}$ and a classical ${\sf dvk}$. The (honest) Lessee outputs a quantum evaluation key $sk$.
        \item[] {${\sf Eval}({\sf msk}, s) \rightarrow t$} is a PPT algorithm. ${\sf msk}$ is a classical master secret key and $s \in D_{\sf prf}$ is an input from the domain. The algorithm evaluates the PRF at point $s$ and outputs $t$.
        \item[]{$LEval(sk,s) \rightarrow t$} is a QPT algorithm. $sk$ is a leased quantum evaluation key and $s \in D_{\sf prf}$ is an input from the domain. The algorithm evaluates the PRF at point $s$ and outputs $t$.
        \item[] {$Del(sk) \rightarrow {\sf cert}$} is a QPT algorithm. $sk$ is a quantum evaluation key. The algorithm destroys the quantum key and outputs a valid classical certificate ${\sf cert}$.
        \item[] {${\sf DelVrfy}({\sf cert}, {\sf dvk}) \rightarrow \top / \bot$} is a PPT algorithm. The algorithm takes as input a classical certificate ${\sf cert}$ and a classical verification key ${\sf dvk}$. The algorithm outputs $\top$ if ${\sf cert}$ is a valid certificate of deletion. It outputs $\bot$ if ${\sf cert}$ is not a valid certificate.
    \end{itemize}
\end{definition}

{\bf Evaluation correctness:} For every $s\in D_{\sf prf}$, we have that
\begin{equation}
    \label{eqn:PRF-cSKL-evaluation-correctness}
    \Pr[t \neq t^\prime \lor res=\bot: 
    \begin{aligned}
        & res \leftarrow \text{\sf PRF-cSKL}.{\sf Setup}{\langle Lessor(1^\lambda), Lessee(1^\lambda) \rangle} \\
        & ({\sf mk},st) \coloneqq res \\
        & (sk, {\sf msk}, {\sf dvk}) \leftarrow \text{\sf PRF-cSKL}.{\sf KG}{\langle Lessor(1^\lambda, {\sf mk}), Lessee(1^\lambda, st) \rangle}  \\
        & t \leftarrow \text{\sf PRF-cSKL}.{\sf Eval}({\sf msk},s) \\
        & t^\prime \leftarrow \text{\sf PRF-cSKL}.LEval(sk,s)
    \end{aligned}
    ] = {\rm negl}(\lambda)
\end{equation}

{\bf Deletion verification correctness:} We have
\begin{equation}
    \Pr[b = \bot \lor res=\bot: 
    \begin{aligned}
        & res \leftarrow \text{\sf PRF-cSKL}.{\sf Setup}{\langle Lessor(1^\lambda), Lessee(1^\lambda) \rangle} \\
        & ({\sf mk},st) \coloneqq res \\
        & (sk, {\sf msk}, {\sf dvk}) \leftarrow \text{\sf PRF-cSKL}.{\sf KG}{\langle Lessor(1^\lambda, {\sf mk}), Lessee(1^\lambda, st) \rangle}  \\
        & {\sf cert} \leftarrow \text{\sf PRF-cSKL}.Del(sk) \\
        & b \leftarrow \text{\sf PRF-cSKL}.{\sf DelVrfy}({\sf cert}, {\sf dvk})
    \end{aligned}
    ] = {\rm negl}(\lambda)
\end{equation}

Then, we will introduce the security definition.

\begin{definition}[PR-VRA security]
    \label{dfn:PRF-PR-VRA}
    The IND-VRA security for a PRF-cSKL scheme is formalized by the experiment ${\sf EXP}^\text{\sf pr-vra}_{\text{PRF-cSKL}, A}(1^\lambda,{\sf coin})$:
    \begin{enumerate}
        \item The challenger $C$ and the adversary $A$ run $res \leftarrow \text{\sf PRF-cSKL.Setup}\langle C(1^\lambda), A(1^\lambda) \rangle$. If $res = \bot$ (${\sf Setup}$ aborted), the experiment ends and the output is $0$. Otherwise, we have $res \coloneqq ( {\sf mk}, st)$ where the challenger has ${\sf mk}$ and the adversary has $st$. The challenger and the adversary run $(sk, {\sf msk}, {\sf dvk}) \leftarrow \text{\sf PRF-cSKL}.{\sf KG}{\langle Lessor(1^\lambda, {\sf mk}), Lessee(1^\lambda, st) \rangle}$ where the challenger obtains {\sf msk} and {\sf dvk}.
        \item The adversary sends a classical string {\sf cert}. If $\text{\sf PRF-cSKL.DelVrfy}({\sf cert}, {\sf dvk}) = \bot$, the challenger outputs $0$ and the experiment ends. Otherwise, the challenger computes $s \leftarrow D_{\sf prf}, t_0 \leftarrow \text{\sf PRF-cSKL}.{\sf Eval}({\sf msk},s), t_1 \leftarrow R_{\sf prf}$ and sends $({\sf dvk}, t_{\sf coint},s)$ to the adversary.
        \item The adversary outputs a guess ${\sf coin}^\prime \in \{0,1\}$.
    \end{enumerate}
    {\sf PRF-cSKL} is PR-VRA secure if and only if for any QPT $A$
    \begin{equation}
        \begin{aligned}
            {\sf Adv}^\text{\sf pr-vra}_{\text{PRF-cSKL},A}(\lambda):= |\Pr[{\sf EXP}^\text{\sf pr-vra}_{\text{PRF-cSKL}, A}(1^\lambda,{\sf 0})=1] \\
            -\Pr[{\sf EXP}^\text{\sf pr-vra}_{\text{PRF-cSKL}, A}(1^\lambda,{\sf 1})=1]| = {\rm negl}(\lambda)
        \end{aligned}
    \end{equation}
\end{definition}

We can also define a one-way variant of the security above.

\begin{definition}[UPF-VRA security]
    \label{dfn:PRF-UPF-VRA}
    The UPF-VRA security for a PRF-cSKL scheme is formalized by the experiment ${\sf EXP}^\text{\sf upf-vra}_{\text{PRF-cSKL}, A}(1^\lambda)$:
    \begin{enumerate}
        \item The challenger $C$ and the adversary $A$ run $res \leftarrow \text{\sf PRF-cSKL.Setup}\langle C(1^\lambda), A(1^\lambda) \rangle$. If $res = \bot$ (${\sf Setup}$ aborted), the experiment ends and the output is $0$. Otherwise, we have $res \coloneqq ( {\sf mk}, st)$ where the challenger has ${\sf mk}$ and the adversary has $st$. The challenger and the adversary run $(sk, {\sf msk}, {\sf dvk}) \leftarrow \text{\sf PRF-cSKL}.{\sf KG}{\langle Lessor(1^\lambda, {\sf mk}), Lessee(1^\lambda, st) \rangle}$ where the challenger obtains {\sf msk} and {{\sf dvk}}.
        \item The adversary sends a classical string {\sf cert}. If $\text{\sf PRF-cSKL.DelVrfy}({\sf cert}, {\sf dvk}) = \bot$, the challenger outputs $0$ and the experiment ends. Otherwise, the challenger computes $s \leftarrow D_{\sf prf}, t \leftarrow \text{\sf PRF-cSKL}.{\sf Eval}({\sf msk},s)$ and sends $({\sf dvk}, s)$ to the adversary.
        \item The adversary sends $t^\prime \in R_{\sf prf}$ to the challenger. The challenger outputs $1$ if $t^\prime = t$. Otherwise, the challenger outputs $0$.
    \end{enumerate}
    {\sf PRF-cSKL} is UPF-VRA secure if and only if for any QPT $A$
    \begin{equation}
        {\sf Adv}^\text{\sf upf-vra}_{\text{PRF-cSKL},A}(\lambda):= \Pr[{\sf EXP}^\text{\sf pr-vra}_{\text{PRF-cSKL}, A}(1^\lambda)=1]= {\rm negl}(\lambda)
    \end{equation}
\end{definition}

\begin{lemma}
    If there exists PRF-cSKL with UPF-VRA security (see \cref{dfn:PRF-UPF-VRA}), there exists PRF-cSKL with PR-VRA security (see \cref{dfn:PRF-PR-VRA}).
\end{lemma}
We can prove the lemma above with almost the same method as Lemma 4.14 in \cite{KMY24}. So, we omit the proof.

\subsection{Digital Signature with Classical Secure Key Leasing}
In this subsection, we define the syntax for a digital signature with classical secure key leasing (DS-cSKL). Our definition is analogous to the definition in \cite{KMY24}.

\begin{definition}[DS-cSKL]
    \label{dfn:DS-CSKL}
    Let ${\cal M}$ be the message space. A scheme of DS-cSKL is a tuple of algorithms $({\sf Setup}, {\sf KG}, Sign, {\sf SignVrfy}, Del, {\sf DelVrfy})$:
    \begin{itemize}
        \item[] {${\sf Setup}{\langle Lessor(1^\lambda), Lessee(1^\lambda) \rangle} \rightarrow ({\sf msk}, st)/\bot$} is an interactive protocol between
            \begin{itemize}
                \item a QPT Lessee
                \item a PPT Lessor
            \end{itemize}
        When the protocol fails, the Lessor and the Lessee output $\bot$. When the protocol succeeds, the Lessor outputs a classical master key ${\sf msk}$. The (honest) Lessee outputs an internal state $st$.
        \item[] {${\sf KG}{\langle Lessor(1^\lambda), Lessee(1^\lambda) \rangle} \rightarrow (sigk, {\sf svk}, {\sf dvk})/\bot$} is an interactive protocol between
            \begin{itemize}
                \item a QPT Lessee
                \item a PPT Lessor
            \end{itemize}
            When the protocol succeeds, the Lessor outputs a classical signature verification key ${\sf svk}$ and a classical deletion verification key ${\sf dvk}$. The (honest) Lessee outputs a quantum signing key $sigk$.
        \item[] {$Sign(sigk, m) \rightarrow (sigk^\prime,\sigma)$} is a QPT algorithm. ${sigk}$ is a quantum signing key and $m$ is a message from the message space. The algorithm outputs a classical signature $\sigma$ and a quantum post-signing key $sigk^\prime$.
        \item[] {${\sf SignVrfy}({\sf svk},\sigma, m) \rightarrow \bot / \top$} is a PPT algorithm. ${\sf svk}$ is a calssical signature verification key, $\sigma$ is a classical signature, and $m$ is a classical message from the message space. The algorithm outputs $\top/\bot$ to indicate whether $\sigma$ is a valid signature for $m$.
        \item[] {$Del(sigk) \rightarrow {\sf cert}$} is a QPT algorithm. $sigk$ is a quantum signing key. The algorithm destroys the quantum key and outputs a valid classical certificate ${\sf cert}$.
        \item[] {${\sf DelVrfy}({\sf cert}, {\sf dvk}) \rightarrow \top / \bot$} is a PPT algorithm. The algorithm takes as input a classical certificate ${\sf cert}$ and a classical verification key ${\sf dvk}$. The algorithm outputs $\top$ if ${\sf cert}$ is a valid certificate of deletion. It outputs $\bot$ if ${\sf cert}$ is not a valid certificate.
    \end{itemize}
\end{definition}

{\bf Signature verification correctness:} For every $m\in {\cal M}$, we have that
\begin{equation}
    \label{eqn:DS-cSKL-signature-verification-correctness}
    \Pr[\text{\sf DS-cSKL}.{\sf SignVrfy}({\sf svk}, \sigma,m) = \bot \lor res =\bot: 
    \begin{aligned}
        & res \leftarrow \text{\sf DS-cSKL}.{\sf Setup}{\langle Lessor(1^\lambda), Lessee(1^\lambda) \rangle} \\
        & ({\sf msk},st) \leftarrow res \\
        & (sigk, {\sf svk}, {\sf dvk}) \leftarrow \text{\sf DS-cSKL}.{\sf KG}\langle 
        \\ 
        &Lessor(1^\lambda, {\sf msk}), Lessee(1^\lambda, st) \rangle\\
        & (sigk^\prime, \sigma) \leftarrow \text{\sf DS-cSKL}.{Sign}(sigk,m) \\
    \end{aligned}
    ] = {\rm negl}(\lambda)
\end{equation}

{\bf Deletion verification correctness:} We have
\begin{equation}
    \Pr[b = \bot \lor res=\bot: 
    \begin{aligned}
        & res \leftarrow \text{\sf DS-cSKL}.{\sf Setup}{\langle Lessor(1^\lambda), Lessee(1^\lambda) \rangle} \\
        & ({\sf msk},st) \leftarrow res \\
        & (sigk, {\sf svk}, {\sf dvk}) \leftarrow \text{\sf DS-cSKL}.{\sf KG}{\langle Lessor(1^\lambda, {\sf msk}), Lessee(1^\lambda, st) \rangle}\\
        & {\sf cert} \leftarrow \text{\sf DS-cSKL}.Del(sigk) \\
        & {\sf b} \leftarrow \text{\sf DS-cSKL}.{\sf DelVrfy}({\sf cert},{\sf dvk})
    \end{aligned}
    ] = {\rm negl}(\lambda)
\end{equation}

{\bf Reusability with static signing key:} Let $sigk$ be the honest generated signing key. Let $m$ be any message from the message space ${\cal M}$. Let $sigk^\prime$ be the signing key after running the sigining algorithm $(sigk^\prime, \sigma) \leftarrow \text{\sf DS-cSKL}.{Sign}(sigk,m)$. We must have
\begin{equation}
    {\sf TD}(sigk,sigk^\prime) ={\rm negl}(\lambda)
\end{equation}

Then, we will introduce the security definition.

\begin{definition}[RUF-VRA security]
    \label{dfn:DS-RUF-VRA}
    The RUF-VRA security for a DS-cSKL scheme is formalized by the experiment ${\sf EXP}^\text{\sf ruf-vra}_{\text{DS-cSKL}, A}(1^\lambda)$:
    \begin{enumerate}
        \item The challenger $C$ and the adversary $A$ run $res \leftarrow \text{\sf DS-cSKL.Setup}\langle C(1^\lambda), A(1^\lambda) \rangle$. If $res = \bot$ (${\sf Setup}$ aborted), the experiment ends and the output is $0$. Otherwise, we have $res \coloneqq ( {\sf msk}, st)$ where the challenger has ${\sf msk}$ and the adversary has $st$. Then, the challenger and the adversary engage in {\sf DS-cSKL.KG}$\langle C(1^\lambda, {\sf msk}), A(1^\lambda, st) \rangle$ where the challenger obtains {\sf svk} and {\sf dvk}.
        \item The adversary sends a classical string {\sf cert} to the challenger. If $\text{\sf DS-cSKL.DelVrfy}({\sf cert}, {\sf dvk}) = \bot$, the challenger outputs $0$ and the experiment ends. Otherwise, the challenger computes $m \leftarrow D_{prf}$ and sends $m, dvk$ to the adversary.
        \item The adversary sends a classical string $\sigma$ to the challenger. The challenger outputs $1$ if $\text{\sf DS-cSKL.SignVrfy}({\sf svk},\sigma, m) = \top$. Otherwise, the challenger outputs $0$.
    \end{enumerate}
    {\sf DS-cSKL} is RUF-VRA secure if and only if for any QPT $A$
    \begin{equation}
        {\sf Adv}^\text{\sf ruf-vra}_{\text{DS-cSKL},A}(\lambda):= \Pr[{\sf EXP}^\text{\sf ruf-vra}_{\text{DS-cSKL}, A}(1^\lambda)=1] = {\rm negl}(\lambda)
    \end{equation}
\end{definition}

\subsection{The construction of PRF-cSKL}
Before introducing our PRF-cSKL, we need a new cryptographic tool.
\begin{definition}[Two-Key Equivocal PRF (TEPRF) \cite{HJO+16,KMY24}]
    \label{dfn:TEPRF}
    A two-key equivocal PRF (TEPRF) with input length $\ell$ (and output length $1$) is a tuple of two algorithms (KG, Eval).
    \begin{itemize}
        \item \text{{\sf KG}($1^\lambda$, s$^\ast$) $\to$ ({\sf key}$_0$, {\sf key}$_1$):} The key generation algorithm is a PPT algorithm that takes as input the security parameter $1^\lambda$ and a string s$^\ast \in \{0, 1\}^\ell$, and outputs two keys key$_0$ and key$_1$.
        \item \text{{\sf Eval}({\sf key}, s) $\to$ b:} The evaluation algorithm is a deterministic classical polynomial-time algorithm that takes as input a key $\sf key$ and an input s $\in \{0, 1\}^\ell$, and outputs a bit b $\in \{0, 1\}$.
    \end{itemize}

    TEPRF satisfies the following properties
    {\bf Equality} For all $\lambda \in \mathbb{N}, s^*\in\{0,1\}^l, \mathsf{(key_0, key_1)} \leftarrow \mathsf{KG}(1^\lambda, s^*), s\in\{0,1\}^l \setminus \{s^*\}$,
    \begin{equation}
        \mathsf{Eval}(\sf{key_0},s) = \sf{Eval}(\sf{key_1},s)
    \end{equation}
    
    {\bf Differs on target}: For all $\lambda \in \mathbb{N}, s^*\in\{0,1\}^l, \mathsf{(key_0, key_1)} \leftarrow \mathsf{KG}(1^\lambda, s^*)$
    \begin{equation}
        \mathsf{Eval}(\sf{key_0},s^*) \neq \sf{Eval}(\sf{key_1},s^*)
    \end{equation}
    
    {\bf Differing point hiding}: For any QPT adversary $\cal A$,
    \begin{align}
        &\left | 
         \Pr \left [\cal{A}(\sf{key}_b)=1:  \begin{array}{l}
             (s_0^*, s_1^*, b) \leftarrow \mathcal{A}(1^\lambda)  \\
             (\sf{key_0, key_1}) \leftarrow \sf{KG}(1^\lambda,s_0^*)
        \end{array} \right ]  \right.
        \\ &\left .  - \Pr \left [\cal{A}(\sf{key}_b)=1:  \begin{array}{l}
             (s_0^*, s_1^*, b) \leftarrow \mathcal{A}(1^\lambda)  \\
             (\sf{key_0, key_1}) \leftarrow \sf{KG}(1^\lambda,s_1^*)
        \end{array} \right ] 
        \right | \leq \sf{negl}(\lambda)
    \end{align}
\end{definition}

\begin{theorem}[Theorem 3.6 from \cite{KMY24}]
    Assuming the existence of OWF, there exists a secure scheme of TEPRFs.
\end{theorem}

We use the following primitives as the building blocks:
\begin{itemize}
    \item A secure TEPRF scheme ${\sf TEPRF = (TEPRF.KG, TEPRF.Eval)}$ with secret key length $l_{sk}$ and input length $l_{in}$.
    \item A secure classical blind quantum computing protocol $\Pi$.
    \item A secure CSG with randomness extraction.
    \item A compiled MSG ${\sf SimBob}$.
\end{itemize}

\begin{table}[hbtp]
    \centering
    \begin{tabular}{c|c|c}
        $XI$ & $IX$ & $XX$ \\
        \hline
        $IZ$ & $ZI$ & $ZZ$ \\
        \hline
        $-XZ$ & $-ZX$ & $YY$ 
    \end{tabular}
    \label{tab:msg-strategy-prf}
    \caption{When Alice (resp. Bob) receives $x$ (resp. $y$) as their question, they measure the observables on $x$-th column (resp. $y$-th row) and outputs the outcomes as their answer.}
\end{table}

{{\sf PRF-cSKL}.{\cal Setup}$\langle$Lessor($1^\lambda$), Lessee($1^\lambda$)$\rangle$} We describe the interactive setup protocol between the Lessor and the honest Lessee below. Note that in the actual protocol, the malicious Lessee may not follow the protocol. The ${\sf Setup}$ algorithm is the same as that for {\sf PKE-cSKL}.

{{\sf PRF-cSKL}.{\cal KG}$\langle$Lessor($1^\lambda$, \sf{msk}), Lessee($1^\lambda$, st)$\rangle$} We describe the interactive key generation protocol between the Lessor and the honest Lessee below. Note that in the actual protocol, the malicious Lessee may not follow the protocol.
\begin{enumerate}
    \item The Lessor samples $s^*_j \in \{0,1\}^{\ell}$ uniformly. Then, the Lessor generates $({\sf TEPRF.key}_{j,0},{\sf TEPRF.key}_{j,1}) \leftarrow \mathsf{TEPRF.KG}(1^\lambda, s^*_j)$.
    \item The Lessor and the Lessee proceed as in ${\sf PKE\text{-}cSKL}.{\sf KG}$ from the $2$nd step, except for the following points:
    \begin{itemize}
        \item The algorithm uses ${\sf TEPRF.key}_{j,0}$ (resp. ${\sf TEPRF.key}_{j,1}$) instead of ${\sf PKE.sk}_{j,0}$ (resp. ${\sf PKE.sk}_{j,1}$).
        \item The algorithm outputs $({\sf TEPRF.key}_{j,0})_{j \in [2n]}$ instead of the public key. Note that this is a private key.
    \end{itemize}
\end{enumerate}
\begin{remark}
    For the sake of clarity, we present the ideal quantum evaluation key $\ket{sk}$ which consists of registers $({\sf A}_j, {\sf SK}_j, {\sf R}_j)$\footnote{In the construction of {\sf PKE-cSKL}, we denote the last register as ${\sf X}_j$. Here, we use ${\sf R}_j$ instead.} as follows:
    \begin{equation}
        \label{eqn:ideal-key-state-prf}
        \begin{aligned}
            &\frac{1}{\sqrt 2}(\ket{0, {\sf TEPRF.key}_{j,0}, x_{j,0}} \pm \ket{1, {\sf TEPRF.key}_{j,1}, x_{j,1}})& q_B^i = 0 \\
            &\ket{0,{\sf TEPRF.key}_{j,0}, x_{j,0}}\ {\rm or}\  \ket{1,{\sf TEPRF.key}_{j,1} , x_{j,1}}& q_B^i = 1 \\
            &\frac{1}{2}(\ket{0, {\sf TEPRF.key}_{j,0}, x_{j,0}}\bra{0, {\sf TEPRF.key}_{j,0}, x_{j,0}}+  \\ &\ket{1, {\sf TEPRF.key}_{j,1}, x_{j,1}}\bra{1, {\sf TEPRF.key}_{j,1}, x_{j,1}})& q_B^i = 2 
        \end{aligned}
    \end{equation}
    We point out that the register $({\sf A}_{2i}, {\sf SK}_{2i}, {\sf R}_{2i})$
    and $({\sf A}_{2i+1}, {\sf SK}_{2i+1}, {\sf R}_{2i+1})$ are entangled for $q_B = 2$. The whole state is
    \begin{equation}
        V_{2i}V_{2i+1} \ket{\Phi_{b[1]b[0]}}_{{\sf A}_{2i} {\sf A}_{2i+1}}
    \end{equation}
    where $V_j$ is an isometry mapping from ${\cal H}_{{\sf A}_j}$ to ${\cal H}_{{\sf A}_j}\otimes {\cal H}_{{\sf SK}_j} \otimes {\cal H}_{{\sf R}_j}$ such that
    \begin{equation}
        V_j \ket{b}_{{\sf A}_j} = \ket{b, {\sf TEPRF.sk}_{j,b},x_{j,b}}_{{\sf A}_j {\sf SK}_j {\sf R}_j}
    \end{equation}
    and $\ket{\Phi_{ab}} \coloneqq \frac{1}{\sqrt 2}(\ket{0a}+(-1)^b\ket{1(1-a)})$ is one of the four Bell states. We present the honest Lessee who outputs a quantum state almost identical to the ideal state, except for ${\rm negl}(\lambda)$ trace distance, in the proof of Evaluation correctness.
\end{remark}

{{\sf PRF-cSKL.LEval}$({\sf msk},s)$}:
\begin{enumerate}
    \item Parse $\mathsf{msk} = ({\sf TEPRF.key}_{j,0})_{j \in [2n]}$ and $s = s_0 || \dots || s_{2n-1}$ where $s_j \in \{0,1\}^{\ell_{in}}$ for each $j\in[2n]$.
    \item Compute 
    \begin{equation}
        t_j \leftarrow \text{\sf TEPRF.Eval}(\text{\sf TEPRF.key}_{j,0},s_j)
    \end{equation}
     for $j \in [2n]$.
    \item Output $t = t_0 || \dots || t_{2n-1}$.
\end{enumerate}

{{\sf PRF-cSKL}.Eval$(sk, s)$}:
\begin{enumerate}
    \item Parse $sk \coloneqq (\{ q_A^i \}_{i \in [n]}, ({\sf A}_j, {\sf SK}_j, {\sf R}_j)_{j \in [2n]})$ and $s = s_0 ||  \dots || s_{2n-1}$. The registers $({\sf A}_j, {\sf SK}_j, {\sf R}_j)_{j \in [2n]}$ are holding the key state.
    \item Let $U_{Dec, j}$ be a unitary on register $({\sf A_j}, {\sf SK}_j, {\sf OUT}_j)$ as follows
    \begin{equation}
        \begin{aligned}
            &U_{Dec,j} \ket{b}_{{\sf A}_j}\ket{{\sf TEPRF.key}_{j,b}}_{{\sf SK}_j}\ket{v}_{{\sf OUT}_j} \\
            = &\ket{b}_{{\sf A}_j}\ket{{\sf TEPRF.key}_{j,b}}_{{\sf SK}_j} \ket{v \oplus \mathsf{TEPRF.Eval(TEPRF.key_{j,b}, s_j)}}_{{\sf OUT}_j}
        \end{aligned}
    \end{equation}
    The algorithm applies the unitary $U_{Dec,j}$ to register $({\sf A}_j, {\sf SK}_j, {\sf OUT}_j)$, where ${\sf OUT}_j$ is initialized to $\ket{0}$. Then, measure the register ${\sf OUT}_j$ in the computational basis and obtain the outcome $t_j^\prime$.
    \item Output $t_0^\prime || \dots || t_{2n-1}^\prime$.
\end{enumerate}

{{\sf PRF-cSKL}.Del$(sk)$} The algorithm is as follows.
\begin{enumerate}
    \item Parse $sk \coloneqq (\{ q_A^i \}_{i \in [n]}, ({\sf A}_j, {\sf SK}_j, {\sf R}_j)_{j \in [2n]})$. The registers $({\sf A}_j, {\sf SK}_j, {\sf R}_j)_{j \in [2n]}$ are holding the key state. 
    \item For $i\in[n]$, measure the register $(A_{2i}, A_{2i+1})$ with $\{\mathcal A_{q_A^i}^a\}$, where $\{\mathcal A_{q_A^i}^a\}$ is Alice's measurement in \cref{tab:msg-strategy} when the question to it is $q_A^i$. Set $a_i$ to be the outcome.
    \item For $j\in[2n]$, measure every qubit of registers ${\sf SK}_j$, ${\sf R}_j$ in Hadamard basis. Let the measurement outcomes be $d_j, d_j^\prime$, respectively.
    \item Output ${\sf cert}= (\{ a_i \}_{i\in [n]}, \{d_j, d_j^\prime \}_{j \in [2n]})$.
\end{enumerate}

{{\sf PRF-cSKL}.{\sf DelVrfy}$({\sf cert}, {\sf dvk})$} The algorithm is as follows.
\begin{enumerate}
    \item Parse 
    \begin{equation}
        {\sf cert} = (\{ a_i \}_{i\in [n]}, \{d_j, d_j^\prime\}_{j \in [2n]})
    \end{equation}
    and 
    \begin{equation}
        \begin{aligned}
            {\sf{dvk}} \coloneqq (\{ q_B^i, b_i^\prime, q_A^i \}_{i \in [n]}, \{ {\sf PRF.key}_{j,0}, {\sf PRF.key}_{j,1}, \\
            e_{z,1}^j, z_j, x_{j,0}, x_{j,1}\}_{j \in [2n]: j \notin J_B})
        \end{aligned}
    \end{equation}
    \item Computes 
    \begin{equation}
        \label{eqn:Z-error-prf}
        \begin{aligned}
            e_{i,0} & = d_{2i} \cdot ({\sf PRF.key}_{2i,0} \oplus {\sf PRF.key}_{2i,1}) \\
            & \oplus (d_{2i}^\prime \oplus e_{z,1}^{2i}) \cdot (x_{2i,0} \oplus x_{2i,1}) \oplus z_{2i} \\
            e_{i,1} & = d_{2i+1} \cdot ({\sf PRF.key}_{2i+1,0} \oplus {\sf PRF.key}_{2i+1,1}) \\
            &\oplus (d_{2i+1}^\prime \oplus e_{z,1}^{2i+1}) \cdot (x_{2i+1,0} \oplus x_{2i+1,1}) \oplus z_{2i+1}
        \end{aligned}
    \end{equation}
    and $a_i^\prime = postprocessing(q_A^i, a_i, e_{i,0}, e_{i,1})$ for $i \in [n]$. We remind the readers that $postprocessing$ is defined in \cref{dfn:postprocessing}.
    \item If $MSG(q_A^i,q_B^i,a_i^\prime, b_i^\prime) = 0$ for some $i\in[n]$, output $\bot$. Otherwise, output $\top$.
\end{enumerate}

\paragraph*{\bf Proof of Evaluation correctness} We first prove that the ideal key state in \cref{eqn:ideal-key-state-prf} satisfies Evaluation correctness. For $s_j \neq s_j^*$, ${\sf TEPRF.Eval}({\sf TEPRF.key}_{j,0},s_j)$ and ${\sf TEPRF.Eval}({\sf TEPRF.key}_{j,1},s_j)$ evaluates to the same value. After applying $U_{Dec,j}$ in $\text{\sf PRF-cSKL}.Eval$, the register ${\sf OUT}_j$ is tensored with other registers. Measuring the register ${\sf OUT}_j$ produces a unique $t_j = {\sf TEPRF.Eval}({\sf TEPRF.key}_{j,0},s_j)$. We point out that for any $s_j$ the probability $s_j^* = s_j$ is ${\rm negl}(\lambda)$, thus the evaluation correctness holds.

Then, the same honest Lessee as PKE-cSKL (see \cref{sec:dfn-cSKL}) outputs a quantum state negligibly close to the ideal state. We complete the proof.

\paragraph*{\bf Proof of deletion verification correctness} The ideal key state in \cref{eqn:ideal-key-state-prf} is the same as the ideal key state for PKE-cSKL, except that it uses ${\sf TEPRF.key}_{j,0}, {\sf TEPRF.key}_{j,1}$ instead of ${\sf PKE.sk}_{j,0}, {\sf PKE.sk}_{j,1}$. Thus, the proof for PKE-cSKL (see \cref{sec:dfn-cSKL}) works for PRF-cSKL as well.

We present the proof of UP-VRA security (see \cref{dfn:PRF-UPF-VRA}) below. 
\begin{theorem}
    \label{thm:PRF-cSKL-secure}
    The 2-party PRF-cSKL above satisfies \cref{dfn:PRF-UPF-VRA}
\end{theorem}
We prove using the following sequence of Hybrids.
\paragraph*{${\sf Hyb}_0(\lambda):$}

\begin{enumerate}
    \item For each $j \in [2n]$, the challenger performs the following steps (as part of step 1 of $\text{\sf PRF-cSKL.KG}$):
    \begin{enumerate}
        \item The challenger samples $s^*_j \in \{0, 1\}^{\ell_{in}}$ uniformly.
        \item The challenger generates $({\sf TEPRF.key}_{j,0}, {\sf TEPRF.key}_{j,1}) \leftarrow {\sf TEPRF.KG}(1^\lambda, s^*_j)$.
        \item The challenger and the adversary take part in the $\Pi_{CSG}$ protocol (see \cref{dfn:CSG}). The challenger obtains $x_{j,0}, x_{j,1} \in \{0, 1\}^{l_{sk}}$ and $z_j \in \{0, 1\}$ after the execution.
    \end{enumerate}

    \item For each $i \in [n]$, the challenger performs the following steps (as part of step 2 of $\text{\sf PRF-cSKL.KG}$):
    \begin{enumerate}
        \item The challenger samples $q_i^B, q_i^A \in \{0, 1, 2\}$ uniformly.
        \item The challenger (as Verifier) and the adversary (as Prover) then engage in the ${\sf SimBob}\langle {\sf V}(1^\lambda, q_i^B), P(1^\lambda) \rangle$ protocol (see \cref{dfn:compiled-msg}), from which the challenger obtains $b_i$. The adversary computes $b_i^\prime$ such that $b_i^\prime[q_A^i] = a_i[q_B^i]$ and the other bits are generated uniformly at random, where ${\rm par}(b_i^\prime) = 1$.
        \item The challenger then sends $q_i^A$ to the adversary.
    \end{enumerate}

    \item For each $j \in [2n]$, the challenger and the adversary repeat the following steps (as part of step 3 of $\text{\sf PRF-cSKL.KG}$):
    \begin{enumerate}
        \item The challenger and the adversary engage in $\Pi_{CBQC}$ (see \cref{dfn:cbqc}), where the challenger acts as the Client (C) and the adversary acts as the Sender (S). If $j \in J_B$, the challenger and the adversary run the protocol
        \begin{equation}
        \Pi_{CBQC} = \langle S(1^\lambda, Q, V), C(1^\lambda, Q, 1) \rangle
        \end{equation}
        Otherwise, the challenger and the adversary run the protocol
        \begin{equation}
        \Pi_{CBQC} = \langle S(1^\lambda, Q, V), C(1^\lambda, Q, 0) \rangle
        \end{equation}
        After the execution, the challenger obtains the classical string $(e_j^x, e_j^z)$. The challenger parses $e_j^x = e_{j,0}^x || e_{j,1}^x$ and $e_j^z = e_{j,0}^z || e_{j,1}^z$.
        \item The adversary sends a classical string $\omega$ to the challenger.
        \item If $j \notin {J_B}$, the challenger checks if $\omega \oplus e_{j,1}^x \neq x_{j,c}$, where $c = b_{\lfloor j/2 \rfloor}[j \bmod 2]$. If this condition is not met, the $\text{\sf PRF-cSKL.KG}$ protocol aborts and the experiment ends and outputs $0$.
    \end{enumerate}

    \item For each $j \in [2n]$, the challenger samples $r_{j,0}, r_{j,1}$. The challenger computes $h_{j,0} = {\sf Ext}(x_{j,0}, r_{j,0}) \oplus {\sf TEPRF.key}_{j,0}$ and $h_{j,1} = {\sf Ext}(x_{j,1}, r_{j,1}) \oplus {\sf TEPRF.key}_{j,1}$ (see \cref{dfn:CSG}). These values $(h_{j,0}, h_{j,1}, r_{j,0}, r_{j,1})$ are then sent by the challenger to the adversary (as part of step 3(c) of $\text{\sf PRF-cSKL.KG}$).

    \item The adversary sends a classical string ${\sf cert} = (\{a_i\}_{i \in [n]}, \{d_j, d'_j\}_{j \in [2n]})$ to the challenger.

    \item For each $i \in [n]$, the challenger computes $e_{i,0}$ and $e_{i,1}$ using the formulas (as defined in $\text{\sf PRF-cSKL.DelVrfy}$ in Appendix B.3, page 25). Then, the challenger computes $a'_i = {postprocessingA}(q_i^A, a_i, e_{i,0}, e_{i,1})$ (see \cref{dfn:postprocessing}). If ${MSG}(q_i^A, q_i^B, a'_i, b_i^\prime) = 0$ for any $i \in [n]$, the $\text{\sf PRF-cSKL.DelVrfy}$ (Appendix B.3, page 37) outputs $\perp$. If $\text{\sf PRF-cSKL.DelVrfy}({\sf cert}, {\sf dvk}) = \perp$, the challenger outputs $0$ and the experiment ends.

    \item The challenger samples an input $s \in {\sf D_{prf}}$ uniformly. The challenger computes $t \leftarrow \text{\sf PRF-cSKL.Eval}({\sf msk}, s)$ (as defined in Appendix B.3, page 37). This involves parsing ${\sf msk} = ({\sf TEPRF.key}_{j,0})_{j \in [2n]}$ and $s = s_0 || \dots || s_{2n-1}$ (where $s_j \in \{0, 1\}^\ell$) and then computing $t_j \leftarrow {\sf TEPRF.Eval}({\sf TEPRF.key}_{j,0}, s_j)$ for each $j \in [2n]$ to form $t = t_0 || \dots || t_{2n-1}$.

    \item The challenger sends the verification key ${\sf dvk}$ and the input $s$ to the adversary.

    \item The adversary sends an output $t' \in {\sf R_{prf}}$ to the challenger.

    \item The challenger outputs $1$ if $t' = t$. Otherwise, the challenger outputs $0$.
\end{enumerate}
The ${\sf Hyb}_0$ is the same as the security game for \cref{dfn:PRF-UPF-VRA}.

{${\sf Hyb}_1(\lambda)$:} We define ${\sf Hyb}_1(\lambda)$ the same as ${\sf Hyb}_0(\lambda)$ except for:
\begin{itemize}
    \item In Step $10$, the challenger checks $t^\prime_j = {\sf TEPRF.Eval}({\sf TEPRF.key}_{j,b[j \bmod 2]}, s_j)$ for $j \in J_B$ instead. If the condition does not hold for some $j$, the challenger outputs $0$ and aborts the experiment.
\end{itemize}

\begin{lemma}
    \label{lem:prf-hyb0-hyb1-negl}
    For any (QPT) adversary ${\sf Adv}^\lambda$, $|\Pr[{\sf Hyb}_0 = 1] - \Pr[{\sf Hyb}_1 = 1]| = {\rm negl}(\lambda)$.
\end{lemma}
\begin{proof}
    By the Equality property (see \cref{dfn:TEPRF}), we see that ${\sf TEPRF.Eval}({\sf TEPRF.key}_{j,0}, s_j) \neq {\sf TEPRF.Eval}({\sf TEPRF.key}_{j,1}, s_j)$ for only $s_j = s_j^*$. The change of Step $10$ in ${\sf Hyb}_1$ affects the output distribution only when $s_j = s_j^*$, which happens with probability $2^{-l_{in}}$. We complete the proof.
    
\end{proof}

{${\sf Hyb}_2(\lambda)$:} We define ${\sf Hyb}_2(\lambda)$ the same as ${\sf Hyb}_1(\lambda)$ except for:
\begin{itemize}
    \item In Step $7$, the challenger samples $s$ such that $s_j = s^*_j$ for $j \in J_B$.
\end{itemize}

\begin{lemma}
    \label{lem:prf-hyb1-hyb2-negl}
    For any QPT adversary ${\sf Adv}^\lambda$, $|\Pr[{\sf Hyb}_1 = 1] - \Pr[{\sf Hyb}_2 = 1]| = {\rm negl}(\lambda)$.
\end{lemma}
\begin{proof}
    By the same argument as in the proof of \cref{lem:pke-hyb1-hyb2-diff}, we can see that the adversary in ${\sf Hyb}_1$ has only information about either ${\sf TEPRF.key}_{j,0}$ or ${\sf TEPRF.key}_{j,1}$ for $j \in J_B$. This can be proved by defining a ${\sf Hyb}_1^\prime$ in which ${\sf TEPRF.key}_{j,1-b_i[j \mod 2]}$ is sent to the adversary with a one-time pad. Then, by the Differing point hiding (see \cref{dfn:TEPRF}), the adversary cannot notice the change in ${\sf Hyb}_2$ and we have $|\Pr[{\sf Hyb}_1 = 1] - \Pr[{\sf Hyb}_2 = 1]| = {\rm negl}(\lambda)$.
    
\end{proof}

{${\sf Hyb}_3(\lambda)$:} We define ${\sf Hyb}_3$ almost the same as ${\sf Hyb}_2$, except for:
\begin{itemize}
    \item ${\sf Hyb}_3$ removes the specific abort condition found in Step $3(c)$ of ${\sf Hyb}_2$. In ${\sf Hyb}_2$, if $j \notin J_B$, the challenger checks $\omega \oplus e_{x,1}^j \neq x_{j,c}$ where $c = b_{\lfloor j/2 \rfloor}[j \mod 2]$. If the condition were not met, the experiment would abort and output $0$. This verification check is entirely omitted in ${\sf Hyb}_3$, ensuring the experiment proceeds regardless of this outcome.
\end{itemize}

\begin{lemma}
    \label{lem:prf-hyb2-hyb3-leq}
    For any (QPT) adversary ${\sf Adv}^\lambda$, $\Pr[{\sf Hyb}_2 = 1] \leq \Pr[{\sf Hyb}_3 = 1]$.
\end{lemma}
\begin{proof}
    In ${\sf Hyb}_3$, the abort condition in Step 3(c) is omitted. Thus, $\Pr[{\sf Hyb}_2 = 1] \leq \Pr[{\sf Hyb}_3 = 1]$.
    
\end{proof}

{${\sf Hyb}_4(\lambda)$:} We define ${\sf Hyb}_4$ almost the same as ${\sf Hyb}_3$, except for:
\begin{itemize}
    \item ${\sf Hyb}_4$ alters the input to the $\Pi_{CBQC}$ protocol in Step $3(a)$. In ${\sf Hyb}_3$, $\Pi_{CBQC}$ is run with input $1$ if $j \notin J_B$ and $0$ otherwise. In ${\sf Hyb}_4$, the $\Pi_{CBQC}$ protocol is uniformly run with input $0$ for all $j \in [2n]$.
\end{itemize}

\begin{lemma}
    \label{lem:prf-hyb3-hyb4-negl}
    For any QPT adversary ${\sf Adv}^\lambda$, $|\Pr[{\sf Hyb}_3 = 1] - \Pr[{\sf Hyb}_4 = 1]| = {\rm negl}(\lambda)$.
\end{lemma}
\begin{proof}[\cref{lem:prf-hyb3-hyb4-negl}]
    By the blindness of $\Pi_{CQBC}$ (\cref{dfn:cbqc}), we can see that changing the classical input from $1$ to $0$ does not affect the output distribution. This proves \cref{lem:prf-hyb3-hyb4-negl}.
    
\end{proof}

Finally, we bound $\Pr[{\sf Hyb}_4(\lambda)=1]={\rm negl}(\lambda)$ using the computational certified deletion property. 
\begin{lemma}
    \label{lem:prf-hyb4-negl}
    For any QPT adversary ${\sf Adv}^\lambda$, $\Pr[{\sf Hyb}_4(\lambda)=1]={\rm negl}(\lambda)$.
\end{lemma}
\begin{proof}
     We bound $\Pr[{\sf Hyb}_4 = 1]$ using the Certified Deletion Property of the Magic Square Game (\cref{dfn:comp-CDP}). We can transform any adversary ${\sf Adv}^\lambda$ (the adversary for ${\sf Hyb}_4$) into an adversary $\tilde P =(A_0, A_1)$ against the Certified Deletion Property of the Magic Square Game \cref{dfn:comp-CDP}.
    
    {$A_0$}: The adversary $A_0$ (acting as Prover $\tilde{P}$ for the CCD game and simulating the challenger for an internal ${\sf Hyb}_4$ adversary ${\sf Adv}^\lambda$) performs the following steps:
    \begin{enumerate}
    \item Initialize the ${\sf Hyb}_4$ adversary ${\sf Adv}^\lambda$. 
    \item For each $j \in [2n]$: 
        \begin{enumerate} 
            \item Sample $s^*_j \leftarrow \{0,1\}^{l_{in}}$.
            \item Sample $({\sf TEPRF.key}_{j,0}, {\sf TEPRF.key}_{j,1}) \leftarrow {\sf TEPRF.KG}(1^\lambda,s^*_j)$. 
            \item Participate in the $\Pi_{CSG}$ protocol with ${\sf Adv}^\lambda$ to obtain $\{x_{j,0}, x_{j,1}\} \in \{0, 1\}^{l_{sk}}$ and $z_j \in \{0, 1\}$. 
        \end{enumerate} 
        \item For each $i \in [n]$: 
        \begin{enumerate} 
            \item Engage in the ${\sf SimBob}\langle {\sf V}(1^\lambda, q_i^B), P(1^\lambda) \rangle$ protocol as the Prover (P). Forward the messages from the Verifier to ${\sf Adv}^\lambda$, the answers from ${\sf Adv}^\lambda$ to the Verifier.
            \item Receive $q_i^A \in \{0, 1, 2\}$ from $V$.
            \item Send $q_i^A$ to ${\sf Adv}^\lambda$. 
        \end{enumerate} 
        \item For each $j \in [2n]$: 
        \begin{enumerate} 
            \item Engage in the $\Pi_{CBQC}$ protocol as Client (C) with ${\sf Adv}^\lambda$ (Sender), using input $0$ for all $j \in [2n]$: $\Pi_{CBQC} = \langle S(1^\lambda, Q, V), C(1^\lambda, Q, 0) \rangle$. Obtain classical string $(e_j^x, e_j^z)$, parsing $e_j^x = e_{j,0}^x || e_{j,1}^x$ and $e_j^z = e_{j,0}^z || e_{j,1}^z$. 
            \item Receive a classical string $\omega$ from ${\sf Adv}^\lambda$.
        \end{enumerate} 
        \item For each $j \in [2n]$, sample $r_{j,0}, r_{j,1}$. Compute $h_{j,0} = {\sf Ext}(x_{j,0}, r_{j,0}) \oplus {\sf TEPRF.key}_{j,0}$ and $h_{j,1} = {\sf Ext}(x_{j,1}, r_{j,1}) \oplus {\sf TEPRF.key}_{j,1}$. Send $(h_{j,0}, h_{j,1}, r_{j,0}, r_{j,1})$ to ${\sf Adv}^\lambda$.
        \item Receive a classical string ${\sf cert} = (\{a_i\}_{i \in [n]}, \{d_j, d'_j\}_{j \in [2n]})$ from ${\sf Adv}^\lambda$.
        \item For each $i \in [n]$:
            \begin{enumerate}
            \item Compute $e_{i,0}$ and $e_{i,1}$ according to \cref{eqn:Z-error-prf}.
            \item Compute $a'_i = { postprocessing_A}(q_i^A, a_i, e_{i,0}, e_{i,1})$.
            \end{enumerate} 
        \item Output the internal state $st$ and $\{a^\prime_i\}_{i \in [n]}$ as the output. 
    \end{enumerate}
    
    {$A_1$}: The adversary $A_1$ (acting as Prover $\tilde{P}$ for the CCD game and continuing to simulate the challenger for ${\sf Adv}^\lambda$) performs the following steps:
    \begin{enumerate}
        \item Receive the internal state $st$ from $A_0$ and the list $\{q_i^B\}_{i \in [n]}$ from the CCD Verifier $V$. 
        \item The adversary computes $b_i^\prime$ such that $b_i^\prime[q_A^i] = a_i[q_B^i]$ and the other bits are generated uniformly at random, where ${\rm par}(b_i^\prime) = 1$, for each $i\in[n]$.
        \item Sample a message $s = s_0 || \dots || s_{2n-1}$ uniformly in a way such that $s_j = s^*_j$ for $j \in J_B$.
        \item Let
        \begin{equation}
            \begin{aligned}
                {\sf dvk} := (\{q_i^B, b_i^\prime, q_i^A\}_{i \in [n]}, \{{\sf TEPRF.key}_{j,0}, {\sf TEPRF.key}_{j,1}, \\
                e_{j,1}^z, z_j, x_{j,0}, x_{j,1}\}_{j \in [2n]: j \notin {J_B}})
            \end{aligned}
        \end{equation}
        Send $s$ and ${\sf dvk}$ to ${\sf Adv}^\lambda$. 
        \item Receive a message $t^\prime = t^\prime_0 || \dots ||t^\prime_{2n-1}$ from ${\sf Adv}^\lambda$. 
        \item Compute $b_i[j \mod 2] = c$ such that $t^\prime_j = {\sf TEPRF.Eval}({\sf TEPRF.key}_{j,c},s^*_j)$ for $j \in J_B$. For $j \notin J_B$, sample random $b_i[j \mod 2] \in \{0,1\}$.
    \end{enumerate}

    We can see that $A_0$ outputs a valid answer for the computational certified deletion property (\cref{dfn:comp-CDP}, \cref{eqn:CCP-first-round}), with the same probability as ${\sf Adv}^\lambda$ outputs a valid certificate ${\sf cert}$ for key revocation in Step 5, ${\sf Hyb}_0(\lambda)$. The proof is the same as that for \cref{lem:pke-hyb4-negl}.

    Then, we can see that whenever the adversary ${\sf Adv}^\lambda$ produces the correct $t_j^\prime$ for $j\in J_B$, the adversary $A_1$ produces the correct $b_i$ for $q_B^i = 1$. In Step $10$ of ${\sf Hyb}_4$, the challenger checks whether $t_j^\prime = {\sf TEPRF.Eval}({\sf TEPRF.key}_{j,b_i[j \mod 2]}, s^*_j)$. When the adversary produces the $t_j^\prime$ correctly, the adversary can check whether $t_j^\prime = {\sf TEPRF.Eval}({\sf TEPRF.key}_{j,0}, s^*_j)$ or $t_j^\prime = {\sf TEPRF.Eval}({\sf TEPRF.key}_{j,1}, s^*_j)$ and produce the value $b_i[j\mod 2]$. By the Differs on Target property (see \cref{dfn:TEPRF}), the adversary uniquely identifies $b_i[j \mod 2]$. For $i$ such that $q_i^B = 1$, the adversary obtains $b_i[0]$ and $b_i[1]$. With the information, the adversary can recover $b_i$ for $i$ such that $q_i^B = 1$. \footnote{Since ${\rm par}(b)=1$ for the valid answer of MSG, it suffices to recover $b$ using only a single bit other than $b[q_A^i]$. We can see that at least one of $b_i[0]$ and $b_i[1]$ differs from $b[q_A^i]$. }
    
    Combining the two facts above, we have
    \begin{equation}
        \Pr[{\sf Hyb}_4 = 1] \leq \Pr[{\sf Adv}^\lambda\ wins\ {\sf CCD}] ={\rm negl}(\lambda)
    \end{equation}
\end{proof}

Combining \cref{lem:prf-hyb0-hyb1-negl}, \cref{lem:prf-hyb1-hyb2-negl}, \cref{lem:prf-hyb2-hyb3-leq}, \cref{lem:prf-hyb3-hyb4-negl}, \cref{lem:prf-hyb4-negl}, we prove \cref{thm:PRF-cSKL-secure}.

Since every component of our PRF-cSKL can be constructed from CSGs, we obtain
\begin{theorem}
    Assuming the existence of CSGs, there exists PRF-cSKL satisfying UP-VRA security (see \cref{dfn:PRF-UPF-VRA}) and PR-VRA security (see \cref{dfn:PRF-PR-VRA})
\end{theorem}

\subsection{The construction of DS-cSKL}

To present our DS-cSKL, we need to introduce the following primitive from \cite{KMY24}.

\begin{definition}[Constrained signatures]
\label{dfn:constrained_signatures}
A constrained signatures (CS) with the message space ${\cal M}$ and constraint class $\mathcal{F} = \{ f : {\cal M} \to \{0, 1\}\}$ is a tuple of four algorithms ({\sf Setup, Constrain, Sign, Vrfy}).
\begin{itemize}
    \item $\text{\sf Setup}(1^\lambda) \to (\text{\sf vk, msk})$: The setup algorithm is a PPT algorithm that takes as input the security parameter $1^\lambda$, and outputs a master signing key ${\sf msk}$ and a verification key ${\sf vk}$.
    \item $\text{\sf Constrain}(\text{{\sf msk}, f}) \to \text{\sf sigk}_f$: The {\sf Constrain} algorithm is a PPT algorithm that takes as input the master signing key {\sf msk} and a constraint $f \in \mathcal{F}$. It outputs a constrained signing key $\text{\sf sigk}_f$.
    \item $\text{\sf Sign}(\text{\sf sigk}_f, \text{m}) \to \sigma$: The ${\sf Sign}$ algorithm is a PPT algorithm that takes as input a constrained key $\text{\sf sigk}_f$ and a message m $\in$ $\cal M$, and outputs a signature $\sigma$.
    \item $\text{\sf Vrfy}(\text{{\sf vk}, m, $\sigma$}) \to \top/\bot$: The {\sf Vrfy} algorithm is a deterministic classical polynomial-time algorithm that takes as input a verification key {\sf vk}, message m $\in$ {\cal M}, and signature $\sigma$, and outputs $\top$ or $\bot$.
\end{itemize}
\end{definition}
The constrained signatures must satisfy the correctness as follows.

\textbf{Correctness:} For any m $\in$ {\cal M} and $f \in \mathcal{F}$ such that $f(\text{m}) = 1$, we have
\begin{equation}
    \Pr[\text{{\sf Vrfy}({\sf vk}, m, $\sigma$) = $\top$ :}
\begin{aligned}
&(\text{\sf vk, msk})\leftarrow \text{\sf Setup}(1^\lambda)\\
&\text{\sf sigk}_f \leftarrow \text{\sf Constrain}(\text{{\sf msk}, f})\\
&\sigma\leftarrow \text{\sf Sign}(\text{\sf sigk}_f, \text{m})
\end{aligned}
]
\geq 1-\text{\rm negl}(\lambda).
\end{equation}

\begin{definition}[Selective single-key security]
\label{dfn:selective_single_key_security}
Let $\mathcal{A}$ be any stateful QPT adversary. We say that a CS scheme satisfies selective single-key security if
\begin{equation}
\Pr
\left[\text{{\sf Vrfy}({\sf vk}, m, $\sigma$) = $\top \land f(\text{m}) = 0$ :}
\begin{aligned}
&\quad f \leftarrow \mathcal{A}(1^\lambda)\\
&\quad (\text{\sf vk, msk})\leftarrow \text{\sf Setup}(1^\lambda)\\
&\quad \text{\sf sigk}_f \leftarrow \text{\sf Constrain}(\text{{\sf msk}, f})\\
&\quad (\text{m, $\sigma$})\leftarrow \mathcal{A}(\text{\sf vk, sigk}_f)
\end{aligned}
\right]
\leq \text{\rm negl}(\lambda).
\end{equation}
\end{definition}

\begin{definition}[Coherent-signability]\label{dfn:coherent_signability}
Let $L = L(\lambda)$ be any polynomial. We say that a CS scheme is coherently-signable if there is a QPT algorithm QSign that takes a quantum state $\ket{\psi}$ and a classical message m $\in$ $\cal M$ and outputs a quantum state $|\psi'\rangle$ and a signature $\sigma$, satisfying the following conditions:
\begin{enumerate}
    \item Let $f \in \mathcal{F}$, $(\text{\sf vk, msk}) \leftarrow  \text{\sf Setup}(1^\lambda)$, and $\text{\sf sigk}_f \leftarrow \text{\sf Constrain}(\text{\sf msk, f})$, the output distribution of $\text{QSign} \left( |z\rangle |\text{\sf sigk}_f\rangle , \text{m} \right)$ is identical to that of $\text{\sf Sign}(\text{\sf sigk}_f, \text{m})$ for any $z \in \{0, 1\}^L$.
    \item For any family $\{f_z \in \mathcal{F}\}$ for $z \in \{0, 1\}^L$, $(\text{\sf vk, msk}) \leftarrow \text{Setup}(1^\lambda)$, $\text{\sf sigk}_{f_z} \leftarrow \text{\sf Constrain}(\text{\sf msk, } f_z)$ for $z \in \{0, 1\}^L$, and m $\in$ $\cal M$ such that $f_z(\text{m}) = 1$ for all $z \in \{0, 1\}^L$, let $|\psi\rangle$ be a state of the form $|\psi\rangle = \sum_{z \in \{0, 1\}^L} \alpha_z |z\rangle |\text{sigk}_{f_z}\rangle$ for $\alpha_z \in \mathbb{C}$ such that $\sum_{z \in \{0, 1\}^L} |\alpha_z|^2 = 1$. Suppose that we run $(|\psi'\rangle , \sigma)\leftarrow \text{QSign}(|\psi\rangle , \text{m})$. Then we have $\| |\psi\rangle \langle\psi| - |\psi'\rangle \langle\psi'| \|_{\text{tr}} = \text{\rm negl}(\lambda)$.
\end{enumerate}
\end{definition}

\begin{lemma}[Theorem 7.5 from \cite{KMY24}]
    Assuming the hardness of the short integer solution (SIS) problem, there exists a secure CS scheme.
\end{lemma}

We use the following primitives as the building blocks:
\begin{itemize}
    \item A secure TEPRF scheme ${\sf TEPRF = (TEPRF.KG, TEPRF.Eval)}$ with secret key length $l_{sk}$ and input length $l_{in}$.
    \item A secure CS scheme ${\sf CS = (CS.Setup, CS.Constrain, CS.Sign, CS.Vrfy)}$ with function class $\cal F$ and the length of constrained signing key $\ell_{csk}$.
    \item The function class $\cal F$ consists of functions $f[{\sf TEPRF.key}]$ which takes as input $(x,y) \in \{0,1\}^{l_{in}} \times \{0,1\}$:
    \begin{equation}
        f[{\sf TEPRF.key}] = \left \{ 
        \begin{aligned}
            &1 & {\sf TEPRF.Eval}({\sf TEPRF.key},x)=y \\
            &0 & {\rm otherwise}
        \end{aligned}
        \right .
    \end{equation}
    \item A secure classical blind quantum computing protocol $\Pi_{CBQC}$.
    \item A secure CSG with randomness extraction.
    \item A compiled MSG ${\sf SimBob}$.
\end{itemize}

\begin{table}[hbtp]
    \centering
    \begin{tabular}{c|c|c}
        $XI$ & $IX$ & $XX$ \\
        \hline
        $IZ$ & $ZI$ & $ZZ$ \\
        \hline
        $-XZ$ & $-ZX$ & $YY$ 
    \end{tabular}
    \label{tab:msg-strategy-DS-cSKL}
    \caption{When Alice (resp. Bob) receives $x$ (resp. $y$) as their question, they measure the observables on $x$-th column (resp. $y$-th row) and output the outcomes as their answer.}
\end{table}

{{\sf DS-cSKL}.{\cal Setup}$\langle$Lessor($1^\lambda$), Lessee($1^\lambda$)$\rangle$} We describe the interactive setup protocol between the Lessor and the honest Lessee below. Note that in the actual protocol, the malicious Lessee may not follow the protocol. The ${\sf Setup}$ algorithm is the same as that for {\sf PKE-cSKL}.

{{\sf DS-cSKL}.{\cal KG}$\langle$Lessor($1^\lambda$, \sf{msk}), Lessee($1^\lambda$, st)$\rangle$} We describe the interactive key generation protocol between the Lessor and the honest Lessee below. Note that in the actual protocol, the malicious Lessee may not follow the protocol.
\begin{enumerate}
    \item The Lessor samples $s^*_j \in \{0,1\}^{\ell}$ uniformly. Then, the Lessor generates (${\sf TEPRF.key}_{j,0},{\sf TEPRF.key}_{j,1}$) $\leftarrow$ $\mathsf{TEPRF.KG}(1^\lambda, s^*_j)$. The Lessor generates (${\sf CS.msk}_j, {\sf CS.vk}_j$) $\leftarrow$ ${\sf CS.Setup}(1^\lambda)$. The Lessor generates ${\sf CS.sigk}_{j,0}$ $\leftarrow$ {\sf CS.Constrain}(${\sf CS.msk}_j, f[{\sf TEPRF.key}_{j,0}])$ and ${\sf CS.sigk}_{j,1} \leftarrow {\sf CS.Constrain}({\sf CS.msk}_j, f[{\sf TEPRF.key}_{j,1}])$.
    \item The Lessor and the Lessee proceed as in ${\sf PKE\text{-}cSKL}.{\sf KG}$ from the $2$nd step, except for the following points:
    \begin{itemize}
        \item The algorithm uses ${\sf TEPRF.key}_{j,0}||{\sf CS.sigk}_{j,0}$ (resp. ${\sf TEPRF.key}_{j,1}||{\sf CS.sigk}_{j,1}$) instead of ${\sf PKE.sk}_{j,0}$ (resp. ${\sf PKE.sk}_{j,1}$). Also, the names of registers change accordingly.
        \item The algorithm outputs $({\sf CS.vk}_j)_{j \in [2n]}$ instead of the public key.
    \end{itemize}
    
\end{enumerate}

\begin{remark}
    For the sake of clarity, we present the ideal quantum signing key $\ket{sigk}$ which consists of registers $({\sf A}_j, {\sf SK}_j,{\sf SIGK}_j, {\sf R}_j)$ as follows:
        \begin{equation}
            \label{eqn:ideal-key-state-ds}
            \begin{aligned}
                &\frac{1}{\sqrt 2}(\ket{0, {\sf TEPRF.key}_{j,0}, {\sf CS.sigk}_{j,0}, x_{j,0}} \pm \ket{1, {\sf TEPRF.key}_{j,1}, {\sf CS.sigk}_{j,1}, x_{j,1}})& q_B^i = 0 \\
                &\ket{0,{\sf TEPRF.key}_{j,0}, {\sf CS.sigk}_{j,0}, x_{j,0}}\ {\rm or}\  \ket{1,{\sf TEPRF.key}_{j,1}, {\sf CS.sigk}_{j,1}, x_{j,1}}& q_B^i = 1 \\
                &\frac{1}{2}(\ket{0, {\sf TEPRF.key}_{j,0}, {\sf CS.sigk}_{j,0}, x_{j,0}}\bra{0, {\sf TEPRF.key}_{j,0}, {\sf CS.sigk}_{j,0}, x_{j,0}}+  \\ &\ket{1, {\sf TEPRF.key}_{j,1}, {\sf CS.sigk}_{j,1}, x_{j,1}}\bra{1, {\sf TEPRF.key}_{j,1}, {\sf CS.sigk}_{j,1}, x_{j,1}})& q_B^i = 2 
            \end{aligned}
        \end{equation}
    We point out that the register $({\sf A}_{2i}, {\sf SK}_{2i}, {\sf R}_{2i})$
    and $({\sf A}_{2i+1}, {\sf SK}_{2i+1}, {\sf R}_{2i+1})$ are entangled for $q_B = 2$. The whole state is $V_{2i}V_{2i+1} \ket{\Phi_{b[1]b[0]}}_{{\sf A}_{2i} {\sf A}_{2i+1}}$ where $V_j$ is an isometry mapping from ${\cal H}_{{\sf A}_j}$ to ${\cal H}_{{\sf A}_j}\otimes {\cal H}_{{\sf SK}_j} \otimes {\cal H}_{{\sf R}_j}$ such that $V_j \ket{b}_{{\sf A}_j} = \ket{b, {\sf TEPRF.sk}_{j,b}, {\sf CS.sigk}_{j,b}, x_{j,b}}_{{\sf A}_j {\sf SK}_j {\sf R}_j}$ and $\ket{\Phi_{ab}} \coloneqq \frac{1}{\sqrt 2}(\ket{0a}+(-1)^b\ket{1(1-a)})$ is one of the four Bell states. We present the honest Lessee who outputs a quantum state almost identical to the ideal state, except for ${\rm negl}(\lambda)$ trace distance, in the proof of Evaluation correctness.
\end{remark}

{{\sf DS-cSKL}.Sign$(sigk,m)$}:
\begin{enumerate}
    \item Parse $sk \coloneqq (\{ q_A^i \}_{i \in [n]}, ({\sf A}_j, {\sf SK}_j, {\sf R}_j)_{j \in [2n]})$ and $m = m_0 ||  \dots || m_{2n-1}$. The registers $({\sf A}_j, {\sf SK}_j, {\sf R}_j)_{j \in [2n]}$ are holding the key state.
    \item Apply $C_{\sf sign,j}$ for $j \in [2n]$ where $C_{\sf sign,j}$ is a quantum circuit  as follows:
    \begin{enumerate}
        \item The circuit initializes the register ${\sf TEPRF.out}_j$ to $\ket{0}$. Then, the circuit applies the unitary $U_{{\sf Eval},j}$ to the register ${\sf TEPRF.SK}_j {\sf TEPRF.out}_j$ and measures ${\sf TEPRF.out}_j$ to obtain $t_j \in \{0,1\}$.
        \item The circuit runs ${\sf CS}.QSign({{\sf A}_j {\sf SK_j} {\sf CSK}_j}, m_j || t_j)$ and obtains a signature $\sigma_j$.
    \end{enumerate}
    \item Output $\sigma \coloneqq (\sigma_j, t_j)_{j \in [2n]}$.
\end{enumerate}

{{\sf DS-cSKL.VrfySign}$({\sf svk}, \sigma, m)$}:
\begin{enumerate}
    \item Parse $\sigma \coloneqq (\sigma_j, t_j)_{j \in [2n]}$ and $\mathsf{svk} \coloneqq ({\sf CS.vk}_{j})_{j \in [2n]}$ and $m = m_0 || \dots || m_{2n-1}$.
    \item The algorithm outputs $\bot$ if ${\sf CS.Vrfy}({\sf CS.vk}_j, m_j||t_j, \sigma_j) = \bot$ for some $j$. Otherwise, the algorithm outputs $\top$.
\end{enumerate}

{{\sf DS-cSKL}.Del$(sk)$} The algorithm is as follows.
\begin{enumerate}
    \item Parse $sk \coloneqq (\{ q_A^i \}_{i \in [n]}, ({\sf A}_j, {\sf SK}_j, {\sf CSK}_j, {\sf R}_j)_{j \in [2n]})$. The registers $({\sf A}_j, {\sf SK}_j, {\sf CSK}_j, {\sf R}_j)_{j \in [2n]})$ are holding the key state.
    \item For $i\in[n]$, measure the register $(A_{2i}, A_{2i+1})$ with $\{\mathcal A_{q_A^i}^a\}$, where $\{\mathcal A_{q_A^i}^a\}$ is Alice's measurement in \cref{tab:msg-strategy-DS-cSKL} when the question to it is $q_A^i$. Set $a_i$ to be the outcome.
    \item For $j\in[2n]$, measure every qubit of registers ${\sf SK}_j$, ${\sf CSK}_j$, ${\sf R}_j$ in Hadamard basis. Let the measurement outcome be $d_j, d_j^\prime, d_j^{\prime\prime}$, respectively.
    \item Output ${\sf cert}= (\{ a_i \}_{i\in [n]}, \{d_j, d_j^\prime, d_j^{\prime\prime}\}_{j \in [2n]})$.
\end{enumerate}

{{\sf DS-cSKL}.{\sf DelVrfy}$({\sf cert}, {\sf dvk})$} The algorithm is as follows.
\begin{enumerate}
    \item Parse 
    \begin{equation}
        {\sf cert} = (\{ a_i \}_{i\in [n]}, \{d_j, d_j^\prime, d_j^{\prime\prime}\}_{j \in [2n]})
    \end{equation}
    and 
    \begin{equation}
        \begin{aligned}
            {\sf{dvk}} \coloneqq (\{ q_B^i, b_i^\prime, q_A^i \}_{i \in [n]}, \{ {\sf TEPRF.key}_{j,0}, {\sf TEPRF.key}_{j,1}, \\
            {\sf CS.sigk}_{j,0}, {\sf CS.sigk}_{j,1}, e_{z,1}^j, z_j, x_{j,0}, x_{j,1}\}_{j \in [2n]: j \notin J_B})
        \end{aligned}
    \end{equation}
    \item Computes 
    \begin{equation}
        \label{eqn:DS-Z-error}
        \begin{aligned}
            e_{i,0}  =& d_{2i} \cdot ({\sf TEPRF.key}_{2i,0} \oplus {\sf TEPRF.key}_{2i,1}) \\
            &\oplus  d_{2i}^\prime \cdot ({\sf CS.sigk}_{2i,0} \oplus {\sf CS.sigk}_{2i,1}) \\ 
            &\oplus (d_{2i}^{\prime\prime} \oplus e_{z,1}^{2i}) \cdot (x_{2i,0} \oplus x_{2i,1}) \oplus z_{2i} \\
            e_{i,1}  =& d_{2i+1} \cdot ({\sf TEPRF.key}_{2i+1,0} \oplus {\sf TEPRF.key}_{2i+1,1}) \\
            &\oplus d_{2i+1}^\prime \cdot ({\sf CS.sigk}_{2i+1,0} \oplus {\sf CS.sigk}_{2i+1,1}) \\ 
            &\oplus (d_{2i+1}^{\prime\prime} \oplus e_{z,1}^{2i+1}) \cdot (x_{2i+1,0} \oplus x_{2i+1,1}) \oplus z_{2i+1}
        \end{aligned}
    \end{equation}
    and $a_i^\prime = postprocessing(q_A^i, a_i, e_{i,0}, e_{i,1})$ for $i \in [n]$. We remind the readers that $postprocessing$ is defined in \cref{dfn:postprocessing}.
    \item If $MSG(q_A^i,q_B^i,a_i^\prime, b_i^\prime) = 0$ for some $i\in[n]$, output $\bot$. Otherwise, output $\top$.
\end{enumerate}

\paragraph*{\bf Proof of signature verification correctness} We first prove that the ideal key state in \cref{eqn:ideal-key-state-ds} satisfies signature verification correctness. By the arguments for PRF-cSKL, we can see that $t_j \neq {\sf TEPRF.Eval}({\sf TEPRF.key}_{j,0}, m_j)$ with negligible probability. Thus we have $f[{\sf TEPRF.key}_{j,b}](m_j||t_j) = 1$ for $b \in \{0,1\}$ with overwhelming probability. By the correctness of ${\sf CS}$, ${\sf CS.Vrfy}({\sf CS.vk}_{j}, m||t_j, \sigma_j) = \top$ with overwhelming probability. The algorithm $\text{\sf DS-cSKL}.{\sf VrfySign}$ outputs $\top$ with overwhelming probability.

Then, the same honest Lessee as PKE-cSKL (see \cref{sec:dfn-cSKL}) outputs a quantum state negligibly close to the ideal state. We complete the proof.

\paragraph*{\bf Proof of deletion verification correctness} The ideal key state in \cref{eqn:ideal-key-state-ds} is the same as the ideal key state for PKE-cSKL, except that it uses ${\sf TEPRF.key}_{j,0}$, ${\sf CS.sigk}_{j,0}$, ${\sf TEPRF.key}_{j,1}$, ${\sf CS.sigk}_{j,1}$ instead of ${\sf PKE.sk}_{j,0}, {\sf PKE.sk}_{j,1}$. Thus, the proof for PKE-cSKL (see \cref{sec:dfn-cSKL}) works for PRF-cSKL as well.
 
We present the proof of RUF-VRA security (see \cref{dfn:DS-RUF-VRA}) below. 
\begin{theorem}
    \label{thm:DS-cSKL-secure}
    The 2-party DS-cSKL above satisfies \cref{dfn:DS-RUF-VRA}.
\end{theorem}

We prove \cref{thm:DS-cSKL-secure} with a sequence of Hybrids.
\paragraph*{${\sf Hyb}_0(\lambda):$}

\begin{enumerate}

\item For each $j \in [2n]$, the challenger performs the following steps (as part of step 1 of $\text{\sf DS-cSKL.KG}$):

\begin{enumerate}

\item The challenger samples $s^*_j \in \{0, 1\}^\ell$ uniformly.

\item The challenger generates $({\sf TEPRF.key}_{j,0}, {\sf TEPRF.key}_{j,1}) \leftarrow {\sf TEPRF.KG}(1^\lambda, s^*_j)$.

\item The challenger generates $({{\sf CS.msk}_j}, {{\sf CS.vk}_j}) \leftarrow {\sf CS.Setup}(1^\lambda)$.

\item The challenger generates ${\sf CS.sigk}_{j,0} \leftarrow {\sf CS.Constrain}({{\sf CS.msk}_j}, F[{\sf TEPRF.key}_{j,0}])$ and ${\sf CS.sigk}_{j,1} \leftarrow {\sf CS.Constrain}({{\sf CS.msk}_j}, F[{\sf TEPRF.key}_{j,1}])$, where $F[{\sf TEPRF.key}]$ is a function that takes $(x, y) \in \{0, 1\}^\ell \times \{0, 1\}$ as input and outputs $1$ if ${\sf TEPRF.Eval}({\sf TEPRF.key}, x) = y$ and $0$ otherwise.

\item The challenger and the adversary take part in the $\Pi_{CSG}$ protocol (see \cref{dfn:CSG}). The challenger obtains $x_{j,0}, x_{j,1} \in \{0, 1\}^{l_{sk}}$ and $z_j \in \{0, 1\}$ after the execution.

\end{enumerate}

\item For each $i \in [n]$, the challenger performs the following steps (as part of step 2 of $\text{\sf DS-cSKL.KG}$):

\begin{enumerate}

\item The challenger samples $q_i^B, q_i^A \in \{0, 1, 2\}$ uniformly.

\item The challenger (as Verifier) and the adversary (as Prover) then engage in the ${\sf SimBob}\langle {\sf V}(1^\lambda, q_i^B), P(1^\lambda) \rangle$ protocol (as defined in Definition 3.1, page 11), from which the challenger obtains $b_i$.

\item The challenger then sends $q_i^A$ to the adversary.

\end{enumerate}

\item For each $j \in [2n]$, the challenger and the adversary repeat the following steps (as part of step 3 of $\text{\sf DS-cSKL.KG}$):

\begin{enumerate}

\item The challenger and the adversary engage in $\Pi_{CBQC}$ (as defined in Definition 2.5, page 10), where the challenger acts as the Client (C) and the adversary acts as the Sender (S). If $j \in J_B$, the challenger and the adversary run the protocol
\begin{equation}
\Pi_{CBQC} = \langle S(1^\lambda, Q, V), C(1^\lambda, Q, 1) \rangle
\end{equation}
Otherwise, the challenger and the adversary run the protocol
\begin{equation}
\Pi_{CBQC} = \langle S(1^\lambda, Q, V), C(1^\lambda, Q, 0) \rangle
\end{equation}
After the execution, the challenger obtains the classical string $(e_j^x, e_j^z)$. The challenger parses $e_j^x = e_{j,0}^x || e_{j,1}^x$ and $e_j^z = e_{j,0}^z || e_{j,1}^z$. The adversary sends a classical string $\omega$ to the challenger.

\item If $j \in {J_B}$, the challenger checks if $\omega \oplus e_{j,1}^x \neq x_{j,c}$, where $c = b_{\lfloor j/2 \rfloor}[j \bmod 2]$. If this condition is not met, the $\text{\sf DS-cSKL.KG}$ protocol aborts and the experiment ends and outputs $0$.

\item For each $j \in [2n]$, the challenger samples $r_{j,0}, r_{j,1}$. The challenger computes $h_{j,0} = {\sf Ext}(x_{j,0}, r_{j,0}) \oplus ({\sf TEPRF.key}_{j,0} || {\sf CS.sigk}_{j,0})$ and $h_{j,1} = {\sf Ext}(x_{j,1}, r_{j,1}) \oplus ({\sf TEPRF.key}_{j,1} || {\sf CS.sigk}_{j,1})$ (${\sf Ext}$ is the randomness extractor defined in \cref{dfn:CSG}). These values $(h_{j,0}, h_{j,1}, r_{j,0}, r_{j,1})$ are then sent by the challenger to the adversary.

\end{enumerate}

\item The adversary sends a classical string ${\sf cert} = (\{a_i\}_{i \in [n]}, \{d_j, d'_j, d''_j\}_{j \in [2n]})$ to the challenge.

\item The challenger runs $\text{\sf DS-cSKL.DelVrfy}({\sf cert}, {\sf dvk})$. This algorithm (as defined in \cref{dfn:DS-CSKL}) performs the following steps:

\begin{enumerate}

\item The challenger parses ${\sf cert}$ and ${\sf dvk}$.

\item For each $i \in [n]$, the challenger computes $e_{i,0}$ and $e_{i,1}$ using the formulas \cref{eqn:DS-Z-error}.
Then, the challenger computes $a'_i = {postprocessingA}(q_i^A, a_i, e_{i,0}, e_{i,1})$ (as defined in \cref{dfn:postprocessing}).

\item If ${\sf MSG}(q_i^A, q_i^B, a'_i, b_i^\prime) = 0$ for some $i \in [n]$, the challenger outputs $0$ and the experiment ends.

\end{enumerate}

\item The challenger samples an input $m^* \in {\cal M}$ uniformly. The challenger then sends $m^*$ and ${\sf dvk}$ to the adversary.

\item The adversary sends a classical string $\sigma'$ to the challenger.

\item The challenger runs $\text{\sf DS-cSKL.VrfySign}({\sf svk}, \sigma', m^*)$. This algorithm performs the following steps:

\begin{enumerate}

\item The challenger parses ${\sf svk} = ({{\sf CS.vk}_j})_{j \in [2n]}$, $m^* = m^*_0 || \dots || m^*_{2n-1}$ (where $m^*_j \in \{0, 1\}^\ell$), and $\sigma' = (t'_j, \sigma'_j)_{j \in [2n]}$.

\item If ${\sf CS.Vrfy}({{\sf CS.vk}_j}, m^*_j || t'_j, \sigma'_j) = \perp$ for some $j \in [2n]$, the challenger outputs $0$. Otherwise, the challenger outputs $1$.

\end{enumerate}

\end{enumerate}

{${\sf Hyb}_1$:} We define ${\sf Hyb}_1$ the same as ${\sf Hyb}_0$ except for:
\begin{itemize}
    \item The challenger outputs $0$ if $t_j^\prime \neq {\sf TEPRF.Eval}({\sf TEPRF.key}_{j,b_i[j \mod 2]}, m_j)$ for some $j \in J_B$.
\end{itemize}

\begin{lemma}
    \label{lem:DS-hyb0-hyb1-negl}
    For any (QPT) adversary, we have $|\Pr[{\sf Hyb}_0=1] - \Pr[{\sf Hyb}_1 = 1] | = {\rm negl}(\lambda)$.
\end{lemma}

\begin{proof}
    By the Equality property (see \cref{dfn:TEPRF}), we see that ${\sf TEPRF.Eval}({\sf TEPRF.key}_{j,0}, m_j) \neq {\sf TEPRF.Eval}({\sf TEPRF.key}_{j,1}, m_j)$ for only $m_j = s_j^*$. By \cref{dfn:selective_single_key_security}, when the adversary produces a valid signature $\sigma$, it holds except for negligible probability that $f[{\sf TEPRF.key}_{j,0}](m_j^*||t^\prime_j)=1$ or $f[{\sf TEPRF.key}_{j,1}](m_j^*||t^\prime_j)=1$. In other words, $t_j^\prime = {\sf TEPRF.Eval}({\sf TEPRF.key}_{j,0}, m_j^*)$ or $t_j^\prime = {\sf TEPRF.Eval}({\sf TEPRF.key}_{j,1}, m_j^*)$ holds except negligible probability. The change of Step $8(b)$ in ${\sf Hyb}_1$ affects the output distribution only when $m_j = s_j^*$, which happens with probability $2^{-l_{in}}$. We complete the proof.
    
\end{proof}

{${\sf Hyb}_2$:} We define ${\sf Hyb}_2$ the same as ${\sf Hyb}_1$ except for:
\begin{itemize}
    \item The challenger samples $m^*_j = s_j^*$ for $j \in J_B$.
\end{itemize}

\begin{lemma}
    \label{lem:DS-hyb1-hyb2-negl}
    For any QPT adversary, we have $|\Pr[{\sf Hyb}_1=1] - \Pr[{\sf Hyb}_2 = 1] | = {\rm negl}(\lambda)$.
\end{lemma}
\begin{proof}
    By the same argument as in the proof of \cref{lem:pke-hyb1-hyb2-diff}, we can see that the adversary in ${\sf Hyb}_1$ has only information about either ${\sf TEPRF.key}_{j,0}$ or ${\sf TEPRF.key}_{j,1}$ for $j \in J_B$. This can be proved by defining a ${\sf Hyb}_1^\prime$ in which ${\sf TEPRF.key}_{j,1-b_i[j \mod 2]}$ is sent to the adversary with a one-time pad. Then, by the Differing point hiding (see \cref{dfn:TEPRF}), the adversary cannot notice the change in ${\sf Hyb}_2$ and we have $|\Pr[{\sf Hyb}_1 = 1] - \Pr[{\sf Hyb}_2 = 1]| = {\rm negl}(\lambda)$.
    
\end{proof}

{${\sf Hyb}_3$:} We define ${\sf Hyb}_3$ the same as ${\sf Hyb}_2$ except for:
\begin{itemize}
    \item The specific abort condition found in Step 3(c) of ${\sf Hyb}_2$ is removed in ${\sf Hyb}_3$.
\end{itemize}

\begin{lemma}
    \label{lem:DS-hyb2-hyb3-leq}
    For any (QPT) adversary, we have $\Pr[{\sf Hyb}_2=1] \leq \Pr[{\sf Hyb}_3 = 1] $.
\end{lemma}
\begin{proof}
    In ${\sf Hyb}_3$, the abort condition in Step 3(c) is omitted. Thus, $\Pr[{\sf Hyb}_2 = 1] \leq \Pr[{\sf Hyb}_3 = 1]$.
     
\end{proof}

{${\sf Hyb}_4$:} We define ${\sf Hyb}_4$ the same as ${\sf Hyb}_3$ except for:
\begin{itemize}
    \item In Step 3(a) of ${\sf Hyb}_4$, the challenger and the adversary engage in $\Pi_{CBQC}$ with input $0$ for {\bf every} $j \in J_B$.
\end{itemize}

\begin{lemma}
    \label{lem:DS-hyb3-hyb4-negl}
    For any (QPT) adversary, we have $|\Pr[{\sf Hyb}_3=1] - \Pr[{\sf Hyb}_4 = 1]| = {\rm negl}(\lambda) $.
\end{lemma}
\begin{proof}[\cref{lem:DS-hyb3-hyb4-negl}]
    By the blindness of $\Pi_{CBQC}$ (\cref{dfn:cbqc}), we can see that changing the classical input from $1$ to $0$ does not affect the output distribution. This proves \cref{lem:DS-hyb3-hyb4-negl}.
    
\end{proof}

Finally, we bound $\Pr[{\sf Hyb}_4 = 1]$ using the computational CDP \cref{dfn:comp-CDP}.
\begin{lemma}
\label{lem:DS-hyb4-negl}
For any QPT adversary ${\sf Adv}^\lambda$, $\Pr[{\sf Hyb}_4(\lambda)=1]={\rm negl}(\lambda)$.
\end{lemma}

\begin{proof}
We bound $\Pr[{\sf Hyb}_4 = 1]$ using the Certified Deletion Property of the Magic Square Game (\cref{dfn:comp-CDP}). We can transform any QPT adversary ${\sf Adv}^\lambda$ (the adversary for ${\sf Hyb}_4$) into an adversary $\tilde P =(A_0, A_1)$ against the Certified Deletion Property of the Magic Square Game \cref{dfn:comp-CDP}.

{$A_0$}: The adversary $A_0$ (acting as Prover $\tilde{P}$ for the CCD game and simulating the challenger for an internal ${\sf Hyb}_4$ adversary ${\sf Adv}^\lambda$) performs the following steps:

\begin{enumerate}
\item Initialize the ${\sf Hyb}_4$ adversary ${\sf Adv}^\lambda$.
\item For each $j \in [2n]$:
\begin{enumerate}
\item Sample $s^*_j \leftarrow \{0,1\}^{l_{in}}$.
\item Sample $({\sf TEPRF.key}_{j,0}, {\sf TEPRF.key}_{j,1}) \leftarrow {\sf TEPRF.KG}(1^\lambda,s^*_j)$.
\item Generate $({{\sf CS.msk}_j},{{\sf CS.vk}_j}) \leftarrow {\sf CS.Setup}(1^\lambda)$.
\item Generate ${\sf CS.sigk}_{j,0} \leftarrow {\sf CS.Constrain}({{\sf CS.msk}_j}, f [{\sf TEPRF.key}_{j,0}])$ and ${\sf CS.sigk}_{j,1} \leftarrow {\sf CS.Constrain}({{\sf CS.msk}_j}, f [{\sf TEPRF.key}_{j,1}])$.
\item Participate in the $\Pi_{CSG}$ protocol with ${\sf Adv}^\lambda$ to obtain $x_{j,0}, x_{j,1} \in \{0,1\}^{l_{sk}}$ and $z_j \in \{0,1\}$.
\end{enumerate}
\item For each $i \in [n]$:
\begin{enumerate}
\item Engage in the ${\sf SimBob}\langle {\sf V}(1^\lambda, q_i^B), P(1^\lambda) \rangle$ protocol as the Prover (P). Forward the messages from the Verifier to ${\sf Adv}^\lambda$, the answers from ${\sf Adv}^\lambda$ to the Verifier.
\item Receive $q_i^A \in \{0, 1, 2\}$ from $V$.
\item Send $q_i^A$ to ${\sf Adv}^\lambda$.
\end{enumerate}
\item For each $j \in [2n]$:
\begin{enumerate}
\item Engage in the $\Pi_{CBQC}$ protocol as Client (C) with ${\sf Adv}^\lambda$ (Sender), using input $0$ for all $j \in [2n]$: $\Pi_{CBQC} = \langle S(1^\lambda, Q, V), C(1^\lambda, Q, 0) \rangle$. Obtain classical string $(e_j^x, e_j^z)$, parsing $e_j^x = e_{j,0}^x || e_{j,1}^x$ and $e_j^z = e_{j,0}^z || e_{j,1}^z$.
\item Receive a classical string $\omega$ from ${\sf Adv}^\lambda$.
\end{enumerate}
\item For each $j \in [2n]$, sample $r_{j,0}, r_{j,1}$. Compute $h_{j,0} = {\sf Ext}(x_{j,0}, r_{j,0}) \oplus ({\sf TEPRF.key}_{j,0}||{\sf CS.sigk}_{j,0})$ and $h_{j,1} = {\sf Ext}(x_{j,1}, r_{j,1}) \oplus ({\sf TEPRF.key}_{j,1}||{\sf CS.sigk}_{j,1})$. Send $(h_{j,0}, h_{j,1}, r_{j,0}, r_{j,1})$ to ${\sf Adv}^\lambda$.
\item Send the classical signature verification key ${\sf svk} := ({{\sf CS.vk}_j})_{j \in [2n]}$ to ${\sf Adv}^\lambda$.
\item Receive a classical string ${\sf cert} = (\{a_i\}_{i \in [n]}, \{d_j, d'_j, d''_j\}_{j \in [2n]})$ from ${\sf Adv}^\lambda$.
\item For each $i \in [n]$:
\begin{enumerate}
\item Compute $e_{i,0}$ and $e_{i,1}$ according to Eq. (118).
\item Compute $a'_i = {postprocessing_A}(q_i^A, a_i, e_{i,0}, e_{i,1})$.
\end{enumerate}
\item Output the internal state $st$ and $\{a^\prime_i\}_{i \in [n]}$ as the output.
\end{enumerate}

{$A_1$}: The adversary $A_1$ (acting as Prover $\tilde{P}$ for the CCD game and continuing to simulate the challenger for ${\sf Adv}^\lambda$) performs the following steps:

\begin{enumerate}
\item Receive the internal state $st$ from $A_0$ and the list $\{q_i^B\}_{i \in [n]}$ from the CCD Verifier $V$.
\item The adversary computes $b_i^\prime$ such that $b_i^\prime[q_A^i] = a_i[q_B^i]$ and the other bits are generated uniformly at random, where ${\rm par}(b_i^\prime) = 1$, for each $i\in[n]$.
\item Sample a message $m^* = m^*_0 || \dots || m^*_{2n-1}$ uniformly in a way such that $m^*_j = s^*_j$ for $j \in J_B$.
\item Let
\begin{equation}
    \begin{aligned}
        {\sf dvk} := (\{q_i^B, b_i^\prime, q_i^A\}_{i \in [n]}, \{{\sf TEPRF.key}_{j,0}, {\sf TEPRF.key}_{j,1}, \\
        {\sf CS.sigk}_{j,0}, {\sf CS.sigk}_{j,1}, e_{j,1}^z, z_j, x_{j,0}, x_{j,1}\}_{j \in [2n]: j \notin {J_B}})
    \end{aligned}
\end{equation}
Send $m^*$ and ${\sf dvk}$ to ${\sf Adv}^\lambda$.
\item Receive a message $\sigma' = (t'_j, \sigma'_j)_{j \in [2n]}$ from ${\sf Adv}^\lambda$.
\item Compute $b_i[j \mod 2] = c$ such that $t'_j = {\sf TEPRF.Eval}({\sf TEPRF.key}_{j,c},s^*_j)$ for $j \in J_B$. For $j \notin J_B$, sample random $b_i[j \mod 2] \in \{0,1\}$. This step is consistent with the implicit winning condition in ${\sf Hyb}_1$ (DS-cSKL), which implies that if ${\sf Adv}^\lambda$ wins, it must have produced a $t'_j$ consistent with the actual $b_i[j \mod 2]$ for $j \in J_B$.
\end{enumerate}

We can see that $A_0$ outputs a valid answer for the computational certified deletion property (\cref{dfn:comp-CDP}), with the same probability as ${\sf Adv}^\lambda$ outputs a valid certificate for key revocation in Step 7 (of $A_0$), by passing the verification check $MSG(q_i^A, q_i^B, a'_i, b_i) = 1$. This part of the proof is the same as that for \cref{lem:pke-hyb4-negl} and \cref{lem:prf-hyb4-negl}.

Then, we can see that whenever the adversary ${\sf Adv}^\lambda$ produces the correct $\sigma' = (t'_j, \sigma'_j)_{j \in [2n]}$ such that $ \text{\sf DS-cSKL.VrfySign(svk, $\sigma'$, $m^*$)} = \top$, it implies that $t'_j = {\sf TEPRF.Eval}({\sf TEPRF.key}_{j,b_i[j \mod 2]}, m^*_j)$ for $j \in J_B$ (as per the winning condition of ${\sf Hyb}_1$ for DS-cSKL, where $m^*_j = s^*_j$ for $j \in J_B$). Thus, the adversary $A_1$ (in its simulation of the challenger to ${\sf Adv}^\lambda$) can determine the correct $b_i[j \mod 2]$ values based on $t'_j$. Moreover, $A_1$ retrieves the original $b_i[q_A^i]$ values from the internal state $st$ (generated by $A_0$). For $i$ such that $q_i^B = 1$, the adversary obtains $b_i[0]$ and $b_i[1]$. With the information, the adversary can recover $b_i$ for $i$ such that $q_i^B = 1$. \footnote{Since ${\rm par}(b)=1$ for the valid answer of MSG, it suffices to recover $b$ using only a single bit other than $b[q_A^i]$. We can see that at least one of $b_i[0]$ and $b_i[1]$ differs from $b[q_A^i]$. } Therefore, whenever ${\sf Adv}^\lambda$ wins ${\sf Hyb}_4$, $\tilde P=(A_0,A_1)$ wins the ${\sf CCD}$ game.

Combining the two facts above, we have
\begin{equation}
\Pr[{\sf Hyb}_4 = 1] \leq \Pr[{\sf Adv}^\lambda\ \text{wins}\ {\sf CCD}] ={\rm negl}(\lambda)
\end{equation}

\end{proof}

Combining \cref{lem:DS-hyb0-hyb1-negl}, \cref{lem:DS-hyb1-hyb2-negl}, \cref{lem:DS-hyb2-hyb3-leq}, \cref{lem:DS-hyb3-hyb4-negl}, \cref{lem:DS-hyb4-negl}, we prove \cref{thm:DS-cSKL-secure}.

Since every component of our DS-cSKL can be constructed from CSGs, we obtain
\begin{theorem}
    Assuming the existence of CSGs and the hardness of the SIS problem, there exists DS-cSKL satisfying RUF-VRA security (see \cref{dfn:DS-RUF-VRA})
\end{theorem}

\section{Construction of \cref{dfn:CSG}}
\label{appdx:CSG}
In this section, we construct CSG with randomness extraction, using Trapdoor Claw-Free functions (TCFs) and their dual-mode variant. First, we introduce the definition of TCFs below.

\begin{definition}[TCF family]
    \label{dfn:TCF}
    Let $\lambda$ be a security parameter. Let ${\cal X}$ and ${\cal Y}$ be two finite sets. Let ${\cal K_F}$ be the set of keys. A family of functions
    \begin{equation}
        {\cal F} = \{f_{k,b}:{\cal X \rightarrow D_Y}\}_{k \in {\cal K_F}, b\in\{0,1\}}
    \end{equation}
    \footnote{$D_{\cal Y}$ is the set of distribution on a finite set ${\cal Y}$} is a TCF family if the following conditions hold:
    \begin{enumerate}
        \item \textbf{Efficient Function Generation.} There exists a QPT algorithm to generate the key $k\in{\cal K_F}$ and the trapdoor $t_k$:
        \begin{equation}
            (k,t_k) \leftarrow GEN_{\cal F}(1^\lambda)
        \end{equation}
        \item \textbf{Trapdoor Injective Pair.} The following conditions holds for all keys $k \in {\cal K_F}$:
        \begin{enumerate}
            \item Trapdoor. There exists a QPT algorithm $INV_{\cal F}$ such that for all $x\in {\cal X}$ and $y\in SUPP(f_{k,b}(x))$, $INV_{\cal F}(t_k, b, y)=x$. Note that this implies $SUPP(f_{k,b}(x))\cap SUPP(f_{k,b}(x^\prime)) = \emptyset$, for any $x \neq x^\prime$ that $x,x^\prime \in {\cal X}$ and $b\in\{0,1\}$.
            \item Injective pair. There exists a perfect matching ${\cal R}_k \in {\cal X \times X}$ such that $f_{k,0}(x_0) = f_{k,1}(x_1)$ iff $(x_0, x_1)\in {\cal R}_k$.
        \end{enumerate}
        \item \textbf{Efficient Range Superposition}. For every $k \in {\cal K_F}$ and $b\in \{0,1\}$, there is a function $f^\prime_{k,b}:{\cal X \rightarrow D_Y}$ such that the following holds:
        \begin{enumerate}
            \item For all $(x_0,x_1) \in {\cal R}_k$ and $y\in SUPP({f^\prime_{k,b}(x_b)})$, $INV_{\cal F}(t_k,b,y) = x_b$ and $INV_{\cal F}(t_k,{1-b},y) = x_{1-b}$
            \item There is an efficient {deterministic} algorithm $CHK$ such that $CHK(k,b,x,y) = 1$ when $y \in SUPP(f^\prime_{k,b}(x))$ and $CHK(k,b,x,y) = 0$ otherwise.
            \item For every $k \in {\cal K_F}$ and $b\in\{0,1\}$
            \begin{equation}
                E_{x \leftarrow U^{\cal X}}[H^2(f_{k,b}(x),f^\prime_{k,b}(x))] = {\rm negl}(\lambda)
            \end{equation}
            Furthermore, there is an efficient algorithm $SAMP_{\cal F}$ such that on input $k$ and $b$, it generates the following superposition
            \begin{equation}
                \frac{1}{\sqrt {\cal |X|}} \sum_{x\in{\cal X}, y\in{\cal Y}} \sqrt{(f^\prime_{k,b}(x))(y)}\ket{x}\ket{y}
            \end{equation}
        \end{enumerate}

        \item \textbf{Claw-free.} For any QPT adversary $\{\text{Adv}^{\lambda} \}_{\lambda \in \mathbb{N}}$,$$\Pr\left[
        f_{k,0}(x_0) = f_{k,1}(x_1) : \begin{array}{l} (k, t_k) \leftarrow \text{\sf Gen}_{\cal F}(1^{\lambda}) , \\({x_0, x_1}) \leftarrow \text{Adv}^{\lambda}(\text{k}) \end{array} 
        \right] = \text{\rm negl}(\lambda). $$ 
        % \item \textbf{Adaptive Hard Core}. The following conditions hold. For every $k \in {\cal K_F}$, there exists $w = poly(\lambda)$
        % \begin{enumerate}
        %     \item For all $b\in\{0,1\}, x\in{\cal X}$, there is a set $G_{k,b,x} \subseteq \{0,1\}^w$ such that $\Pr_{d \leftarrow {\{0,1\}^w}}[d \notin G_{k,b,x}] = {\rm negl}(\lambda)$. In addition, there is an efficient algorithm that checks for the membership in $G_{k,b,x}$ given $k,b,x$ and the trapdoor $t_k$.
        %     \item There is an efficient computable injection $I: {\cal X} \rightarrow \{0,1\}^w $, such that it can be inverted efficiently on its range, and the following holds. If
        %     \begin{equation}
        %         \begin{aligned}
        %             & H_k = \{(b,x_b,d,d\cdot (J(x_0)\oplus J(x_1))) | b\in\{0,1\}, (x_0,x_1)\in {\cal R}_k, d\in G_{k,0,x_0} \cap G_{k,1,x_1} \} \\
        %             & \bar H_k = \{ (b,x_b, d, c) | (b,x_b, d, 1-c) \in H_k\}
        %         \end{aligned}
        %     \end{equation}
        %     Then, for any QPT adversary ${\cal A}$
        %     \begin{equation}
        %         |\Pr_{(k,t_k) \leftarrow GEN_{\cal F}(1^\lambda)}[A(k) \in H_k]-\Pr_{(k,t_k) \leftarrow GEN_{\cal F}(1^\lambda)}[A(k) \in \hat H_k]|={\rm negl}(\lambda)
        %     \end{equation}
        % \end{enumerate}
    \end{enumerate}
\end{definition}

\begin{remark}
    We adopt the definition of NTCF from \cite{BCM+18} and remove the requirement for the adaptive hard-core property. Compared to the definition in \cite{BK24}, we require the function $f_{k,b}$ to be indexed by an additional bit $b \in \{0,1\}$. And we require the ${ INV}_{\cal F}$ algorithm to succeed with probability $1 - {\rm negl}(\lambda)$, which is a stronger requirement than the inverse polynomial probability. However, we inspect the prior works \cite{BCM+18,AMR22,GV24,PWY+25} have realized the strengthened requirements. Thus, the stronger requirement does not strengthen the cryptographic assumption.
\end{remark}

Let $\lambda$ be the security parameter, $l_r$ be the length of the extracted randomness. Then, we state the construction of CSGs with randomness extraction below.
\begin{itemize}
    \item The sender $S$ samples $(k_i,t_{k_i}) \leftarrow GEN_{\cal F}(1^\lambda)$ for $i \in [l_r]$. It sends $\{k_i\}_{i\in[l_r]}$ to the receiver $R$.
    \item The {\bf honest} receiver prepares $\frac{1}{\sqrt 2}(\ket{0}+\ket{1})$ in register $\sf A$. Then, for $i \in [l_r]$, the receiver runs $SAMP_{\cal F}$ where it uses $k_i$ as one classical input and the register $\sf A$ in superposition as the other input and outputs to register ${\sf X}_i {\sf Y}_i$. The receiver measures ${\sf Y}_i$ and sends the measurement outcome $y$ to the sender $S$.
    \item The sender computes $x_{i,0} \leftarrow INV_{\cal F}(t_{k_i},0,y_i)$ and $x_{i,1} \leftarrow INV_{\cal F}(t_{k_i},1,y_i)$. It outputs ${\bf x_0} = x_{1,0} || \dots || x_{l_r,0}$ and ${\bf x_1} = x_{1,1} || \dots || x_{l_r,1}$. The receiver outputs register $({\sf A}, {\sf X}_1, \dots, {\sf X}_{l_r})$.
\end{itemize}

\begin{lemma}
    The CSGs construction above satisfies {\sf Correctness} and {\sf Search Security} in \cref{dfn:CSG}.
\end{lemma}
\begin{proof}
    The proof is the same as that in \cite{BK24}.
    
\end{proof}

The {\sf Extract} algorithm is as follows
\begin{itemize}
    \item The algorithm takes as input $x \in \{0,1\}^{l_r n(\lambda)}$ and $r \in \{0,1\}^{l_r n(\lambda)}$.
    \item The algorithm parses $x \coloneqq x_1 || \dots || x_{l_r}$ and $r \coloneqq r_1 || \dots || r_{l_r}$. It computes $s_i = x_i \cdot r_{i}$ , where $\cdot$ represents the dot product, for $i \in [l_r]$.
    \item The algorithm outputs $s_1 || s_2 || \dots || s_{l_r}$.
\end{itemize}

\begin{lemma}
    The construction satisfies {\sf Randomness Extraction} in \cref{dfn:CSG}.
\end{lemma}
\begin{proof}
    Without loss of generality, we assume the adversary outputs ${\bf x}_0$. Then the adversary cannot output ${\bf x}_1$ except for negligible probability. By the Goldreich-Levin lemma, the adversary should not be able to distinguish $s_i$ and a random bit, for $i \in [l_r]$.
\end{proof}

\end{appendices}


\bibliography{refs}% common bib file

@book{IntroToModernCrypt,
  title={Introduction to modern cryptography: principles and protocols}}

@inproceedings{KN22,
  title={Functional encryption with secure key leasing},
  author={Kitagawa, Fuyuki and Nishimaki, Ryo},
  booktitle={International Conference on the Theory and Application of Cryptology and Information Security},
  pages={569--598},
  year={2022},
  organization={Springer},
  doi={10.1007/978-3-031-30545-0_20}
}

@misc{AKN+23,
      title={{Public Key Encryption with Secure Key Leasing}}, 
      author={Shweta Agrawal and Fuyuki Kitagawa and Ryo Nishimaki and Shota Yamada and Takashi Yamakawa},
      year={2023},
      eprint={2302.11663},
      archivePrefix={arXiv},
      primaryClass={quant-ph},
      doi ={10.1007/978-3-031-30545-0_20}
}

@inproceedings{CGJL23,
  title={{Quantum key leasing for PKE and FHE with a classical lessor}},
  author={Chardouvelis, Orestis and Goyal, Vipul and Jain, Aayush and Liu, Jiahui},
  booktitle={Annual International Conference on the Theory and Applications of Cryptographic Techniques},
  pages={248--277},
  year={2025},
  organization={Springer},
  doi={10.1007/978-3-031-91131-6_9}
}

@inproceedings{APV23,
  title={Revocable cryptography from learning with errors},
  author={Ananth, Prabhanjan and Poremba, Alexander and Vaikuntanathan, Vinod},
  booktitle={Theory of Cryptography Conference},
  pages={93--122},
  year={2023},
  organization={Springer},
  doi = {10.1007/978-3-031-48624-1_4}
}

@inproceedings{KMY24,
  title={A Simple Framework for Secure Key Leasing},
  author={Kitagawa, Fuyuki and Morimae, Tomoyuki and Yamakawa, Takashi},
  booktitle={Annual International Conference on the Theory and Applications of Cryptographic Techniques},
  pages={217--247},
  year={2025},
  organization={Springer},
  doi={10.1007/978-3-031-91131-6_8}
}

@inproceedings{PWY+25,
  title={{Adaptive Hardcore Bit and Quantum Key Leasing over Classical Channel from LWE with Polynomial Modulus}},
  author={Phan, Duong Hieu and Wen, Weiqiang and Yan, Xingyu and Zheng, Jinwei},
  booktitle={International Conference on the Theory and Application of Cryptology and Information Security},
  pages={185--214},
  year={2024},
  organization={Springer},
  doi = {10.1007/978-981-96-0947-5_7},
}

@article{WBMS16,
  title = {{Device-independent parallel self-testing of two singlets}},
  author = {Wu, Xingyao and Bancal, Jean-Daniel and McKague, Matthew and Scarani, Valerio},
  journal = {Phys. Rev. A},
  volume = {93},
  issue = {6},
  pages = {062121},
  numpages = {5},
  year = {2016},
  month = {Jun},
  publisher = {American Physical Society},
  doi = {10.1103/PhysRevA.93.062121}
}

@article{ABG07,
   title={{Device-Independent Security of Quantum Cryptography against Collective Attacks}},
   volume={98},
   ISSN={1079-7114},
   DOI={10.1103/physrevlett.98.230501},
   number={23},
   journal={Physical Review Letters},
   publisher={American Physical Society (APS)},
   author={Acín, Antonio and Brunner, Nicolas and Gisin, Nicolas and Massar, Serge and Pironio, Stefano and Scarani, Valerio},
   year={2007},
   month=jun }

@article{VV14,
  title = {{Fully Device-Independent Quantum Key Distribution}},
  author = {Vazirani, Umesh and Vidick, Thomas},
  journal = {Phys. Rev. Lett.},
  volume = {113},
  issue = {14},
  pages = {140501},
  numpages = {6},
  year = {2014},
  month = {Sep},
  publisher = {American Physical Society},
  doi = {10.1103/PhysRevLett.113.140501}
}

@article{ARV19,
author = {Arnon-Friedman, Rotem and Renner, Renato and Vidick, Thomas},
title = {{Simple and Tight Device-Independent Security Proofs}},
journal = {SIAM Journal on Computing},
volume = {48},
number = {1},
pages = {181-225},
year = {2019},
doi = {10.1137/18M1174726},
eprint = {https://doi.org/10.1137/18M1174726}
}

@ARTICLE{JMS20,
  author={Jain, Rahul and Miller, Carl A. and Shi, Yaoyun},
  journal={IEEE Transactions on Information Theory}, 
  title={{Parallel Device-Independent Quantum Key Distribution}}, 
  year={2020},
  volume={66},
  number={9},
  pages={5567-5584},
  doi={10.1109/TIT.2020.2986740}}

@inproceedings{KLVY22,
  title={{Quantum advantage from any non-local game}},
  author={Kalai, Yael and Lombardi, Alex and Vaikuntanathan, Vinod and Yang, Lisa},
  booktitle={Proceedings of the 55th Annual ACM Symposium on Theory of Computing},
  pages={1617--1628},
  year={2023},
  doi={10.1145/3564246.3585164}
}

@misc{NZ23,
      title={{Bounding the quantum value of compiled nonlocal games: from CHSH to BQP verification}}, 
      author={Anand Natarajan and Tina Zhang},
      year={2023},
      eprint={2303.01545},
      archivePrefix={arXiv},
      primaryClass={quant-ph},
      doi = {10.48550/arXiv.2303.01545}
}

@article{KT20,
  title={{Composably secure device-independent encryption with certified deletion}},
  author={Srijita Kundu and Ernest Y.-Z. Tan},
  journal={Quantum},
  year={2020},
  volume={7},
  pages={1047},
  doi={10.22331/q-2023-07-06-1047}
}

@article{FM18,
   title={{Local randomness: Examples and application}},
   volume={97},
   ISSN={2469-9934},
   DOI={10.1103/physreva.97.032324},
   number={3},
   journal={Physical Review A},
   publisher={American Physical Society (APS)},
   author={Fu, Honghao and Miller, Carl A.},
   year={2018},
   month=mar }

@inproceedings{KNP25,
  title={{PKE and ABE with collusion-resistant secure key leasing}},
  author={Kitagawa, Fuyuki and Nishimaki, Ryo and Pappu, Nikhil},
  booktitle={Annual International Cryptology Conference},
  pages={35--68},
  year={2025},
  organization={Springer},
  doi={10.1007/978-3-032-01881-6_2}
}

@article{BCM+18,
  title={A cryptographic test of quantumness and certifiable randomness from a single quantum device},
  author={Brakerski, Zvika and Christiano, Paul and Mahadev, Urmila and Vazirani, Umesh and Vidick, Thomas},
  journal={Journal of the ACM (JACM)},
  volume={68},
  number={5},
  pages={1--47},
  year={2021},
  publisher={ACM New York, NY},
  doi={10.1145/3441309}
}

@inproceedings{BK24,
  title={{On the power of oblivious state preparation}},
  author={Bartusek, James and Khurana, Dakshita},
  booktitle={Annual International Cryptology Conference},
  pages={575--607},
  year={2025},
  organization={Springer},
  doi = {10.1007/978-3-032-01878-6_19}
}

@inproceedings{KMP+25,
author = {Kulpe, Alexander and Malavolta, Giulio and Paddock, Connor and Schmidt, Simon and Walter, Michael},
title = {{A Bound on the Quantum Value of All Compiled Nonlocal Games}},
year = {2025},
isbn = {9798400715105},
publisher = {Association for Computing Machinery},
address = {New York, NY, USA},
doi = {10.1145/3717823.3718237},
booktitle = {Proceedings of the 57th Annual ACM Symposium on Theory of Computing},
pages = {222–233},
numpages = {12},
location = {Prague, Czechia},
series = {STOC '25}
}

@misc{CFNZ25,
      title={{A convergent sum-of-squares hierarchy for compiled nonlocal games}}, 
      author={David Cui and Chirag Falor and Anand Natarajan and Tina Zhang},
      year={2025},
      eprint={2507.17581},
      archivePrefix={arXiv},
      primaryClass={quant-ph},
      doi = {10.48550/arXiv.2507.17581}
}

@article{MP14,
   title={{Stronger Uncertainty Relations for All Incompatible Observables}},
   volume={113},
   ISSN={1079-7114},
   DOI={10.1103/physrevlett.113.260401},
   number={26},
   journal={Physical Review Letters},
   publisher={American Physical Society (APS)},
   author={Maccone, Lorenzo and Pati, Arun K.},
   year={2014},
   month=dec }

@article{Maz17,
   title={{A relação de incerteza de Maccone-Pati}},
   volume={39},
   ISSN={1806-1117},
   DOI={10.1590/1806-9126-rbef-2017-0014},
   number={4},
   journal={Revista Brasileira de Ensino de Física},
   publisher={FapUNIFESP (SciELO)},
   author={Maziero, Jonas},
   year={2017},
   month=may }

@misc{KPR＋25,
      title={{Quantitative Quantum Soundness for Bipartite Compiled Bell Games via the Sequential NPA Hierarchy}}, 
      author={Igor Klep and Connor Paddock and Marc-Olivier Renou and Simon Schmidt and Lucas Tendick and Xiangling Xu and Yuming Zhao},
      year={2025},
      eprint={2507.17006},
      archivePrefix={arXiv},
      primaryClass={quant-ph},
      url={https://arxiv.org/abs/2507.17006}, 
}

@misc{CMM+24,
      title={{A Computational Tsirelson's Theorem for the Value of Compiled XOR Games}}, 
      author={David Cui and Giulio Malavolta and Arthur Mehta and Anand Natarajan and Connor Paddock and Simon Schmidt and Michael Walter and Tina Zhang},
      year={2024},
      eprint={2402.17301},
      archivePrefix={arXiv},
      primaryClass={quant-ph},
      doi={10.48550/arXiv.2402.17301}
}

@misc{MPW24,
      title={{Self-testing in the compiled setting via tilted-CHSH inequalities}}, 
      author={Arthur Mehta and Connor Paddock and Lewis Wooltorton},
      year={2025},
      eprint={2406.04986},
      archivePrefix={arXiv},
      primaryClass={quant-ph},
      doi={10.48550/arXiv.2406.04986}
}

@misc{BVB+24,
      title={{Quantum bounds for compiled XOR games and $d$-outcome CHSH games}}, 
      author={Matilde Baroni and Quoc-Huy Vu and Boris Bourdoncle and Eleni Diamanti and Damian Markham and Ivan Šupić},
      year={2024},
      eprint={2403.05502},
      archivePrefix={arXiv},
      primaryClass={quant-ph},
      doi = {10.48550/arXiv.2403.05502}
}

@inproceedings{HJO+16,
  title={{Adaptively secure garbled circuits from one-way functions}},
  author={Hemenway, Brett and Jafargholi, Zahra and Ostrovsky, Rafail and Scafuro, Alessandra and Wichs, Daniel},
  booktitle={Annual International Cryptology Conference},
  pages={149--178},
  year={2016},
  organization={Springer},
  doi={10.1007/978-3-662-53015-3_6}
}

@misc{HK25,
  author = {Andrew Huang and Yael Tauman Kalai},
  title = {{Parallel Repetition for Post-Quantum Arguments}},
  howpublished = {Cryptology {ePrint} Archive, Paper 2025/1027},
  year = {2025},
  url = {https://eprint.iacr.org/2025/1027}
}

@inproceedings{BQSY24,
  title={{An efficient quantum parallel repetition theorem and applications}},
  author={Bostanci, John and Qian, Luowen and Spooner, Nicholas and Yuen, Henry},
  booktitle={Proceedings of the 56th Annual ACM Symposium on Theory of Computing},
  pages={1478--1487},
  year={2024},
  doi = {10.1145/3618260.3649603}
}

@article{GKK19,
  title={{Verification of quantum computation: An overview of existing approaches}},
  author={Gheorghiu, Alexandru and Kapourniotis, Theodoros and Kashefi, Elham},
  journal={Theory of computing systems},
  volume={63},
  number={4},
  pages={715--808},
  year={2019},
  publisher={Springer},
  doi={10.1007/s00224-018-9872-3}
}

@article{RUV12,
  title={Classical command of quantum systems},
  author={Reichardt, Ben W and Unger, Falk and Vazirani, Umesh},
  journal={Nature},
  volume={496},
  number={7446},
  pages={456--460},
  year={2013},
  publisher={Nature Publishing Group UK London},
  doi={10.1038/nature12035}
}

@article{McK16, 
    title ={{Interactive Proofs for BQP
 via Self-Tested Graph States}},
    volume={12},
   ISSN={1557-2862},
   DOI={10.4086/toc.2016.v012a003},
   number={1},
   journal={Theory of Computing},
   publisher={Theory of Computing Exchange},
   author={McKague, Matthew},
   year={2016},
   pages={1–42} }

@article{GKW15,
   title={{Robustness and device independence of verifiable blind quantum computing}},
   volume={17},
   ISSN={1367-2630},
   DOI={10.1088/1367-2630/17/8/083040},
   number={8},
   journal={New Journal of Physics},
   publisher={IOP Publishing},
   author={Gheorghiu, Alexandru and Kashefi, Elham and Wallden, Petros},
   year={2015},
   month=aug, pages={083040} }

@article{FHM18,
   title={{Post-hoc
 Verification of Quantum Computation}},
   volume={120},
   ISSN={1079-7114},
   DOI={10.1103/physrevlett.120.040501},
   number={4},
   journal={Physical Review Letters},
   publisher={American Physical Society (APS)},
   author={Fitzsimons, Joseph F. and Hajdušek, Michal and Morimae, Tomoyuki},
   year={2018},
   month=jan }

@inproceedings{FV14,
  title={{A multiprover interactive proof system for the local Hamiltonian problem}},
  author={Fitzsimons, Joseph and Vidick, Thomas},
  booktitle={Proceedings of the 2015 Conference on Innovations in Theoretical Computer Science},
  pages={103--112},
  year={2015},
    doi={10.1145/2688073.2688094}
}

@inproceedings{NV16,
   title={{A quantum linearity test for robustly verifying entanglement}},
   DOI={10.1145/3055399.3055468},
   booktitle={Proceedings of the 49th Annual ACM SIGACT Symposium on Theory of Computing},
   publisher={ACM},
   author={Natarajan, Anand and Vidick, Thomas},
   year={2017},
   month=jun, pages={1003–1015},
   collection={STOC ’17} }

@inproceedings{MNZ24,
  title={{Succinct arguments for QMA from standard assumptions via compiled nonlocal games}},
  author={Metger, Tony and Natarajan, Anand and Zhang, Tina},
  booktitle={2024 IEEE 65th Annual Symposium on Foundations of Computer Science (FOCS)},
  pages={1193--1201},
  year={2024},
  organization={IEEE},
  doi={10.1109/FOCS61266.2024.00078}
}

@inproceedings{AHH24,
  title={{Quantum Key-Revocable Dual-Regev Encryption, Revisited}},
  author={Ananth, Prabhanjan and Hu, Zihan and Huang, Zikuan},
  booktitle={Theory of Cryptography Conference},
  pages={257--288},
  year={2024},
  organization={Springer},
  doi = {10.1007/978-3-031-78020-2_9}
}

@article{MPV23,
  title={{Revocable quantum digital signatures}},
  author={Morimae, Tomoyuki and Poremba, Alexander and Yamakawa, Takashi},
  journal={arXiv preprint arXiv:2312.13561},
  year={2023}
}

@inproceedings{AMR22,
  title={Candidate trapdoor claw-free functions from group actions with applications to quantum protocols},
  author={Alamati, Navid and Malavolta, Giulio and Rahimi, Ahmadreza},
  booktitle={Theory of Cryptography Conference},
  pages={266--293},
  year={2022},
  organization={Springer},
  doi = {10.1007/978-3-031-22318-1_10}
}

@inproceedings{GV24,
  title={How to construct quantum fhe, generically},
  author={Gupte, Aparna and Vaikuntanathan, Vinod},
  booktitle={Annual International Cryptology Conference},
  pages={246--279},
  year={2024},
  organization={Springer},
  doi = {10.1007/978-3-031-68382-4_8}
}

@inproceedings{KLYY25,
  author       = {Fuyuki Kitagawa and
                  Jiahui Liu and
                  Shota Yamada and
                  Takashi Yamakawa},
  editor       = {Nadia Heninger and
                  Mike Rosulek},
  title        = {A Unified Approach to Quantum Key Leasing with a Classical Lessor},
  booktitle    = {Advances in Cryptology - {CRYPTO} 2026 - 46th Annual International
                  Cryptology Conference, Santa Barbara, CA, USA, August 17-20, 2026,
                  Proceedings, Part {V}},
  series       = {Lecture Notes in Computer Science},
  volume       = {16804},
  pages        = {90--121},
  publisher    = {Springer},
  year         = {2026},
  url          = {https://doi.org/10.1007/978-3-032-35409-9\_4},
  doi          = {10.1007/978-3-032-35409-9\_4},
  bibsource    = {dblp computer science bibliography, https://dblp.org}
}

\end{document}